\documentclass[a4paper,11pt]{article}
\usepackage{jheppub}
\usepackage{amsmath,amssymb}
\usepackage{natbib}
\usepackage{graphicx,subfigure}
\usepackage{hyperref}
\usepackage{braket}
\usepackage{comment}
\usepackage{leftidx}
\usepackage[dvipsnames]{xcolor}
\usepackage[varg]{txfonts}
\usepackage[normalem]{ulem}
\usepackage[utf8]{inputenc}
\usepackage[T1]{fontenc}

\newcommand{\be}{\begin{equation}}
 \newcommand{\ee}{\end{equation}}
 \newcommand{\bse}{\begin{subequations}}
 \newcommand{\ese}{\end{subequations}}
\newcommand{\ba}{\begin{eqnarray}}
\newcommand{\ea}{\end{eqnarray}}
\newcommand{\at}{\tilde{A}}
\newcommand{\bt}{\tilde{B}}
\newcommand{\ct}{\tilde{C}}

\newcommand{\rhoL}{\varrho}
\newcommand{\rL}{v}

\usepackage[normalem]{ulem}

\begin{document}


\title{Holographic Gubser flow: A combined analytic and numerical study}

\author[a,b]{Toshali Mitra,} \author[c,d]{Sukrut Mondkar,}\author[e,f]{Ayan Mukhopadhyay} \author[g]{and Alexander Soloviev}

\affiliation[a]{Asia Pacific Center for Theoretical Physics, Pohang, 37673, Korea}
\affiliation[b]{Department of Physics, Pohang University of Science and Technology, Pohang, 37673, Korea}
\affiliation[c]{Harish-Chandra Research Institute, A CI of Homi Bhabha National Institute, Chhatnag Road, Jhunsi, Prayagraj (Allahabad) 211019, India}
\affiliation[d]{Homi Bhabha National Institute, Training School Complex, Anushakti Nagar, Mumbai 400094, India}
\affiliation[e]{Instituto de F\'{\i}sica, Pontificia Universidad Cat\'{o}lica de Valpara\'{\i}so, Avenida Universidad 330, Valpara\'{\i}so, Chile}
\affiliation[f]{Department of Physics, Indian Institute of Technology Madras, Chennai 600036, India}
\affiliation[g]{Faculty of Mathematics and Physics, University of Ljubljana, Jadranska ulica 19, SI-1000 Ljubljana, Slovenia}
\emailAdd{toshali.mitra@apctp.org}
\emailAdd{sukrutmondkar@hri.res.in} 
\emailAdd{ayan.mukhopadhyay@pucv.cl}
\emailAdd{alexander.soloviev@fmf.uni-lj.si}
\date{\today}

\abstract{Gubser flow is an evolution with cylindrical and boost symmetries, which can be best studied by mapping the future wedge of Minkowski space (R$^{3,1}$) to dS$_3$ $\times$ $\mathbb{R}$ in a conformal relativistic theory. Here, we sharpen our previous analytic results and validate them via the first numerical exploration of the Gubser flow in a holographic conformal field theory. 

Remarkably, the leading generic behavior at large de Sitter time is free-streaming in transverse directions and the sub-leading behavior is that of a color glass condensate. We also show that Gubser flow can be smoothly glued to the vacuum outside the future Minkowski wedge generically given that the energy density vanishes faster than any power when extrapolated to early proper time or to large distances from the central axis. 
We find that at intermediate times the ratio of both the transverse and longitudinal pressures to the energy density converge approximately to a fixed point which is hydrodynamic only for large initial energy densities. 
We argue that our results suggest that the Gubser flow is better applied to collective behavior in jets rather than the full medium in the phenomenology of heavy ion collisions and can reveal new clues to the mechanism of confinement. 
}

\maketitle

\section{Introduction}

Gubser flow \cite{Gubser:2010ui,Gubser:2010ze} is a primary example of an evolution in a relativistic quantum field theory in which an expanding plasma cannot remain in a hydrodynamic regime (i.e. close to local thermal equilibrium) for a long time. The Gubser flow is boost invariant with rotational symmetry about an axis. These symmetries are further enhanced in a conformal field theory. As a consequence of the symmetries governing the expansion, the Knudsen number, the ratio of the temperature gradients (say in a putative perfect fluid regime) to the mean free path, is larger than unity at late time in the central region of the flow, implying that the gradient expansion of hydrodynamics breaks down. 

Although Gubser flow is distinguished by special symmetries, it is theoretically interesting to ask how such an evolution can be created in a quantum gauge theory, and whether such an evolution can lead to a new way to probe the quantum dynamics of the gauge theory at scales where the gauge coupling is strong, perhaps revealing new insights into confinement. A specific question in this regard is whether the final non-hydrodynamic stage of evolution of Gubser flow in the central region is independent of the possible initial conditions, and if so, then whether this non-equilibrium phase is determined by the fundamental infrared dynamics of the gauge theory.

In fact, the original context of the Gubser flow has been heavy ion collisions in which the collision axis can naturally be the axis of symmetry of the Gubser flow. However, generic initial conditions should lead to hydrodynamization of the expanding matter produced by the collisions at least when the multiplicity is sufficiently high. It is therefore a valid question to ask which initial conditions may produce a Gubser flow not only in heavy ion collisions, but also perhaps in high multiplicity small systems collisions (such as p-p or p-Pb collisions). Furthermore, it is possible that Gubser flow can be relevant to substructures in the bulk flow such as jets. The other phenomenologically relevant questions are whether hydrodynamics can always describe the Gubser flow at intermediate times, and if we can indeed probe the final non-hydrodynamic phase of evolution in the central region of the Gubser flow. 

The goal of this work is to investigate these questions in holographic strongly interacting large N conformal gauge theories, such as $\mathcal{N} =4$ supersymmetric Yang-Mills (SYM) theory at large N and large 't Hooft coupling. For such theories (in $3+1$-dimensions), we can investigate the Gubser flow via solving the classical Einstein's equations for gravity in $4+1$ dimensions with a negative cosmological constant. 
Thus, in the present work,
\begin{enumerate}
    \item we establish analytic results for the generic late time behavior of the Gubser flow in the central region,
    \item determine the possible initial conditions of the Gubser flow and also generic peripheral behaviour of the flow analytically,
    \item  confirm these analytic results via numerical solutions of the dual gravitational equations,
    \item and study numerically under which conditions can the flow be described by hydrodynamics at intermediate times.
\end{enumerate}

In \cite{Banerjee:2023djb}, the methodology of analyzing Gubser flow in holographic large N strongly interacting theories was developed via the analysis of regular solutions of pure gravity in $4+1$ dimensions with a negative cosmological constant and with dS$_{3}$ $\times$ $\mathbb{R}$ set as the boundary metric. We develop the results of this work further to obtain general analytic results about the Gubser flow, especially regarding the large de Sitter time behavior and the possible initial conditions. We also utilize the methodology of \cite{Chesler:2010bi} to study the holographic Gubser flow via numerical relativity, and confirm our analytic results, and also examine whether we obtain hydrodynamics or novel attractors (for reviews, see \cite{Soloviev:2021lhs,Jankowski:2023fdz}) generally at intermediate times in holographic Gubser flow. 

Furthermore, we  will be able to extract some important insights for quantum chromodynamics (QCD), the fundamental theory of the strong nuclear force, which are of phenomenological relevance for realistic heavy ion collisions. Additionally, as discussed in Sec.~\ref{Sec:Disc}, our results are relevant for fresh insights into the fundamental nature of the holographic correspondence, particularly in the context of bulk (wedge) reconstruction from boundary data and for understanding quantum thermodynamics of holographic quantum matter in de Sitter space. 

\subsection{A statement of results}
In this paragraph, we provide a quick summary of conformal Gubser flow. In a conformal field theory, the symmetries of the Gubser flow are enhanced to SO(3) $\times$ SO(1,1) \cite{Gubser:2010ze,Gubser:2010ui}. These symmetries can be made manifest by mapping the future wedge\footnote{In the context of heavy ion collisions, one is interested in the evolution of matter formed by the collisions in the future wedge. Let $z$ be the longitudinal axis of collisions. Idealizing the colliding nuclei as planar shocks, the collision happens on the plane $z=0$ at $t =0$. The future wedge is the region $\vert z\vert < t$, the causal future of the plane of collisions. The proper time is $\tau = \sqrt{t^2 - z^2}$ and the rapidity coordinate is $\eta = {\rm arctanh} (z/t)$.}  of the Minkowski space R$^{3,1}$ to dS$_{3}$ $\times$ $\mathbb{R}$ (the product of three-dimensional de Sitter space and the real line) via a combination of Weyl and coordinate transformations. The de Sitter factor, dS$_3$, involves a contracting and then expanding S$^2$ which manifests the SO(3) symmetries. The $\mathbb{R}$ factor is the rapidity coordinate on which the SO(1,1) boost transformation is manifested simply as a translation. The symmetries imply that the energy density can only be a function of $\rho/L$, the ratio of the de Sitter time $\rho$ to $L$, the minimal radius of S$^2$. Furthermore, when translated to Minkowski space, $\rho/L$ is a function of $qx_\perp$ and $q\tau$, where $x_\perp$ is the transverse distance from the central axis, $\tau$ is the proper time in the future wedge and $q$ is a state-dependent scale (which plays the role of the dimensionful constant $\lim_{\tau\rightarrow\infty}\tau^{4/3}\varepsilon(\tau)$ for Bjorken flow). It is useful to employ the dimensionless de Sitter time $\sigma = \exp(\rho/L)$ to readily capture the physical features of the Gubser flow.

The first result of our work is that Gubser flow in strongly interacting conformal large N holographic field theories always evolves to a transverse free-streaming phase at late de Sitter time (i.e. at large proper time in the central region of the future wedge of R$^{3,1}$) in which the ratio of the transverse pressure ($P_T$) to the energy density ($\varepsilon$) vanishes while the ratio of the longitudinal pressure ($P_L$) to the energy density goes to unity. This phase is attained irrespective of the initial conditions. We establish this result analytically and confirm this numerically (see Fig.~\ref{fig:late_time_2} for numerically obtained evolutions of $P_T/\varepsilon$ and $P_L/\varepsilon$ for various initial conditions with a fixed initial energy density)\footnote{In de Sitter space, the conformal anomaly affects the energy momentum tensor but it disappears in flat Minkowski space. With the conformal anomaly subtracted from the energy-momentum tensor in de Sitter space, we obtain the energy-momentum tensor in Minkowski space just by Weyl scaling. When we refer to the energy-momentum tensor in de Sitter space, we mean what we obtain after the state-independent conformal anomaly is subtracted. In this case, $P_T/\varepsilon$ and $P_L/\varepsilon$ are the same in de Sitter and Minkowski spaces.}. Furthermore, we analytically determine all possible sub-leading corrections to this behavior in the late de Sitter time expansion, which is an expansion in powers of $\sigma^{-1}$. The generic sub-leading behavior is that of a color glass condensate.\footnote{If the first sub-leading term was the dominant contribution, then we would have obtained $\varepsilon = P_T = -P_L$ at late de Sitter time as in the color glass condensate phase which can be described by the perturbative physics of saturated gluons \cite{Epelbaum:2013ekf,Muller:2019bwd,Berges:2020fwq}.} 

In \cite{Banerjee:2023djb}, it has been shown that one can solve pure Einstein's equations perturbatively in powers of $\sigma^{-1}$ at large $\sigma$ (i.e. large $\rho/L$) such that we obtain a regular future horizon at each order in this expansion. The regularity of the future horizon determines the possible powers of $\sigma^{-1}$  but allows the coefficients of this expansion to be arbitrary. However, the analytic (normalizable) solution corresponding to the leading power was not found in \cite{Banerjee:2023djb} (only a pure gauge solution was found). The subleading solutions explicitly found in \cite{Banerjee:2023djb} assuming the leading coefficient to be zero were rational functions of the radial coordinate in ingoing Eddington-Finkelstein gauge. Here, we find the leading physical solution explicitly which interestingly is not a rational function of the radial coordinate. This leading solution gives the generic transverse free-streaming behavior at late de Sitter time. Numerically, we confirm not only the leading transverse free-streaming but also the subleading color glass condensate like behavior for generic states.

Curiously, the late de Sitter time transverse free-streaming also occurs generically for Gubser flow in kinetic theories in the relaxation time approximation ignoring particle production \cite{Denicol:2014xca,Denicol:2014tha,Behtash:2017wqg,Martinez:2017ibh,Denicol:2018pak,Chattopadhyay:2018fzy,Behtash:2019qtk,Dash:2020zqx}. However, kinetic theory does not provide a good approximation to the quantum dynamics in this regime as the expanding plasma has diluted sufficiently. Furthermore, our derivation of the transverse free-streaming for holographic conformal field theories indicates that the behavior of the expanding plasma of QCD in this late de Sitter time regime (to the extent that one can use holographic models for QCD) should be determined by the fundamental strongly coupled infrared dynamics and not the ultraviolet behavior of the theory for which kinetic theory provides a better approximation. We expect that the late de Sitter time behavior of the Gubser flow in QCD to be different from what we have found in holographic conformal theories, and the study of holographic confining and non-conformal theories \cite{Craps:2013iaa,Gursoy:2015nza,Attems:2016ugt,Attems:2016tby} and semi-holographic setups \cite{Iancu:2014ava,Ecker:2018ucc} could provide further insights (c.f. Sec.~\ref{Sec:Gub-IC} and Sec.~\ref{Sec:Disc}). 

The second key result of our work is to determine the possible initial conditions of the Gubser flow. The Gubser flow is distinguished because the symmetries imply that the behavior at a large transverse distance ($x_\perp$) from the central axis on the future wedge of R$^{3,1}$ at any fixed proper time $\tau$ is related to the behavior both at early de Sitter time $\sigma \sim 0$ (i.e. large negative $\rho/L$) and early proper time $\tau\sim 0$ at any fixed distance  ($x_\perp$) from the central axis. (Such a feature where symmetry connects early proper time behavior with peripheral behavior is absent in Bjorken flow which is homogeneous in transverse directions.) 

Therefore, demanding that the energy density should decay at large $x_\perp$ (which should be a feature of any physically realizable state), we can analytically extend the behavior of the energy density in the late de Sitter time regime determined via bulk regularity to the entire future wedge of R$^{3,1}$. We find that
\begin{itemize}
    \item the energy density decays at large $x_\perp$ on the future wedge faster than any power of $x_\perp^{-1}$, and
    \item the energy density vanishes in the limit $\tau \rightarrow 0^+$ on the future wedge or in the limit $\rho/L \rightarrow -\infty$ (i.e. $\sigma \rightarrow 0^+$) faster than any power of $\tau$ and $\sigma$, respectively.
\end{itemize}
These imply that the holographic Gubser flow can be smoothly glued to the vacuum outside of the future wedge, and that it cannot be created from asymptotic states (e.g. by colliding two gravitational shock waves \cite{Chesler:2010bi} as in the case of holographic Bjorken flow). The non-trivial evolution in the Gubser flow on the future wedge is sustained by self-consistent pressure gradients. These results had been already established earlier in \cite{Banerjee:2023djb} assuming that the leading coefficient vanishes, and here we validate them more generally with a choice of bases for which the perturbative expansion converges on the entire future wedge. 

We also gather strong numerical support for the first statement above that the energy density decays at large $x_\perp$ on the future wedge faster than any inverse power of $x_\perp$ at any fixed proper time $\tau$. Although we cannot directly verify the other two statements regarding the behavior at early de Sitter time and at early proper time, these follow from the large $x_\perp$ behavior utilizing the symmetries of the Gubser flow as mentioned above. As discussed primarily in Sec.~\ref{Sec:Num}, the numerical results which establish these features for generic initial conditions are significant because our analytic arguments for the behavior on the entire future wedge (although not for the late de Sitter time regime) depends on the choice of a class of bases which may not capture all states exhibiting Gubser flow.

Our results suggest that the Gubser flow cannot be applied to the whole medium created in heavy-ion collisions. Nevertheless, we argue in Sec.~\ref{Sec:Gub-jet} that the Gubser flow can be relevant for sub-structures such as jets, and particularly for understanding collective flow in high charge multiplicity jets, discovered by the ALICE collaboration, as suggested in \cite{Taghavi:2019mqz} and reinforced by analysis of experimental data in \cite{ALICE:2023lyr}. We will specifically argue that if we insist that the Gubser flow provides a good approximation  only for parts of the full evolving medium, then it can be embedded within non-trivial phases created by shock wave collisions. In Sec.~\ref{Sec:Disc}, we argue that further studies in this direction can lead to new non-perturbative understanding of jets in quantum gauge theories.

Finally, we investigate whether a hydrodynamic regime appear in the holographic Gubser flow at intermediate time generally. We study a class of initial conditions which can be specified only by the initial values of $\varepsilon$ and $P_T/\varepsilon$ (or $P_L/\varepsilon$ which is related to $P_T/\varepsilon$ by a Ward identity) at a fixed negative value of the de Sitter time. As shown in Fig.~\ref{fig:late_time_2}, we find that remarkably $P_T/\varepsilon$ (and also $P_L/\varepsilon$) for different evolutions converge approximately to a common value at a certain $\rho/L$ if we start with a fixed value of the energy density. As discussed in Sec.~\ref{Sec:Non-Hydro}, this convergence implies that the Gubser flow hydrodynamizes at intermediate de Sitter times when the initial energy density is sufficiently large at least for a class of initial conditions. We further develop a more general phenomenological description of the holographic Gubser flow at intermediate de Sitter times, and discuss its relevance for the phenomenology of jets.

\subsection{Plan of the paper}
The plan of the paper is as follows. We provide some general preliminaries of Gubser flow, including a discussion of the symmetries and hydrodynamics, in Sec.~\ref{sec:setup}. We then provide the background for the holographic description of Gubser flow in Sec.~\ref{sec:holographic-setup}. Next, as a warm up, in Sec.~\ref{sec:scalar} we consider a toy computation of the evolution of a massless scalar field preserving the symmetries of the Gubser flow. We then turn our attention to the full case, providing analytic results in Sec.~\ref{Sec:Gen} and numerical verification of those results in Sec.~\ref{Sec:Num}. With these results in hand, we are able to comment on the possible intermediate hydrodynamization of the strongly coupled Gubser flow in Sec.~\ref{Sec:Non-Hydro}. In Sec.~\ref{Sec:Entropy}, we study entropy generation in holographic Gubser flow. Finally, in Sec.~\ref{Sec:Disc}, we conclude with a discussion on the fundamental implications of our results.

\section{Preliminaries}\label{sec:setup}
\subsection{The future wedge and dS$_3\times$ $\mathbb{R}$}

Let the axis of symmetry of the Gubser flow be $z$, and let $x$ and $y$ be the transverse spatial coordinates. It is useful to use the Milne coordinates which are namely, the proper time $\tau = \sqrt{t^2- z^2}$, the rapidity $\eta = {\rm arctanh}(z/t)$, the radial coordinate $x_\perp = \sqrt{x^2 + y^2}$ and the angular coordinate $\phi =\arctan (y/x)$ on the spatial plane transverse to the $z$-axis. 

In the context of heavy ion collisions, one is interested in the evolution of matter on the future wedge of R$^{3,1}$. Idealizing the colliding nuclei/protons as shocks, we can assume that the collision takes place on the plane $z=0$ at $t=0$. The region $\vert z\vert < t$, i.e. $\tau >0$ which is in the causal future of the plane of collisions is the future wedge of R$^{3,1}$. The Milne coordinates cover the entire future wedge. In Milne coordinates, the Minkowski metric takes the form
\begin{equation}\label{Eq:dsM}
    {\rm d}s_{M}^2 = - {\rm d}\tau^2 + \tau^2 {\rm d}\eta^2+ {\rm d}x_\perp^2+x_\perp^2 {\rm d}\phi^2.
\end{equation}
Although we will be agnostic about how the Gubser flow is initiated to determine the possible initial conditions without prejudice, we will still be studying the Gubser flow in the future wedge.

The \textit{entire} future wedge can be mapped to dS$_3\times$ $\mathbb{R}$ via a combination of coordinate and Weyl transformations. These coordinates make the full symmetry of the Gubser flow manifest. The rapidity coordinate $\eta$ parametrizes the  $\mathbb{R}$ factor. The three dimensional de Sitter space dS$_3$ is a contracting and then expanding two dimensional sphere S$^2$ as a function of the de Sitter time  $\rho$ extending from $-\infty$ to $\infty$. The radius of the sphere S$^2$ takes its minimum value $L$ at $\rho =0$. Furthermore, $\theta$ and $\phi$ are the coordinates of the sphere (recall $0 \leq \theta \leq \pi $ and $0\leq \phi < 2\pi$). The polar angle $\phi$ of S$^2$ is the same as the polar angle of the Milne coordinates, while the de Sitter time $\rho$ and the azimuthal angle $\theta$ of S$^2$ are related to the Milne coordinates via
 \begin{eqnarray}
      \frac{\rho}{L} &=& -  \,\,{\rm arcsinh}\left(\frac{1 - q^2\tau^2 + q^2 x_\perp^2}{2q\tau}\right), \label{Eq:rho}\\ \theta &=& {\rm arctan}\left(\frac{2qx_\perp}{1 + q^2\tau^2 - q^2 x_\perp^2}\right).\label{Eq:theta}
 \end{eqnarray}
Above, the parameter $q$ has mass dimension one. All physical quantities in dS$_3\times$ $\mathbb{R}$ when expressed as functions of the dimensionless time $\rho/L$ and in units $L=1$ can be mapped to physical quantities in the future wedge of R$^{3,1}$ when expressed as functions of the dimensionless coordinates $q\tau$ and $qx_\perp$, and in units $q=1$. This will be clear explicitly in the following subsection.

Using the coordinate transformations \eqref{Eq:rho} and \eqref{Eq:theta}, we can readily see that we obtain the dS$_3\times$ $\mathbb{R}$ metric from the Minkowski metric up to a Weyl scaling. Explicitly, the line element ${\rm d}{s}^2$ on dS$_3\times$ $\mathbb{R}$ is related to the Minkowski line element ${\rm d}s_M^2$ via
\begin{align}\label{Eq:dSds2}
    {\rm d}{s}^2  &= -{\rm d}\rho^2 + L^2 \cosh^2\left(\dfrac{\rho}{L}\right) \left({\rm d}\theta^2 + \sin^2\theta{\rm d}\phi^2\right) +L^2 {\rm d}\eta^2
    =  \frac{L^2}{\tau^2}{\rm d}s_M^2.
\end{align}
The full SO(3) $\times$ SO(1,1) $\times$ $\mathbb{Z}_2$ symmetry of the conformal Gubser flow is manifest in dS$_3\times$ $\mathbb{R}$ as should be clear from the explicit metric above. SO(3) forms the isometry group of the sphere S$^2$ covered by the $\theta$ and $\phi$ coordinates, while the boost symmetry SO(1,1) act as translations of the rapidity coordinate $\eta$. Finally $\mathbb{Z}_2$, the reflection over the plane of collisions (i.e. $z\rightarrow -z$), acts simply as $\eta \rightarrow - \eta$ in dS$_3\times$ $\mathbb{R}$. 

\subsection{Ward identities}

Consider a conformal quantum field theory in dS$_3$ $\times$ $\mathbb{R}$ where the spacetime metric is \eqref{Eq:dSds2}. The expectation value of the energy-momentum tensor in an arbitrary quantum state assumes the form
\begin{equation}\label{Eq:TmndS}
    T^\mu_{\,\,\nu} = t^\mu_{\,\,\nu} + \mathcal{A}^\mu_{\,\,\nu},
\end{equation}
where $t^\mu_{\,\,\nu}$ depends on the state (and also vanishes for the vacuum state). It is traceless
\begin{equation}\label{Eq:trt}
    t^\mu_{\,\,\mu} = 0.
\end{equation}
$\mathcal{A}^\mu_{\,\,\nu}$ is the vacuum expectation value of the energy-momentum tensor which is determined by the Weyl anomaly. Explicitly, 
\begin{align}\label{Eq:A}
 \mathcal{A}^\mu_{\,\,\nu} = \frac{{\rm N}^2}{32\pi^2 L^4}  {\rm diag}\left( 1,  1 ,   1 ,   -3\right)
\end{align}
for $\mathcal{N} =4$ super-Yang Mills theory with gauge group SU(N) \cite{Henningson:1998gx, Balasubramanian:1999re, Banerjee:2023djb}. Note that $\mathcal{A}^\mu_{\,\,\mu}$ is traceless in dS$_3$ $\times$ $\mathbb{R}$, and is separately conserved, i.e. $\nabla_\mu \mathcal{A}^\mu_{\,\,\nu} = 0$ with $\nabla$ being the covariant derivative constructed from the metric \eqref{Eq:dSds2}. Therefore, the conservation of energy and momentum implies that
\begin{equation}\label{Eq:consv}
  \nabla_\mu  t^\mu_{\,\,\nu} = 0.
\end{equation}

The SO(3) $\times$ SO(1,1) isometries imply that $t^\mu_{\,\,\nu}$ should assume the form
\begin{equation}\label{Eq:TmndS2}
    t^\mu_{\,\,\nu} = {\rm diag}\left(\varepsilon\left(\rho\right), P_T\left(\rho\right), P_T\left(\rho\right), P_L\left(\rho\right) \right)
\end{equation}
in Gubser flow. Note in the dS$_3$ $\times$ $\mathbb{R}$ coordinates, the fluid is {homogeneously} static  
at all times. The conformal Ward identities, namely \eqref{Eq:trt} and \eqref{Eq:consv} then give
\begin{align}
    &  {P}_T(\rho) =  -\frac{L}{2} \coth \left(\dfrac{\rho}{L}\right) {\varepsilon} '\left(\rho\right)-{\varepsilon} \left(\rho\right),\label{Eq:PT}\\
    &{P}_L(\rho) = L\coth \left(\frac{\rho}{L}\right) {\varepsilon}'\left(\rho\right)+3 {\varepsilon}\left(\rho\right)\label{Eq:PL}.
\end{align}
Here, $'$ denotes the derivative of the function w.r.t.~to its argument. Thus the Ward identities imply that the expectation value of the energy-momentum of a Gubser flow can be specified solely in terms of just the energy density $\varepsilon(\rho)$. The evolution of the energy density as a function of the de Sitter time is determined by the exact microscopic theory or a suitable phenomenological approximation of the latter. 

Let us denote the energy-momentum tensor of the Gubser flow in the future wedge of R$^{3,1}$ as $T^\mu_{M\nu}$. The conformal anomaly vanishes in flat space, and therefore $t^\mu_{\,\,\nu}$ in dS$_3$ $\times$ $\mathbb{R}$ to $T^\mu_{M\nu}$ in R$^{3,1}$ simply via
\begin{equation}\label{Eq:TMt}
    T^\mu_{M\nu} = \frac{L^4}{\tau^4}\, {\mathcal{P}t}^\mu_{\,\,\nu}
\end{equation}
with $\mathcal{P}X$ denoting the pullback of the tensor $X$ from dS$_3$ $\times$ $\mathbb{R}$ to R$^{3,1}$ under the coordinate transformations given by \eqref{Eq:rho} and \eqref{Eq:theta} while $L^4/\tau^4$ is the Weyl scaling derived from \eqref{Eq:dSds2}. 

It is convenient to define the energy density in  R$^{3,1}$ as 
\begin{align}
    \varepsilon_M =T^\mu_{M\nu} u_{d\mu}u_{d}^\nu,
\end{align} where $u_d^\mu$ has norm $-1$ and explicitly
\begin{align}\label{ud}
u_d^\mu \partial_\mu  = \frac{\partial}{\partial\rho},
\end{align} thereby generating translations along the de Sitter time. This is a natural definition since the fluid is static in a local inertial frame co-moving with $u_d$ (where the subscript $d$ is to remind the reader that this is a de Sitter vector). Note that we are not assuming any hydrodynamic description although the definition of $\varepsilon_M$ and $u_d$ are similar to the Landau-Lifshitz definition of the hydrodynamic energy density and fluid velocity. Here, our definition of $\varepsilon_M$ is guided simply by the symmetries of the Gubser flow.

Weyl transformation relates the energy density $\varepsilon_M(\tau, x_\perp)$ in  R$^{3,1}$ to the energy density $ \varepsilon(\rho)$ in dS$_3$ $\times$ $\mathbb{R}$ via
\begin{equation}\label{Eq:eMe}
    \varepsilon_M(\tau, x_\perp) =  \frac{L^4}{\tau^4} \varepsilon(\rho).
\end{equation}
Since conformal field theory has no intrinsic scale and $L$ is the only scale of the background, the dimensionless combination $L^4\varepsilon$ should be a function of only $\rho/L$, i.e.
\begin{equation}
     \varepsilon = L^{-4}f\left(\dfrac{\rho}{L}\right),
\end{equation}
and therefore from \eqref{Eq:rho} and \eqref{Eq:TMt} we obtain that
\begin{align}
    \varepsilon_M = \frac{1}{\tau^4} f\left(\dfrac{\rho}{L}(q\tau, qx_\perp)\right) &= q^4 \tilde{f}(q\tau, qx_\perp), \nonumber\\\tilde{f}(q\tau, qx_\perp) &:= \frac{1}{(q\tau)^4} f\left(\dfrac{\rho}{L}(q\tau, qx_\perp)\right).
\end{align}
Thus the arbitrary length scale $L$ is replaced by the arbitrary energy scale $q$ when physical observables are mapped from dS$_3$ $\times$ $\mathbb{R}$ to R$^{3,1}$. 

As an aside, it is interesting to note that the dS$_3$ $\times$ $\mathbb{R}$ metric \eqref{Eq:dSds2} is a solution of the four-dimensional gravitational equations
\begin{equation}
    R_{\mu\nu} - \frac{1}{2}R g_{\mu\nu} + \Lambda^{(4)} g_{\mu\nu} = 8\pi G_N^{(4)} \mathcal{A}_{\mu\nu}
\end{equation}
with $\Lambda^{(4)}$ and $G_N^{(4)}$ being the four-dimensional cosmological constant and Newton's constant, respectively, and $\mathcal{A}_{\mu\nu}$ given by \eqref{Eq:A}, provided
\begin{align}
    \Lambda^{(4)} = \frac{3}{2 L^2}, \qquad \text{and} \qquad 
  \frac{G_N^{(4)}}{L^2} = \frac{2 \pi}{{\rm N}^2}.
\end{align}
Therefore, the  dS$_3$ $\times$ $\mathbb{R}$ metric \eqref{Eq:dSds2} can be sustained by the combination of a positive cosmological constant and the vacuum energy momentum tensor of a conformal field theory which has a non-vanishing Weyl anomaly. Since ${G_N^{(4)}}/{L^2}\propto$ N$^{-2}$, this solution can be trusted in the semi-classical approximation when N is large in the conformal field theory.

\subsection{Gubser hydrodynamics}\label{Sec:CHydro}
It is useful to see how Gubser flow behaves under the assumption that it is governed by hydrodynamics. In this case the dynamics of $\varepsilon$ in dS$_3$ $\times$ $\mathbb{R}$ is simply given by the constitutive relations of hydrodynamics and the Ward identities  \eqref{Eq:PT} and \eqref{Eq:PL}. In a perfect conformal fluid, the energy density in dS$_3$ $\times$ $\mathbb{R}$ is simply  $$ \varepsilon(\rho)=\varepsilon_0 \cosh^{-8/3}(\rho/L),$$but including viscous corrections we obtain that $\varepsilon(\rho)$ goes to a constant in the limit $\rho\rightarrow\infty$ \cite{Gubser:2010ui, Gubser:2010ze}. Clearly the viscous correction dominates over the perfect fluid behavior at large $\rho$. This is simply because the Knudsen number is given by $\tanh \rho$ and thus becomes $\mathcal{O}(1)$ as $\rho\rightarrow\infty$. However, the Knudsen number is small at $\rho\sim 0$, and therefore the hydrodynamic derivative expansion can be trusted. Explicitly, we obtain that the energy density in a conformal fluid should take the form 
\begin{equation}\label{Eq:hydroexp}
    \varepsilon(\rho) =\varepsilon_0 - \frac{4}{3}\varepsilon_0 \frac{\rho^2}{L^2}+ \frac{16}{27} H_0 \varepsilon_0^{3/4} \frac{\rho^3}{L^4}+ \mathcal{O}(\rho^4),
\end{equation}
near $\rho \sim 0$ with $H_0 = \eta/\varepsilon_0^{3/4}$, a constant ($\eta$ is the shear viscosity and $\varepsilon_0 = \varepsilon(\rho=0)$). Note that the coefficient of the $\rho^2$ term above is determined by perfect fluidity and the viscosity gives sub-leading $\rho^3$ correction.

If a hydrodynamic description is indeed valid at $\rho\sim 0$, then it follows from \eqref{Eq:hydroexp} (and using \eqref{Eq:PL} and \eqref{Eq:PT}) that $P_L/\varepsilon = P_T/\varepsilon = 1/3$ should hold \textit{exactly} at $\rho = 0$. The departure of the ratios $P_L/\varepsilon$ and $P_T/\varepsilon$ from $1/3$ at $\rho = 0$ should not be attributed to viscous corrections, but to non-hydrodynamic effects. 

\section{Holographic setup}\label{sec:holographic-setup}
\subsection{The form of the bulk metric}

Any state in the universal sector of a holographic large $N$ strongly interacting gauge theory in $D$ dimensions has a dual representation in the form of a \textit{regular} asymptotically anti-de Sitter Einstein spacetime in $D+1$ dimensions  \cite{Maldacena:1997re,Witten:1998qj,Gubser:1998bc}. The asymptotic region of the $D+1$-dimensional geometry {contains} the microscopic data of the dual state, such as expectation value of single trace operators, which can be extracted from the on-shell gravitational action via a procedure called holographic renormalization \cite{Balasubramanian:1999re,Henningson:1998gx, deHaro:2000vlm,Skenderis:2002wp}. The ultraviolet divergences of the field theory are captured by the infrared divergences of the gravitational action, and can be removed by local counterterms on a radial cutoff which plays the role an ultraviolet cutoff in the field theory \cite{Balasubramanian:1999re,Henningson:1998gx,Skenderis:2002wp}. Indeed the emergent radial coordinate $r$ can be interpreted as the energy scale of the gauge theory \cite{Freedman:1999gp,Heemskerk:2010hk,Bredberg:2010ky,Faulkner:2010jy,Kuperstein:2011fn,Kuperstein:2013hqa,Behr:2015aat,Behr:2015yna,Mukhopadhyay:2016fre}. 

Let the boundary of the dual $D+1$-dimensional spacetime be at $r=0$. According to the holographic dictionary, the boundary metric, $g^{(b)}$, which is defined in terms of the bulk metric $G$ via
\begin{equation}
    \lim_{r\rightarrow 0} \frac{r^2}{l^2}G_{\mu\nu} \equiv g^{(b)}_{\mu\nu},
\end{equation}
with $\mu$ and $\nu$ denoting indices corresponding to the $D$ boundary coordinates, should always be the physical background metric where the dual gauge theory lives. We will consider $D=4$ in this paper and therefore study regular asymptotically AdS$_5$ (aAdS$_5$) spacetimes with appropriate boundary metrics.  The bulk cosmological constant is $\Lambda = - 6/l^2$, where $l$ is the asymptotic radius of the aAdS$_5$ spacetime. 

The spacetimes dual to the Gubser flow should be solutions of Einstein's equations
\begin{equation}\label{Eq:Einstein}
    R_{MN} - \frac{1}{2}R G_{MN} - \frac{6}{l^2} G_{MN} =0,
\end{equation}
such that they are aAdS$_5$, devoid of naked singularities and also have the global SO(3) $\times$ SO(1,1) $\times$ $\mathbb{Z}_2$ isometries. To study the Gubser flow in the future wedge of R$^{3,1}$, we can choose the five dimensional coordinates to be $r$ and the Milne coordinates $\tau$, $x_\perp$, $\phi$ and $\eta$. The boundary metric should be the Minkowski metric \eqref{Eq:dsM}. However, as discussed before, it will be convenient to study the Gubser flow by mapping the future wedge of R$^{3,1}$ to dS$_3$ $\times$ $\mathbb{R}$. In the holographic setup, we can choose the boundary metric to be \eqref{Eq:dSds2} and the dual bulk spacetime coordinates to be $r$, $\rho$, $\theta$, $\phi$ and $\eta$. The Weyl and coordinate transformations relating the two boundary metrics, namely the Minkowski metric \eqref{Eq:dsM} and the dS$_3$ $\times$ $\mathbb{R}$ metric \eqref{Eq:dSds2}, can be lifted to a bulk diffeomorphism which is always an improper residual gauge transformation for any choice of gauge fixing of the diffeomorphism symmetry of the gravitational theory \cite{deHaro:2000vlm,Skenderis:2002wp}\footnote{An improper gauge transformation affects boundary data whereas a proper gauge transformation does not. More generally, any global symmetry at the boundary is gauged in the bulk. A global symmetry transformation can be lifted to a improper residual gauge transformation for any choice of the gauge fixing in the dual bulk theory.}. 

The metric which is dual to the vacuum state of the gauge theory in dS$_3$ $\times$ $\mathbb{R}$ should be locally AdS$_5$ with boundary metric given by \eqref{Eq:dSds2}. Explicitly, this metric takes the following form in the ingoing Eddington-Finkelstein coordinates
\begin{align}\label{Eq:Vac-metric}
    &{\rm d}s^2 = -\frac{2 l^2}{r^2} {\rm d} r {\rm d} \rho -\frac{l^2}{r^2} \left( 1 -\frac{r^2}{L^2} \right) {\rm d}\rho^2 \nonumber\\&\qquad+\frac{l^2}{r^2} (L \cosh(\rho/L) + r \sinh (\rho/L))^2 \times \,\Big( {\rm d} \theta^2 + \sin^2 \theta {\rm d} \phi^2 \Big)+ \frac{l^2}{r^2} \,L^2{\rm d}\eta^2.
\end{align}
It can be checked that the above metric is a solution of \eqref{Eq:Einstein} and is indeed a maximally symmetric 5-dimensional spacetime with a constant negative scalar curvature. Nevertheless, this metric has a horizon at $r=L$ with a monotonically growing entropy as discussed in Sec.~\ref{Sec:Entropy}.\footnote{In fact, \eqref{Eq:Vac-metric} is similar to the topological black hole discussed in \cite{Casini:2011kv}. The horizon $r=L$ has a constant Hawking temperature $(2\pi L)^{-1}$.} Although the boundary metric \eqref{Eq:dSds2} is symmetric under de Sitter time-reversal $\rho \rightarrow -\rho$, the bulk metric does not have this symmetry due to the $\sinh(\rho/L)$ term in the $\theta\theta$ and $\phi\phi$ components of the bulk metric. The monotonically growing entropy is actually possible due to the absence of the time reversal symmetry. From the point of view of the future wedge of R$^{3,1}$, the monotonically growing entropy should naturally capture the entropy associated to observers co-moving with $u_d^\mu$ (see \eqref{ud}) which generates translations along de Sitter time. This entropy is thus similar to that associated with the accelerated observers comoving with Rindler time in Rindler spacetime. For more discussions see Sec.~\ref{Sec:Entropy}.

We choose the ingoing Eddington-Finkelstein coordinates as it is helpful to analyze the regularity of the spacetime in these coordinates. The general 5-dimensional metrics dual to the Gubser flow should preserve only SO(3) $\times$ SO(1,1) $\times$ $\mathbb{Z}_2$ symmetry unlike \eqref{Eq:Vac-metric} which has the full SO(4,2) symmetry. Therefore, the spacetime metric which captures a generic Gubser flow should take the form
\begin{align}\label{Eq:ABC-metric}
    &{\rm d}s^2 = -\frac{2 l^2}{r^2} {\rm d} r {\rm d} \rho -\frac{l^2}{r^2} \left( 1 -\frac{r^2}{L^2}  + A\left(\frac{r}{L},\frac{\rho}{L}\right)\right) {\rm d}\rho^2 \nonumber\\ &+\frac{l^2}{r^2}(L \cosh(\rho/L) + r \sinh (\rho/L))^2 \times e^{B\left(\frac{r}{L},\frac{\rho}{L}\right)}\,\Big( {\rm d} \theta^2 + \sin^2 \theta {\rm d} \phi^2 \Big)\nonumber\\&+ \frac{l^2}{r^2}  e^{C\left(\frac{r}{L},\frac{\rho}{L}\right)-2B\left(\frac{r}{L},\frac{\rho}{L}\right)}\,L^2{\rm d}\eta^2
\end{align}
in the ingoing Eddington-Finkelstein coordinates\footnote{Note that $l$, the asymptotic AdS radius, always appears as an overall multiplicative factor $l^2$ in the general spacetime metric as in \eqref{Eq:ABC-metric}. This is evident from Einstein's equations \eqref{Eq:Einstein} in which $l$ disappears when written as an equation for $G_{MN}/l^2$ instead of as an equation for $G_{MN}$.}. The above bulk metric manifestly inherits the SO(3) $\times$ SO(1,1) $\times$ $\mathbb{Z}_2$ symmetry from the boundary, and has three functions, namely $A$, $B$ and $C$ of $\rho/L$ which should be determined by solving Einstein's equations \eqref{Eq:Einstein}. Of course, $A = B = C=0$ gives the locally AdS$_5$ metric \eqref{Eq:Vac-metric} which is dual to the vacuum state in dS$_3$ $\times$ $\mathbb{R}$.
In what follows, we introduce the following notation for brevity
\begin{align}\label{notation}
    \varrho=\rho/L, \quad \text{and} \quad \rL=r/l.
\end{align}

With the metric \eqref{Eq:ABC-metric} dual to any arbitrary Gubser flow in dS$_3$ $\times$ $\mathbb{R}$, Einstein's equations \eqref{Eq:Einstein} can be readily solved in a radial expansion about the boundary $r = 0$ (i.e. $v=0$). The functions $A$, $B$ and $C$ which specify this metric assume a power series expansion
\begin{align}\label{Eq:ABC-rad}
    A(\rL,\rhoL) = \sum_{n=0}^\infty a_{(n)}(\rhoL) \rL^n,\,\, B(\rL,\rhoL) = \sum_{n=0}^\infty b_{(n)}(\rhoL) \rL^n,\,\,  C(\rL,\rhoL) = \sum_{n=0}^\infty c_{(n)}(\rhoL) \rL^n
\end{align}
with no log terms (unlike in Fefferman-Graham coordinates) although the boundary metric is curved (and the Weyl anomaly is non-vanishing). Explicitly, by solving Einstein's equations in the radial expansion, we obtain
\begin{align}\label{Eq:near_boundaryExpA}
 &A(\rL,\rhoL) =  a_{(1)}\left(\rhoL\right)\rL +\frac{1}{4} \left(a_{(1)}^2\left(\rhoL\right) -4 a_{(1)}'\left(\rhoL\right)\right) \rL^2 + a_{(4)}\left(\rhoL\right) \rL^4 + \ldots, \\
\label{Eq:near_boundaryExpB}
&B(\rL,\rhoL) = a_{(1)}(\rhoL) \cosh ^2(\rhoL) ~ \rL  -  \frac{1}{4} a_{(1)}(\rhoL) \left( a_{(1)}(\rhoL)  + 4 L^{-1} \tanh (\rhoL)\right) ~ \rL^2 \nonumber \\  & + \frac{1}{12} a_{(1)}(\rhoL) \Big( \left(a_{(1)}(\rhoL)  + 3 L^{-1} \tanh \left(\rhoL \right) \right)^2  + 3 L^{-2} \tanh ^2\left(\rhoL\right) \Big) ~ \rL^3
 + b_{(4)}(\rhoL) ~ \rL^4 +  \ldots ,\\
\label{Eq:near_boundaryExpS}
& C(\rL,\rhoL) = 3  a_{(1)}(\rhoL) ~ \rL - \frac{1}{4  }  a_{(1)}(\rhoL)\Big(8 L^{-1} \tanh (\rhoL) + 3 a_{(1)}(\rhoL) \Big) ~ \rL^2 \nonumber \\ & + \frac{1}{4} a_{(1)}(\rhoL) \left( \left(a_{(1)}(\rhoL) + 2 L^{-1}\tanh (\rhoL) \right)^2 + 4 L^{-2} \tanh ^2(\rhoL) \right) ~ \rL^3  + s_{(4)}(\rhoL) ~ \rL^4 + \ldots ,
\end{align}
where the $\ldots$ terms are simply polynomials of $a_{(1)}$, $a_{(4)}$, $b_{(4)}$, $s_{(4)}$ and their derivatives. Furthermore, $b_{(4)}$ and $s_{(4)}$ are determined by $a_{(4)}$ and $a_{(1)}$ \textit{algebraically} via the constraints of Einstein's equations:
\begin{align}\label{Eq:constraint}
& s_{(4)}\left(\rhoL\right) = -\frac{1}{32} a_{(1)}\left(\rhoL\right)  \Big(  a_{(1)}^2\left(\rhoL\right) + 4 L^{-2} \tanh ^2\left(\rhoL\right) \Big) \Big( 3  a_{(1)}\left(\rhoL\right) + 16 L^{-1} \tanh \left(\rhoL\right) \Big) \nonumber\\& \qquad\quad+ \frac{9}{8} a_{(1)}^2\left(\rhoL\right) L^{-2} \tanh^2 \left(\rhoL\right),
\nonumber \\
& a_{(4)}'\left(\rhoL\right) =  \frac{8 L^{-1}}{3}\tanh \left(\rhoL\right) \Big(b_{(4)}\left(\rhoL\right)-a_{(4)}\left(\rhoL\right)\Big)    + \frac{L^{-1}}{12} \Big(a_{(1)}^3\left(\rhoL\right) + 4 a_{(1)}\left(\rhoL\right)L^{-1}\tanh^2 \left(\rhoL\right) \Big) \Big( a_{(1)}\left(\rhoL\right)\nonumber\\&\qquad \quad + 8 L^{-1} \tanh \left(\rhoL\right) \Big)+ \frac{5}{6} a_{(1)}^2\left(\rhoL\right) L^{-3} \tanh^2\left(\rhoL\right).
\end{align}

Therefore, $A$, $B$ and $C$ and thus the metric \eqref{Eq:ABC-metric} can be specified fully by just two inputs: 
\begin{itemize}
    \item $a_{(1)}(\rhoL)$ which is a parameter for a proper residual gauge transformation, $v\rightarrow v + f(\rhoL)$, of the ingoing Eddington-Finkelstein gauge that does not affect boundary data and thus is not physical.
    \item $a_{(4)}(\rhoL)$ which determines the energy density $\varepsilon(\rhoL)$ in the dual state and whose form is determined via the absence of naked singularities in the bulk spacetime.
\end{itemize}
 We will show in the following subsection that indeed $a_{(1)}(\rhoL)$ does not affect boundary data and that $a_{(4)}(\rhoL)$ gives the dual energy density. 

\subsection{Holographic renormalization}

The expectation value of the energy momentum tensor in any state of the holographic field theory can be extracted from the asymptotic behavior of the dual bulk spacetime via holographic renormalization \cite{Henningson:1998gx,Balasubramanian:1999re,deHaro:2000vlm,Skenderis:2002wp}. The on-shell gravitational action with the boundary metric of the spacetime set as the physical background metric of the field theory is identified with the partition function of the field theory \cite{Witten:1998qj,Gubser:1998bc}. This action is divergent. Therefore, it is necessary to regularize by setting a radial cut-off which acts as the ultraviolet cut-off in the dual field theory. Renormalization is then performed by adding a counterterm action which is a local functional of the induced metric on the cut-off hypersurface (and other bulk fields evaluated on the hypersurface more generally) and then taking the radial cut-off to the boundary. The energy-momentum tensor of the dual state can be obtained by differentiating the renormalized gravitational action with respect to the background metric of the dual field theory (i.e. with respect to the boundary metric). Explicitly, this yields that $\langle T_{\mu\nu}\rangle$, the expectation value of the energy momentum tensor in the dual state is given by \cite{Balasubramanian:1999re}
\begin{eqnarray}\label{Eq:hol-ren}
    \langle T_{\mu\nu}\rangle = - \lim_{\epsilon \rightarrow 0} \frac{l^2}{8\pi G_N\epsilon^2}\left(K_{\mu\nu} - K \gamma_{\mu\nu} +\frac{3}{l}\gamma_{\mu\nu}  -\frac{l}{2}\left(\leftidx{^\gamma}{R}{_{\mu\nu}} -\frac{1}{2} \leftidx{^\gamma}{R}{}\gamma_{\mu\nu}\right) \right),
\end{eqnarray}
where $\gamma_{\mu\nu}$ is the induced metric and the $K_{\mu\nu}$ is the extrinsic curvature of the cut-off hypersurface $r = \epsilon$, and $\leftidx{^\gamma}{R}{_{\mu\nu}}$ is the Ricci tensor built out of the induced metric. The combination $\epsilon^{-2}(K_{\mu\nu} - K \gamma_{\mu\nu})$, which is essentially the Brown-York stress tensor of the cutoff hypersurface $r=\epsilon$, can be interpreted as the \textit{bare} energy-momentum tensor of the dual state. The remaining terms in \eqref{Eq:hol-ren} involves a volume counterterm proportional to the induced metric and a counterterm proportional to the Einstein tensor (built out of the induced metric), which cancel the $\epsilon^{-4}$ and $\epsilon^{-2}$ divergences of the bare energy-momentum tensor, respectively. (Note that both these counterterms are conserved with respect to the induced metric.) After adding the counterterms, we obtain a finite $\varepsilon \rightarrow 0$ limit. Here, we are using the minimal subtraction scheme which can be mapped to appropriate regularization schemes in the field theory \cite{Balasubramanian:1999re}.\footnote{Both in the field theory and in the holographic renormalizaiton scheme, the anomalous term in the trace of the energy-momentum tensor has a $\Box R$ ambiguity. In both cases, we can choose a regularization where this term is absent.}

In the Eddington-Finkelstein coordinates, there is no $\ln \epsilon$ divergence unlike in Fefferman-Graham coordinates. Note that \eqref{Eq:hol-ren} is invariant under proper diffeomorphisms which do not affect the boundary, so $ \langle T_{\mu\nu}\rangle$ is independent of the choice of bulk coordinates (Eddington-Finkelstein vs Fefferman-Graham) as long as the boundary metric is not affected \cite{Balasubramanian:1999re,deHaro:2000vlm}. 

Using the radial expansion of the bulk metric \eqref{Eq:ABC-metric} given by \eqref{Eq:near_boundaryExpA}, \eqref{Eq:near_boundaryExpB} and \eqref{Eq:near_boundaryExpS}, and the constraints \eqref{Eq:constraint}, we can readily obtain $\langle T_{\mu\nu}\rangle$ of the dual holographic Gubser from \eqref{Eq:hol-ren}. Explicitly, we obtain
\begin{align}\label{Eq:boundarygubser_stress_tensor}
\langle T^{\mu }_{\,\nu} \rangle =  {\rm diag}(\varepsilon(\rhoL), P_T(\rhoL), P_T(\rhoL), P_L(\rhoL))
+ \mathcal{A}^\mu_{\,\nu}
\end{align}
where 
\begin{align}\label{Eq:holeP}
  & \varepsilon(\rhoL) = - \frac{3l^3}{16 \pi G_N } a_{(4)}\left(\rhoL\right), \nonumber \\
 & P_T(\rhoL) = \frac{3l^3}{16 \pi G_N }  \left( \frac{1}{2} L\coth \left(\rhoL\right) a_{(4)}'\left(\rhoL\right)+  a_{(4)}\left(\rhoL\right)\right), \nonumber \\
 & P_T(\rhoL) =  - \frac{3l^3}{16 \pi G_N }\Big( L\coth \left(\rhoL\right) a_{(4)}'\left(\rhoL\right)+ 3 a_{(4)}\left(\rhoL\right) \Big).
\end{align} 
and $ \mathcal{A}^\mu_{\,\nu}$ is the vacuum expectation value of the energy-momentum given by
\begin{eqnarray}
{\mathcal{A}}^{\mu}_{\,\nu } = \frac{l^3}{64 \pi G_N } {\rm diag}\left( 1,  1 , 1,   -3 \right) L^{-4},
\end{eqnarray} 
which is a result of the Weyl anomaly. Firstly, we note that $a_{(1)}$ does not affect $\langle T^{\mu }_{\,\nu} \rangle$ which is determined solely by $a_{(4)}$ (the application of the constraints \eqref{Eq:constraint} eliminates $a_{(1)}$ from \eqref{Eq:hol-ren}). This is expected as $a_{(1)}$ is related to a (residual) diffeomorphism. Secondly, we note that we reproduce the form of the energy-momentum tensor given by \eqref{Eq:TmndS}, \eqref{Eq:TmndS2}, \eqref{Eq:PT} and \eqref{Eq:PL} once $l^3/G_N$ is identified with N$^2$ of the dual large N strongly interacting gauge theory up to an appropriate proportionality factor. For instance, in the case of $\mathcal{N} =4$ SYM with SU(N) gauge group, we should use the identification $$l^3/G_N = {\rm 2 N}^2/\pi$$ to reproduce \eqref{Eq:A}.\footnote{In a general background metric, the Weyl anomaly of $\mathcal{N} =4$ SYM with SU(N) gauge group gives \cite{Henningson:1998gx}
\begin{eqnarray}
{\mathcal{A}}^{\mu}_{\,\nu } = \frac{l^3}{128 \pi G_N } \Big(\frac{4}{3} R^\mu_{\,\nu} R - 2 R^{\mu}_{\,\kappa} R^\kappa_{\,\nu} - \delta^{\mu}_{\,\nu } \Big( \frac{1}{2} R^2 - R_{\kappa \sigma} R^{\kappa \sigma}\Big) \Big),
\end{eqnarray} 
where the curvatures are those of the physical background metric. It reduces to \eqref{Eq:A} when the background metric is set to the metric \eqref{Eq:dSds2} of dS$_3$ $\times$ $\mathbb{R}$.} With the latter identification, we obtain from \eqref{Eq:holeP} that
\begin{equation}
    \varepsilon(\rhoL) = - \frac{3 {\rm N}^2}{8 \pi^2} a_{(4)}\left(\rhoL\right)
\end{equation}
for the Gubser flow in $\mathcal{N} =4$ SU(N) SYM at large N and large 't Hooft coupling.

As mentioned before, we obtain the energy-momentum tensor of the Gubser flow of the field theory on the future wedge of R$^{3,1}$ by appropriate Weyl and coordinate transformations from the flows in dS$_3$ $\times$ $\mathbb{R}$ which preserve the SO(3) $\times$ SO(1,1) isometries. This transformation is represented holographically by an improper bulk diffeomorphism which transforms the boundary metric from \eqref{Eq:dSds2} to \eqref{Eq:dsM} whose explicit form is not necessary for present purposes. We recall that the anomalous term disappears when the background metric is \eqref{Eq:dsM}.

\section{The toy case of the massless scalar field}\label{sec:scalar}
\subsection{Generic behavior}
Before setting out to find physical regular solutions of non-linear Einstein's equations with SO(3) $\times$ SO(1,1) $\times$ $\mathbb{Z}_2$ symmetries corresponding to the Gubser flow in dual holographic conformal field theories, it is instructive to understand the general physical regular solutions of the linear Klein-Gordon equation for a massless scalar field exhibiting these symmetries. The Klein-Gordon equation for a massless scalar field in the background \eqref{Eq:Vac-metric} corresponding to the vacuum of the dual theory in dS$_3$ $\times$ $\mathbb{R}$ is explicitly
\begin{align}\label{Eq:KGMassless}
    r\left(1 - \frac{r^2}{L^2}\right)\partial_r^2\Phi(r, \rho) &-2 r \partial_r\partial_\rho \Phi(r, \rho) -\left(3 + \frac{r^2}{L^2} \right)\partial_r \Phi(r, \rho) \nonumber\\&+ \left(1 + \frac{2}{1 + \frac{r}{L}\tanh\left(\dfrac{\rho}{L}\right)}\right)\partial_\rho\Phi(r, \rho) = 0,
\end{align}
when $\Phi$ is restricted to be a function of $r$ and $\rho$ 
. We would be interested in solutions of this equation with the following properties:
\begin{itemize}
    \item the solutions should be regular in the sense that all derivatives should be finite at the horizon $r = L$,
    \item the solutions are normalizable, i.e. $\Phi(r =0, \rho) =0$, so that they represent intrinsic (sourceless) excitations of the dual theory, and
    \item the v.e.v. in the dual theory (which is essentially $\partial_r^4\Phi(r=0, \rho)$) decays at large $x_\perp$ when mapped to the future wedge of R$^{3,1}$ so that the dual state can be prepared via physically realizable operations on the ground state.
\end{itemize}
The methodology to find such solutions which is valid globally in dS$_3$ $\times$ $\mathbb{R}$ and the future wedge of R$^{3,1}$ in the dual theory has been developed in \cite{Banerjee:2023djb}. Here, we will review this. In the following subsection, we will show with examples that the solutions obtained via the perturbative method developed in  \cite{Banerjee:2023djb} focusing on early and late de Sitter time behavior can indeed converge globally at all values of the de Sitter time, and therefore on the entire future Minkowski wedge.

We first examine the large de Sitter time regime. One can readily see from \eqref{Eq:rho} that it corresponds to large $
\tau$ and the central region $x_\perp \ll \tau$ region in the future wedge of R$^{3,1}$ (see also top left figure in Fig.~\ref{Fig:Scalarlog}). Given that the background metric \eqref{Eq:Vac-metric} is a rational function of $\sigma \equiv \exp(\varrho)$, we can expect that the massless scalar field should behave as (recall $\varrho = \rho/L$ and $v = r/L$)
\begin{equation}\label{Eq:AnsSc}
    \Phi(r,\rho) = \sigma^{-\alpha}\sum_{n=0}^\infty \phi_n(v)\sigma^{-2n}
\end{equation}
at large $\rho$ with $\alpha >0$ (we expect the v.e.v. to decay as a result of the expansion of the S$^2$ factor). If the solution is $C^\infty$ at the horizon $r=L$, then near $r =L$ (i.e. $v = 1$) we should have
\begin{equation}\label{Eq:AnsHor}
    \phi_n(v) = \sum_{m=0}^\infty \lambda_{n,m} (1-v)^m,
\end{equation}
with $\lambda_{n,m}$ being pure numbers. We can choose $\lambda_{0,m}$ freely. Substituting \eqref{Eq:AnsSc} and \eqref{Eq:AnsHor} in \eqref{Eq:KGMassless}, we can readily obtain $\lambda_{n,m}$ in terms of $\lambda_{0,m}$ for all $m$ and $n>0$. Normalizability requires
\begin{equation}
    \phi_n(v=0) =0.
\end{equation}
To obtain the leading coefficient $\alpha$ in the late de Sitter time expansion, we need to solve $\lambda_{0,n}$ to high orders in $n$ and then compute $\phi_0(v=0)$. Explicitly, $\phi_0(v=0)$ takes the form
\begin{equation}
    \phi_0(v=0) =\lambda_{0,0} f_0(\alpha).
\end{equation}
Clearly, the possible values of $\alpha$ are the roots of $f_0(\alpha)$.\footnote{The late de Sitter time expansion \eqref{Eq:AnsSc} itself makes sense if $f_0(\alpha)$ has no negative real root. However it is in principle possible that $\alpha = \alpha_R \pm \iota \alpha_I$, with $\alpha_R$ real and positive, and $\alpha_I$ real. We do not find such complex roots of $f_0(\alpha)$.} Stable roots, $\alpha = 4 + 2k$ with $k = 0,1,2, \ldots$ are obtained to high numerical precision by solving $\lambda_{0,n}$ to orders $n >$ 200. This implies that for each non-negative integral value $k$, we should find a normalizable regular solution with an arbitrary coefficient setting $\alpha = 4 + 2k$. As the Klein-Gordon equation \eqref{Eq:KGMassless} reduces to ordinary differential equations for $\phi_n(v)$ when expanded in powers of $\sigma^{-2}$, we should obtain solutions of $\phi_n(v)$ which are normalizable ($\phi_{n}(v=0) =0$), regular at the horizon $v =1$ and furthermore have an independent coefficient at each order in $n$ corresponding to setting $\alpha = 4 + 2 n$. Explicitly,

\begin{align}\label{Eq:phi0-phi1}
   &\phi_0(v) ={o}_0 \frac{v^4}{(1+v)^4} ,\nonumber\\
   &\phi_1(v) = {o}_1 \frac{v^4(3+v^2)}{3(1+v)^6}+{o}_0 \frac{4v^6}{3(1+v)^6}  , \,\, {\rm etc.}
\end{align}
where $o_0$, $o_1$, etc. are arbitrary integration constants. It is convenient to choose the integration constants $o_n$ such that they are the corresponding coefficients of the $v^4$ term (the leading term in the asymptotic near-boundary expansion) of $\phi_n(v)$.

The vacuum expectation value (v.e.v.) of the marginal operator ${O}$ dual to the massless scalar field is the $v^4$ coefficient of $\Phi(r,\rho)$ up to a numerical factor times N$^2$. It follows that at large de Sitter time $\rho \gg L$ (i.e. $\varrho\gg 1$), the v.e.v. should behave as 
\begin{equation}\label{Eq:Ogen}
    \hat{O}(\rho) \sim \sigma^{-4} \sum_{k =0}^\infty {o}_k\sigma^{-2k},
\end{equation}
where ${o}_k$ are the integration constants corresponding to $\phi_k(v)$.  

It turns out that for $\alpha = 2$ in \eqref{Eq:AnsSc}, there exists the leading order solution
\begin{equation}
    \phi_0(v) = \frac{{o}_0}{2} \left( \frac{v^2}{(1+v)^2}  + \frac{(1-v)}{(1+v)} \ln(1-v^2)\right),
\end{equation}
which is regular (but not $C^\infty$) at the horizon and also normalizable. It is easy to check that if the above is the leading behavior at late de Sitter time, then the kinetic energy of the scalar field blows up at the horizon (as $\ln(1 - r)$) implying that the backreaction on the metric would lead to a naked singularity. Therefore, the $\exp(-2\varrho)$ behavior of the v.e.v. at late time is ruled out.

In Sec.~\ref{Sec:Num-1}, we will indeed see that numerical simulations of the massless scalar field in the background metric \eqref{Eq:Vac-metric} (dual to the vacuum) with generic regular initial conditions for the massless scalar field given by radial profiles that are $C^\infty$ functions in the domain $0\leq v<\infty$ at an early de Sitter time reproduce the predicted behavior \eqref{Eq:Ogen} at late de Sitter time. 

However, not all such solutions which follow the expansion \eqref{Eq:AnsSc} at late de Sitter time with $\alpha =4$ and resulting in the behavior \eqref{Eq:Ogen} of the v.e.v. are physical. To see this explicitly, we can examine some \textit{exact} solutions of the Klein-Gordon equation \eqref{Eq:KGMassless} obtained by summing over the infinite series giving the late de Sitter time expansion with proper choices of integration constants $o_k$. An example of such an exact normalizable solution of \eqref{Eq:KGMassless} with leading behavior $\sigma^{-4}$ at late de Sitter time is
\begin{equation}\label{Eq:Sol1}
    \Phi(r,\rho) = \Gamma_4 \frac{\sigma^{-4}v^4}{(1+v)^5}\frac{(1+v)^2\sigma^2 - 3-v^2}{1+\sigma^2 + v(\sigma^2 -1)},
\end{equation}
where $\Gamma_4$ is an arbitrary coefficient. Another such exact solution of \eqref{Eq:KGMassless} with leading behavior $\sigma^{-6}$ at late de Sitter time is
\begin{equation}\label{Eq:Sol2}
    \Phi(r,\rho) = \Gamma_6 \frac{\sigma^{-6}v^4}{(1+v)^7}\frac{(1+v)^2(3+v^2)\sigma^2 - 6-8v^2-v^4}{1+\sigma^2 + v(\sigma^2 -1)},
\end{equation}
with an arbitrary coefficient $\Gamma_6$, etc.

We observe that although both \eqref{Eq:Sol1} and \eqref{Eq:Sol2} are regular at the horizon $v =1$ for any finite value of $\rho$ (i.e. $\sigma >0$), they diverge at the horizon at $\sigma = 0$, i.e. $\rho = -\infty$. We may presume that this problem can be rectified by summing over a sequence of exact solutions. Nevertheless, the symmetries of Gubser flow tie the singular behavior at $\rho \rightarrow-\infty$ to singular behavior at any finite value of the proper time at $\tau$ on the future wedge (at the boundary of AdS$_5$). We note from \eqref{Eq:rho} that the limit of large $x_\perp$ for any fixed value of $\tau$ corresponds to $\rho\rightarrow-\infty$ (see also top left figure in Fig.~\ref{Fig:Scalarlog}). Therefore, if an exact solution is singular as $\rho \rightarrow -\infty$, we can expect that it should have singular behavior as $x_\perp \rightarrow\infty$ for any fixed value of $\tau$ in the future wedge at the boundary. This is indeed the case. The v.e.v. of the dual marginal operator corresponding to the exact solution \eqref{Eq:Sol1} is explicitly
\begin{equation}
    {O}(\sigma) \sim \sigma^{-4}\frac{\sigma^2 -3}{\sigma^2 -1}
\end{equation}
 The v.e.v. ${O}_M(\sigma)$ on the future wedge of R$^{3,1}$ in the dual theory can be obtained after the following necessary Weyl scaling of the v.e.v. in dS$_3$ $\times$ $\mathbb{R}$ : 
 \begin{equation}\label{Eq:O-OM}
 {O}_M(\tau,x_\perp) = L^4 \tau^{-4}{O}(\sigma(\tau,x_\perp)).
 \end{equation}
 Utilizing \eqref{Eq:rho}, we find that at large $x_\perp$, $$O_M\sim \frac{x_\perp^8}{\tau^8},$$implying that the v.e.v. diverges at large $x_\perp$ for any fixed $\tau$ on the entire future wedge. Furthermore as $\tau \rightarrow 0$, we obtain that $$O_M\sim - \frac{(1+ x_\perp^2)^4}{\tau^8},$$implying that the v.e.v. diverges as we approach $\tau \rightarrow 0$ at any fixed $x_\perp$. Both of these features also hold for the exact solution \eqref{Eq:Sol2}. 
 
 The lesson from the examination of the exact solutions \eqref{Eq:Sol1} and \eqref{Eq:Sol2} is that although the late de Sitter time expansion \eqref{Eq:AnsSc} is valid at large de Sitter time (large $\tau$ and $x_\perp \ll \tau$), it does not allow us to extrapolate to large $x_\perp$ at any fixed $\tau$ or to small $\tau$ at any fixed $x_\perp$ in R$^{3,1}$, since both of these lead us to the domain where $\varrho$ is large and negative (i.e. where $\sigma \rightarrow 0$). The issue is that the lack of reliable extrapolation does not allow us to determine whether physically realizable evolutions with SO(3) $\times$ SO(1,1) $\times$ $\mathbb{Z}_2$ symmetries require additional constraints on the coefficients \eqref{Eq:Ogen} of the late de Sitter time expansion. Especially if the v.e.v. does not decay at large $x_\perp$ in the future wedge of R$^{3,1}$, such evolutions cannot be created by physically realizable operations on the ground state where the v.e.v. vanishes. We expect that controlling the large $x_\perp$ behavior should also lead us to constrain the behavior of the flow at large and negative $\rho$ ($\sigma \rightarrow 0$), and hence also the initial conditions set in the limit $\tau\rightarrow 0^+$. 

These issues can be addressed if one does an alternative expansion of the bulk scalar field (proposed in \cite{Banerjee:2023djb}) which is as follows
\begin{align}\label{Eq:Phi-new-expansion}
    \Phi(r, \rho) = \sum_{n=0}^\infty \frac{ \phi^a_n(v)}{\cosh{(\varrho)}^{4+ 2 n}} + \sum_{n=0}^\infty  \frac{ \phi^b_n(v)}{\left(2 \cosh{(\varrho)}+\sinh{(\varrho)} \right)^{4 + 2 n}} 
\end{align}
Extracting the $v^4$ coefficient from above, we readily see that the above leads to the following expansion for the v.e.v. on the full dS$_3$ $\times$ $\mathbb{R}$:
\begin{align}\label{Eq:New-O-Expansion}
   {O}(\rho) = \sum_{k=0}^\infty  \left(\frac{o^a_k}{\cosh{(\varrho)}^{4+ 2 k}} + \frac{o^b_k}{\left(2 \cosh{(\varrho)}+\sinh{(\varrho)} \right)^{4 + 2 k}} \right) 
\end{align}
We note that the above expansions has the following features:
\begin{itemize}
    \item The expansions \eqref{Eq:Phi-new-expansion} and \eqref{Eq:New-O-Expansion} are compatible with the large de Sitter time expansions \eqref{Eq:AnsSc} (with $\alpha =4$) and \eqref{Eq:Ogen}, respectively, since each term in these expansion behaves as $\sigma^{-(4+2n)}\equiv \exp(-(4+2n)\varrho)$ with $n =0,1,2,\ldots$ at large $\varrho$. Explicitly, the coefficients of the late de Sitter time expansion are related to the expansion \eqref{Eq:New-O-Expansion} as follows:
     \begin{equation}\label{Eq:Late-time-coefficients}
 {o}_0 = 16 \left(o^a_0 + \frac{o^b_0 }{81}\right), \quad {o}_1 = 64\left(o^a_1 + \frac{o^b_1 }{729}-o^a_0 - \frac{o^b_0 }{243}\right), \,\, {\rm etc.}
\end{equation}
    \item Each term in the expansions \eqref{Eq:Phi-new-expansion} and \eqref{Eq:New-O-Expansion} decays at large $x_\perp$ for any fixed $\tau$, as ${\rm sech}(\varrho) \sim x_\perp^{-2}$ at large $x_\perp$ for any fixed $\tau$ , etc.
   \item We note that
 \begin{equation}
      0\leq \left(2 \cosh{(\varrho)}+\sinh{(\varrho)} \right)^{-1}\leq \frac{1}{\sqrt{3}}, \quad 0\leq {\rm sech}(\varrho) \leq 1.
  \end{equation}
  Therefore, \eqref{Eq:New-O-Expansion} should converge for \textit{for all} $\varrho$ if
  \begin{equation}
      \lim_{n\rightarrow\infty}\left\vert \frac{o^a_{n+1}}{o^a_n}\right\vert < 1, \quad \lim_{n\rightarrow\infty}\left\vert \frac{o^b_{n+1}}{o^b_n}\right\vert < 3.
  \end{equation}
  and may be expected to converge if 
  \begin{equation}
      \lim_{n\rightarrow\infty}\left\vert \frac{o^a_{n+1}}{o^a_n}\right\vert = 1, \quad \lim_{n\rightarrow\infty}\left\vert \frac{o^b_{n+1}}{o^b_n}\right\vert = 3.
  \end{equation}
  We will show that global convergence is indeed realized. We also expect that \eqref{Eq:Phi-new-expansion} should converge at least if $v$ is sufficiently small.
\end{itemize}
The upshot of these three properties is that the expansions \eqref{Eq:Phi-new-expansion} and \eqref{Eq:New-O-Expansion} are reliable at all $\varrho$ and can describe physically realizable evolutions with SO(3) $\times$ SO(1,1) $\times$ $\mathbb{Z}_2$ isometries in the future wedge of R$^{3,1}$. 

We will show in the next subsection that the expansion \eqref{Eq:Phi-new-expansion} does not describe all physically realizable evolutions as we can obtain other such evolutions with different choices of bases elements. The choice of bases in \eqref{Eq:Phi-new-expansion} and \eqref{Eq:New-O-Expansion} which satisfy the three desired properties mentioned above is indeed not unique. In fact, we can replace $$\left(2 \cosh{(\varrho)}+\sinh{(\varrho)} \right)^{-(4 + 2k)}$$in \eqref{Eq:Phi-new-expansion} with another bounded function such as $$\left(3\cosh{(\varrho)}+\sinh{(\varrho)} \right)^{-(4+2k)}.$$
We will explicitly show in the following subsection that different choices of bases give different physically realizable evolutions corresponding to the same and arbitrarily chosen coefficients of the late de Sitter time expansion \eqref{Eq:Ogen}. Furthermore, in Sec.~\ref{Sec:Num-1} we will show that the general lessons learnt from the expansions with a class of bases of the form \eqref{Eq:Phi-new-expansion} are indeed borne out by numerical simulations of the massless Klein-Gordon equation in the background \eqref{Eq:Vac-metric} (dual to the vacuum) with regular initial conditions set at a finite value of the de Sitter time.

Crucially, the expansion \eqref{Eq:New-O-Expansion} implies that at early de Sitter time $\sigma \rightarrow 0$ (i.e. $\rho \rightarrow -\infty$), the behavior of the v.e.v. should be
\begin{align}\label{Eq:Ogenearly}
   {O}(\rho) = \sigma^4 \sum_{k=0}^\infty  {o}^e_k \sigma^{2k},
\end{align}
if the coefficients $o^e_k$ are non-vanishing. Explicitly,

\begin{equation}\label{Eq:Early-time-coefficients}
    {o}^e_0= 16 \left(o^a_0 + o^b_0 \right),\quad {o}^e_1 = 64\left(o^a_1  +o^b_1  -o^a_0 - 3 o^b_0 \right), \,\, {\rm etc.}
\end{equation}
Each ${o}^e_k$ is a linear combination of $o^a_k$ and $o^b_k$, and lower order terms. Similarly, the behavior of the scalar field is
\begin{equation}\label{Eq:Phi-early}
    \Phi(r,\rho) = \sigma^4\sum_{n=0}^\infty \phi^e_n(v) \sigma^{2n},
\end{equation}
in the limit $\sigma \rightarrow 0$  with

\begin{equation}\label{Eq:Phi-early-coeffs}
    \phi^e_0(v)= 16 \left(\phi^a_0(v) + \phi^b_0(v) \right),\quad \phi^e_1(v) = 64\left(\phi^a_1(v)  +\phi^b_1(v)  -\phi^a_0(v) - 3 \phi^b_0(v) \right), \,\, {\rm etc.}
\end{equation}
The doubling of terms in \eqref{Eq:Phi-new-expansion} and \eqref{Eq:New-O-Expansion} compared to the corresponding late time expansions is simply because they parameterize also the early de Sitter time expansion. However, we will soon show by solving the Klein-Gordon equation in the late and early de Sitter time expansions that there is actually only one independent coefficient at each order in \eqref{Eq:Phi-new-expansion} and \eqref{Eq:New-O-Expansion}.

Explicitly, the large $x_\perp$ behavior of the v.e.v. on the future wedge of R$^{3,1}$ which follows from \eqref{Eq:New-O-Expansion} is (with $x = qx_\perp$ and $t= q\tau$):
\begin{align}\label{Eq:Olargex}
O_M(x_\perp, \tau)  &= q^4 L^4\Bigg( \frac{16 (o^a_0 +o^b_0)}{x^8} +  \frac{64 (o^a_0 +o^b_0)(-1 +t^2)}{x^{10}} \nonumber\\
& + \frac{32 (5(o^a_0 +o^b_0)+2 (- 7 o^a_0 - 9 o^b_0 + o^a_1 +o^b_1)t^2 + 5(o^a_0 +o^b_0)t^4)}{x^{12}} \nonumber\\
&+ \frac{64(-1 + t^2) (5(o^a_0 +o^b_0)+2 (- 11 o^a_0 - 17 o^b_0 +3 o^a_1 +3 o^b_1)t^2 + 5(o^a_0 +o^b_0)t^4)}{x^{14}} + \mathcal{O}(x^{-16})\Bigg)
\end{align}
provided the coefficients are non-vanishing. (Above we have used \eqref{Eq:rho}.) Indeed the v.e.v. decays at large $x_\perp$ at any fixed $\tau$. Since the large and negative $\varrho$ domain is the  large $x_\perp$ region at fixed $\tau$, and also small $\tau$ at fixed $x_\perp$, we can expect that the v.e.v. should not diverge at small $\tau$ at fixed $x_\perp$ given that the v.e.v. decays at large and negative $\varrho$ following \eqref{Eq:Ogenearly}. Indeed at early proper time $\tau\sim 0$, we find from \eqref{Eq:New-O-Expansion} that (with $x = qx_\perp$ and $t= q\tau$):

\begin{align}\label{Eq:Oearlytau}
    O_M (x_\perp, \tau)  &= q^4 L^4\Bigg( \frac{16 (o^a_0 +o^b_0)}{(x^2+1)^4} 
    \nonumber\\ &+ \frac{64(o^a_1 +o^b_1  + o^a_0(-1+x^2)+o^b_0 (-3 +x^2))}{(x^2+1)^6} t^2 + \mathcal{O}(t^3)\Bigg),
\end{align}
and therefore the v.e.v. should remain finite or vanish in the limit $\tau\rightarrow 0^+$. At late proper time $\tau\sim\infty$, we similarly obtain (with $x = qx_\perp$ and $t= q\tau$):

\begin{align}\label{Eq:Olatetau}
    O_M (x_\perp, \tau)  &= q^4 L^4\Bigg( \frac{16 (81 o^a_0 +o^b_0)}{81 t^8} 
   \nonumber\\ &+ \frac{64(729o^a_1 +o^b_1  + 729 o^a_0(-1+x^2) + 3 o^b_0 (-1 +3x^2))}{729 t^{10}}  + \mathcal{O}(t^{-12})\Bigg).
\end{align}

To obtain $\phi_n^a(v)$ and $\phi_n^a(v)$ in the expansion \eqref{Eq:Phi-new-expansion}, which can be demonstrated to converge globally for all $\varrho$, we simply solve the Klein-Gordon equation \eqref{Eq:KGMassless} in both early de Sitter time expansion (in power series of $\sigma^2$) and at late de Sitter time expansion (in power series of $\sigma^{-2}$). At each order in these expansions, we obtain two coupled second order ordinary differential equations for $ \phi^a_n(v)$ and $ \phi^b_n(v)$ whose solutions have four integration constants. Imposing these conditions, namely
\begin{itemize}
    \item $\Phi$ is normalizable so that $ \phi^a_n(0) = 0$ and $\phi^b_n(0) = 0$, and
    \item $\Phi$ is regular at the horizon $v = 1$ at late time, 
\end{itemize}
at each order, we can eliminate three integration constants, so that there is only one integration constant $\Gamma_n$ at each order. Remarkably, imposing that $\Phi$ is regular at the horizon at late de Sitter time automatically ensures that it is regular at the horizon for all de Sitter time, including in the limit $\rho\rightarrow -\infty$. We recall that this feature is not exhibited by the exact solutions \eqref{Eq:Sol1} and \eqref{Eq:Sol2}. Explicitly,

\begin{align}\label{Eq:phi-a-b}
  &  \phi^a_0(v) = \Gamma_0 \frac{v^4}{\left(1+ v\right)^4}, \,\,\,\,\,\,\,\,\,\,   \phi^b_0(v) = -\Gamma_0 \frac{v^4}{\left(1+ v\right)^4} \nonumber \\ 
  &  \phi^a_1(v) = \frac{v^4 \left(2 \Gamma_0 \left(v^2+728 v-1085\right)+729 \Gamma_1 \left(v^2+3\right)\right)}{728 (v+1)^6},  \nonumber \\     
  &\phi^b_1(v) = -\frac{3 v^4 \left(2 \Gamma_0 \left(243 v^2+728 v-119\right)+243 \Gamma_1 \left(v^2+3\right)\right)}{728 (v+1)^6}, \,\, {\rm etc.}
\end{align}

Note that $\Phi(r,\rho)$ is determined uniquely by the choices of the integration constants $\Gamma_n$. Obviously the v.e.v. is given by the globally valid expansion \eqref{Eq:New-O-Expansion} with
\begin{align}\label{Eq:oa-ob-explicit}
  &  o^a_0 = \Gamma_0, \,\,\,\,\,\,\, o^b_0 = -\Gamma_0 \nonumber \\ 
  &  o^a_1 = \frac{1}{728} (2187 \Gamma_1 - 2170 \Gamma_0), \,\,\,\,\,\,\,  o^b_1 = -\frac{3 (729 \Gamma_1 -238 \Gamma_0 )}{728}, \,\, {\rm etc.}
\end{align}
Of course, at late de Sitter time we can recover the expansion \eqref{Eq:AnsSc} with $\alpha =4$ and the coefficients $\phi_n(v)$ as determined before in \eqref{Eq:phi0-phi1} with $o_n$ (the coefficients of the late de Sitter expansion of the v.e.v. \eqref{Eq:Ogen}) explicitly being

\begin{align}\label{Eq:Late-O-coeffs}
 {o}_0= \frac{1280 \Gamma_0}{81}, \quad {o}_1 = 192 \Gamma_1 -\frac{20608 \Gamma_0}{81},\,\, {\rm etc.}
\end{align}

Remarkably, it follows from the explicit solutions \eqref{Eq:phi-a-b} of $\phi_n^a(v)$ and $\phi_n^b(v)$ that all the coefficients of the early de Sitter time expansion \eqref{Eq:Phi-early} vanish. We have checked that
\begin{equation}\label{Eq:phi-e-vanish}
    \phi^e_n(v) = 0, \,\, {\rm for} \,\, n= 0,1,2, \ldots, 
\end{equation}
to high orders using the relations \eqref{Eq:Phi-early-coeffs}. It follows that
\begin{equation}\label{Eq:phi-e-vanish-1}
     \lim_{\sigma\rightarrow 0^+} \sigma^{-n} \Phi(r,\sigma) =   \lim_{\rho\rightarrow -\infty} \exp(-n(\rho/L)) \Phi(r,\rho)   = 0, \,\, \,\, {\rm for}\,\, n = 0,1, 2, \ldots,
\end{equation}
i.e. $\Phi(r,\sigma)$ and all its derivatives w.r.t. $\sigma$ vanish in the limit $\sigma\rightarrow 0$ ($\varrho \rightarrow - \infty$). Therefore, 
\begin{align}\label{Eq:Remarkable-O}
   {o}^e_n = 0, \,\, \,\, {\rm for}\,\, n = 0,1, 2, \ldots,
\end{align}
as each $o^e_n$ is the $v^4$ coefficient of $\phi^e_n$, and thus the v.e.v. and all its derivatives vanish in the limit $\sigma \rightarrow 0$, i.e.
\begin{eqnarray}\label{Eq:Remarkable-O-n}
    \lim_{\sigma\rightarrow 0^+} \sigma^{-n} {O}(\sigma) =   \lim_{\rho\rightarrow -\infty} \exp(-n(\rho/L)) {O}(\rho)   = 0, \,\, \,\, {\rm for}\,\, n = 0,1, 2, \ldots.
\end{eqnarray}

It follows that the full global expansion of $\Phi$ given by \eqref{Eq:Phi-new-expansion} and that of the v.e.v. given by \eqref{Eq:New-O-Expansion} are fully determined by the corresponding late de Sitter time expansions which are \eqref{Eq:Phi-new-expansion} and \eqref{Eq:New-O-Expansion}, respectively.  To see this, we can note from \eqref{Eq:Early-time-coefficients} that the vanishing of all $o^e_n$, the coefficients of the early de Sitter time expansion, requires that $o^b_n$ are related to $o^a_m$ with $m\leq n$ via linear relations, which explicitly are 
\begin{equation}\label{Eq:ob-oa-rel}
    o^b_0 =- o^a_0, \quad o^b_1 = - o^a_1  - o^a_0, \,\, {\rm etc.}
\end{equation}
We can then interchange $o^a_n$ with $\Gamma_n$ via \eqref{Eq:oa-ob-explicit} and then $\Gamma_n$ with $o_n$, the coefficients of the late de Sitter time expansion \eqref{Eq:Ogen}, via \eqref{Eq:Late-O-coeffs}, and therefore the full function \eqref{Eq:New-O-Expansion} is indeed determined by the late de Sitter time expansion. The same is true for the bulk scalar field as \eqref{Eq:Phi-new-expansion} is determined uniquely by the coefficients $\Gamma_n$. However, the function $O(\rho)$ (and thus $\Phi(r,\rho)$) depends on the choice of bases in \eqref{Eq:New-O-Expansion} (and \eqref{Eq:Phi-new-expansion}). We will illustrate this explicitly in the next subsection by showing that although global convergence is achieved with a different choice of bases (e.g. by replacing $(2\cosh\varrho + \sinh\varrho)^{-(4+2k)}$ with $(3\cosh\varrho + \sinh\varrho)^{-(4+2k)}$ in \eqref{Eq:New-O-Expansion}), we can get a different $O(\rho)$ with the \textit{same} choices of coefficients of the late de Sitter time expansion \eqref{Eq:Ogen} of the v.e.v..

Since the $\sigma \rightarrow 0$ limit corresponds to $\tau\rightarrow 0^+$ for any fixed $x_\perp$, and also $x_\perp \rightarrow\infty$ for any fixed $\tau$ in R$^{3,1}$, we expect that the v.e.v. should decay in the future wedge of R$^{3,1}$ in both the limits $\tau\rightarrow 0^+$ and $x_\perp \rightarrow\infty$ given that $O(\sigma)$ and all its derivatives vanish in the limit $\sigma\rightarrow 0$. Indeed by substituting the relations between $o^b_n$ and $o^a_n$ given in \eqref{Eq:ob-oa-rel}, which are obtained from the vanishing of the coefficients $o^e_n$ of the early de Sitter time expansion \eqref{Eq:Ogenearly} of the v.e.v., into \eqref{Eq:Oearlytau} and \eqref{Eq:Olargex}, we readily find that 
\begin{eqnarray}\label{Eq:Remarkable-O-1}
    \lim_{\tau\rightarrow 0^+} \tau^{-n} O_M(\tau, x_\perp) =  \lim_{x_\perp\rightarrow \infty} x_\perp^{n} O_M(\tau, x_\perp) = 0, \,\, \,\, {\rm for}\,\, n = 0,1, 2, \ldots,
\end{eqnarray}
implying that the v.e.v. vanishes in the future wedge of R$^{3,1}$ as we extrapolate to $\tau = 0$ and to $x_\perp = \infty$ faster than any positive power of $\tau$ and $x_\perp^{-1}$, respectively. 

Repeating explicit calculations with different choices of bases (e.g. by replacing $(2\cosh\varrho + \sinh\varrho)^{-(4+2k)}$ with $(3\cosh\varrho + \sinh\varrho)^{-(4+2k)}$ in \eqref{Eq:New-O-Expansion}), we find that the results \eqref{Eq:Remarkable-O-n} and \eqref{Eq:Remarkable-O-1} remain the same. Therefore, these features are independent of the choice of bases and we have validated this by actual numerical simulations of the massless Klein-Gordon equation \eqref{Eq:KGMassless} in Sec.~\ref{Sec:Num-1}.

As $O_M(\tau, x_\perp)$ and all its derivatives w.r.t. $\tau$ vanish in the limit $\tau\rightarrow 0^+$, it follows that \textit{any evolution in the future wedge of R$^{3,1}$ with SO(3) $\times$ SO(1,1) $\times$ $\mathbb{Z}_2$ symmetries can be smoothly glued to the vacuum outside the future wedge in remaining R$^{3,1}$.}\footnote{More precisely, the reduced density matrix on the complement of the future wedge should be identical to that of the vacuum state.} Note that smooth gluing to the vacuum solution $\Phi =0$ on the edge of the future wedge $\tau =0$ (corresponding to $z = \pm t$ with $t >0$), requires not only $O_M(\tau, x_\perp)$ to vanish in the limit $\tau\rightarrow 0^+$ but all $\tau$-derivatives of $O_M(\tau, x_\perp)$ to vanish in this limit too. Another way to explicitly see this is to note from \eqref{Eq:phi-e-vanish-1} that $\Phi(r,\sigma)$ and all its $\sigma$-derivatives vanish in the limit $\sigma\rightarrow 0$ ($\varrho\rightarrow -\infty$). Since the Weyl and coordinate transformation which maps dS$_3$ $\times$ $\mathbb{R}$ to the future wedge of R$^{3,1}$ at the boundary can be lifted to (an improper) bulk diffeomorphism, it follows that the non-trivial solution of $\Phi(r,\rho(\tau, x_\perp))$ can be smoothly glued to the vacuum solution $\Phi =0$ outside the bulk future domain of influence of the future wedge at the boundary.

To summarize, we have studied solutions of the massless bulk scalar field with SO(3) $\times$ SO(1,1) $\times$ $\mathbb{Z}_2$ symmetries which are both normalizable and regular at the horizon, and also correspond to physical states in the dual theory where the v.e.v. decays at large distance from the central axis in the future wedge of R$^{3,1}$. We have found that the v.e.v. should have the generic behavior \eqref{Eq:Ogen} at late  de Sitter time and decay faster than any power of $x_\perp^{-1}$ as we extrapolate to large distances from the central axis in the future wedge of R$^{3,1}$, and also faster than any power of $\tau$ when we extrapolate to early proper time taking the limit $\tau\rightarrow 0^+$. The last feature implies that the dual state can be glued smoothly to the vacuum outside the future wedge.

\subsection{Global convergence and basis dependence}\label{Sec:Basis}

We have argued that the v.e.v. of the marginal operator dual to the massless bulk scalar field in evolutions with SO(3) $\times$ SO(1,1) $\times$ $\mathbb{Z}_2$ symmetries can be expressed in the form of the series \eqref{Eq:New-O-Expansion} in full dS$_3$ $\times$ $\mathbb{R}$, and thus on the entire future wedge of R$^{3,1}$ after necessary Weyl and coordinate transformations. Here we will study whether the expansion \eqref{Eq:New-O-Expansion} indeed converges for all $\varrho$ and thus on the entire future wedge of R$^{3,1}$. Also, we discuss how the v.e.v. depends on the choice of bases.

To begin, we get a unique function $O(\varrho)$ giving the v.e.v. on full dS$_3$ $\times$ $\mathbb{R}$ once we choose the coefficients $o_n$ of the late de Sitter time expansion \eqref{Eq:Ogen} of the v.e.v. and a basis for expansion in the full future wedge. For a concrete example, let us choose the behavior of the v.e.v. to be 
\begin{equation}
    O(\varrho) \sim \exp(- 4 \varrho) = \sigma^{-4},
\end{equation}
at late de Sitter time, i.e.
\begin{equation}\label{Eq:Late-coeffs-example}
    o_0 = 1, \quad o_n = 0, \,\, {\rm for \, all} \,\, n>0.
\end{equation}
Equivalently, our choice amounts to
\begin{equation}\label{Eq:Late-time-specify}
    \lim_{\sigma\rightarrow\infty} \sigma^n\left(O(\sigma) - \sigma^{-4}\right) = 0, \,\, {\rm for} \,\, n =0, 1, 2, \ldots, 
\end{equation}
As discussed in the previous subsection, normalizability and regularity of $\Phi$ implies that $o^e_n$, the coefficients of the early de Sitter time expansion \eqref{Eq:Ogenearly} should vanish. Using \eqref{Eq:Late-coeffs-example} and that $o^e_n$ vanish, we can determine all $o^a_n$ and $o^b_n$, the coefficients of \eqref{Eq:New-O-Expansion} as discussed in the previous subsection. Explicitly, we obtain 
\begin{align}
    & o^a_0 = \frac{81}{1280},\,\,o^b_0 = -\frac{81}{1280}, \,\,  o^a_1 = \frac{81}{1280},\,\,o^b_1 = -\frac{243}{1280},\nonumber\\& o^a_2 = \frac{567}{10240},\,\,o^b_2 = -\frac{5103}{10240},\,\, {\rm etc.}
\end{align}
It turns out that the values of the ratios $o^a_{n+1}/o^a_n$ and $o^b_{n+1}/o^b_n$ converge to 1 and 3 respectively in the limit $n\rightarrow \infty$ (recall the discussion on the convergence criteria in the previous subsection). The expansion \eqref{Eq:New-O-Expansion} indeed converges in this case, leading to
\begin{align}\label{Eq:O-specify}
   & O(\sigma) = F_1(\sigma)
   =\begin{cases} 
   \frac{1}{\sigma^{4}}, \,\, {\rm for} \,\, \sigma > 1,\\
   \frac{-1 + 81 \sigma^8}{80 \sigma^4}  \,\, {\rm for} \,\, \frac{1}{\sqrt{3}}< \sigma \leq 1,\\
   0 \,\, {\rm for} \,\, 0\leq\sigma \leq \frac{1}{\sqrt{3}}.
   \end{cases}
\end{align}
Note that $O(\sigma)$ is continuous everywhere. There are two kinks, one at $\sigma =1$ (i.e. $\varrho =0$) where $(\cosh\varrho)^{-1}$ takes its maximum value and another at $\sigma = 1/\sqrt{3}$ (i.e. $\varrho \sim- 0.549$) where $(2\cosh\varrho + \sinh\varrho)^{-1}$ takes its maximum value. Also $F_1(\sigma)$ and all its derivatives vanish at $\sigma =0$ as argued in the previous subsection.

Next, instead of \eqref{Eq:New-O-Expansion}, let us choose the following expansion
\begin{align}\label{Eq:New-O-Expansion-2}
   {O}(\rho) =\sum_{k=0}^\infty  \left(\frac{o^a_k}{\cosh{(\varrho)}^{4 + 2 k}} + \frac{o^b_k}{\left(3 \cosh{(\varrho)}+\sinh{(\varrho)} \right)^{4 + 2 k}} \right).
\end{align}
It turns out then for the same large de Sitter time behavior \eqref{Eq:Late-time-specify}, we obtain that instead of \eqref{Eq:O-specify} the above expansion converges to the following continuous function
\begin{align}\label{Eq:O-specify-1}
& O(\sigma)= F_2(\sigma) = \begin{cases} 
   \frac{1}{\sigma^{4}}, \,\, {\rm for} \,\, \sigma > 1,\\
   \frac{-1 + 16 \sigma^8}{15 \sigma^4}  \,\, {\rm for} \,\, \frac{1}{\sqrt{2}}< \sigma \leq 1,\\
   0 \,\, {\rm for} \,\, 0\leq \sigma \leq \frac{1}{\sqrt{2}}.
   \end{cases}
\end{align}
There are two kinks, one at $\sigma =1$ (i.e. $\varrho =0$) where $(\cosh\varrho)^{-1}$ takes its maximum value and another at $\sigma = 1/\sqrt{2}$ (i.e. $\varrho \sim -0.347$) where $(3\cosh\varrho + \sinh\varrho)^{-1}$ takes its maximum value. Note that $F_1(\sigma)$ and $F_2(\sigma)$ are identical for $\sigma >1$. Also $F_1(\sigma)$ and all its derivatives vanish at $\sigma =0$ as argued in the previous subsection. $F_1(\sigma)$ and $F_2(\sigma)$ are however different in the regime $1/\sqrt{2}<\sigma < 1$.

\begin{figure}[ht]
   \centering
        \includegraphics[width=0.6 \linewidth]{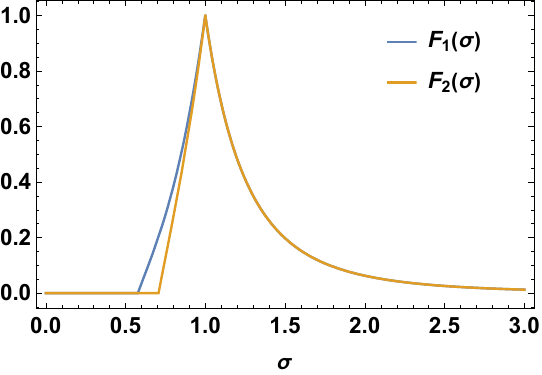}
   \caption{The functions $F_1(\sigma)$ and $F_2(\sigma)$ defined in \eqref{Eq:O-specify} and \eqref{Eq:O-specify-1}, respectively. }
   \label{Fig:Basis}
 \end{figure}

In Fig.~\ref{Fig:Basis}, we compare the function $F_2(\sigma)$ above with the function $F_1(\sigma)$. Clearly, the global function $O(\sigma)$ depends on the choice of bases used for extrapolating the late de Sitter time behavior to the entire future wedge.

We note that both $F_1(\sigma)$ and $F_2(\sigma)$ have cusps. Nevertheless, the bulk scalar field profiles corresponding to $  O(\sigma)=F_1(\sigma)$ and $ O(\sigma)= F_2(\sigma)$ do make sense as distributions, i.e. as an infinite sum of a sequence of smooth (sourceless) profiles obtained by solving the Klein-Gordon equation in the corresponding bulk expansions at early and late de Sitter time (thus determining $\phi_n^a(v)$ and $\phi_n^b(v)$) as discussed in the previous subsection.\footnote{Note that even the 1+1-dimensional wave equation $(-\partial_t^2 +\partial_x^2) \chi =0$ can have kinks, e.g.~in a sawtooth waveform. At an initial time such a waveform is a superposition of an infinite sum of sine and cosine waves. This waveform propagates following the wave equation. The full solution is a superposition of sourceless solutions.} Note that in both cases, the bulk scalar field has non-vanishing sub-leading $\sigma^{-(4+2k)}$ terms with $k= 1, 2,\ldots$ (see \eqref{Eq:phi0-phi1}), although these sub-leading terms vanish for the v.e.v. The coefficients $\phi_n(v)$ of the late-time expansion \eqref{Eq:AnsSc} (with $\alpha =4$) of the bulk scalar field are identical in both cases. However, globally these are two distinct sourceless solutions. These examples indeed illustrate a general lesson that there are many physically realizable evolutions with the same late de Sitter time expansions of the v.e.v. and the bulk scalar field. We will see in the next subsection that these results are also applicable to generic solutions of the bulk scalar field obtained numerically from smooth initial conditions.

More generally, we can consider the expansion
 \begin{align}\label{Eq:New-O-Expansion-k}
   {O}(\rho) =\sum_{k=0}^\infty  \left(\frac{o^a_k}{\cosh{(\varrho)}^{4 + 2 k}} + \frac{o^b_k}{\left(\kappa \cosh{(\varrho)}+\sinh{(\varrho)} \right)^{4 + 2 k}} \right)
\end{align}
with $\kappa >1$ (the bases functions are bounded if $\kappa >1$). It turns out that the above global expansion that reproduces the late de Sitter time expansion \eqref{Eq:Ogen} and early de Sitter time expansion \eqref{Eq:Ogenearly} with all vanishing coefficients $o^e_n$, is equivalent to the following function for all $\sigma =\exp \varrho$:
\begin{align}\label{Eq:O-specify-k}
   & O(\sigma) = F_\kappa (\sigma)
   =\sum_{k=0}^\infty o_k \sigma^{-(4 +2k)}\theta\left(\sigma - 1\right)\nonumber\\
   &\qquad\quad +  \sum_{k=0}^\infty o_k\frac{(\kappa+1)^{(4+2k)}\sigma^{8+4k}-(\kappa-1)^{(4+2k)}}{\left((\kappa+1)^{(4+2k)}-(\kappa-1)^{(4+2k)}\right)\sigma^{4+2k}} \theta\left(1- \sigma\right)\theta\left(\sigma - \sqrt{\frac{\kappa-1}{\kappa +1}}\right),
\end{align}
which is also continuous for all $\sigma$. (Note that the above reproduces \eqref{Eq:O-specify} and \eqref{Eq:O-specify-1} for $\kappa =2$ and $\kappa =3$, respectively when the late de Sitter time behavior is \eqref{Eq:Late-time-specify}.) All $F_\kappa(\sigma)$ agree for $\sigma >1$. Moreover, all of them generically have kinks at $\sigma = 1$ and $\sigma = \sqrt{(\kappa -1)/(\kappa +1)}$, and vanish for $0 \leq \sigma <  \sqrt{(\kappa -1)/(\kappa +1)}$.

\begin{figure}[ht]
   \centering
        \includegraphics[width=0.6 \linewidth]{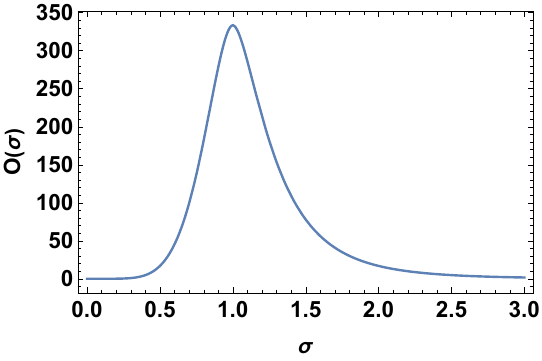}
   \caption{The function $O(\sigma) = F_\kappa(\sigma)$ with $F_\kappa(\sigma)$ defined in \eqref{Eq:O-specify-k}, and with the choices of coefficients \eqref{Eq:coeffs-example}, $o_0 = -1$ and $\kappa =1.001$.}
   \label{Fig:Basis-2}
 \end{figure}

It is easy to see that we can make $F_\kappa(\sigma)$ as smooth as possible. Let us choose 
\begin{equation}\label{Eq:coeffs-example}
    o_1 = -o_0\frac{2\kappa(3 + 10\kappa^2 +3\kappa^4)}{3(-1- \kappa +\kappa^4 +\kappa^5)}, \,\, o_2 = o_0\frac{1 +6\kappa^2 +\kappa^4}{(\kappa -1)(\kappa +1)^3},\,\, o_n = 0\,\, \forall n\geq 3.
\end{equation}
 Then $F_\kappa(\sigma)$ and $F_\kappa'(\sigma)$ are both continuous at all $\sigma$, i.e. $F_\kappa(\sigma)$ is a $C^1$ function of $\sigma$ for $0\leq \sigma < \infty$ when the coefficients are chosen according to \eqref{Eq:coeffs-example}. Furthermore,  this implies that $F_\kappa'(\sigma)$ vanishes at $\sigma = \sqrt{(\kappa -1)/(\kappa +1)}$. Clearly,
 \begin{itemize}
     \item by including 2n+1 non-vanishing coefficients we can make $F_\kappa(\sigma)$ and all its first n-derivatives continuous for all $\sigma$ (i.e. make $F_\kappa(\sigma)$ a $C^n$ function for all $\sigma$), and
     \item by also taking $\kappa$ closer to 1, we can make all derivatives of $F_\kappa(\sigma)$ vanish closer to 0.
 \end{itemize}
We demonstrate this in Fig.~\ref{Fig:Basis-2} by choosing a sufficiently small value of $\kappa$. 
 
We also want to emphasize that even if $O(\sigma)$ is not a $C^\infty$ function, we can obtain a corresponding bulk solution $\Phi$ which is an appropriate distribution. For the bulk scalar field $\Phi$, we should consider the general expansion ($\kappa>1$)
\begin{align}\label{Eq:New-Phi-Expansion-k}
   \Phi(r,\rho) =\sum_{k=0}^\infty  \left(\frac{\phi^a_k(v)}{\cosh{(\varrho)}^{4 + 2 k}} + \frac{\phi^b_k(v)}{\left(\kappa \cosh{(\varrho)}+\sinh{(\varrho)} \right)^{4 + 2 k}} \right),
\end{align}
and obtain the coefficients $\phi^a_k(v)$ and $\phi^b_k(v)$ by solving the Klein-Gordon equation at early and late de Sitter time systematically as demonstrated in the case $\kappa =2$ in the previous subsection. We recall that at each order we find one integration constant. We can also verify that the coefficients of the early de Sitter time expansion $\phi^e_n(v)$ in \eqref{Eq:Phi-early} vanish order by order. Furthermore, resumming  the full series \eqref{Eq:New-Phi-Expansion-k} we obtain that
\begin{align}\label{Eq:Phi-specify-k}
   & \Phi(r,\rho)
   =\sum_{k=0}^\infty \phi_k(v) \sigma^{-(4 +2k)}\theta\left(\sigma - 1\right)\nonumber\\
   &\qquad +  \sum_{k=0}^\infty \phi_k(v)\frac{(\kappa+1)^{(4+2k)}\sigma^{8+4k}-(\kappa-1)^{(4+2k)}}{\left((\kappa+1)^{(4+2k)}-(\kappa-1)^{(4+2k)}\right)\sigma^{4+2k}} \theta\left(1- \sigma\right)\theta\left(\sigma - \sqrt{\frac{\kappa-1}{\kappa +1}}\right),
\end{align}
where $\phi_k(v)$ are the coefficients of the late de Sitter time expansion \eqref{Eq:AnsSc} with $\alpha = 4$. Note that $\Phi(r,\rho)$ is continuous at all $\rho$ and for all $r>0$. The $v^4$ coefficient corresponds to the v.e.v. which is explicitly $F_\kappa(\sigma)$ defined in \eqref{Eq:O-specify-k}.

Also note that corresponding to any given late de Sitter time expansion we can obtain infinitely many global completions of $O(\sigma)$ (and similarly $\Phi(r,\rho)$) simply by superimposing the functions $F_\kappa(\sigma)$ defined in \eqref{Eq:O-specify-k} as follows
\begin{equation}
    F(\sigma) = \int_1^\infty{\rm d}\kappa\,\, c(\kappa) F_\kappa(\sigma) \,\, {\rm with} \,\, \int_1^\infty {\rm d}\kappa\,\,c(\kappa) = 1 \,\, {\rm and} \,\, c(1) = 0.
\end{equation}
Note for any choice of $c(\kappa)$ satisfying the requirements above, we get the same late de Sitter time expansion (since all $F_\kappa(\sigma)$ agree for $\sigma >1$ in this case). Therefore, the generic late de Sitter time expansion needs to be supplemented by infinite data to recover information of the initial state.

Of course, we can choose different set of bases which are bounded functions of $\varrho$, compatible with the late de Sitter time expansion \eqref{Eq:Ogen} but \textit{not} of the type \eqref{Eq:New-O-Expansion-k}. It is therefore natural to ask the following:
\begin{itemize}
    \item Are our conclusions, especially the vanishing of the v.e.v. and all its derivatives in the limit $\sigma \rightarrow 0$ as stated in \eqref{Eq:Remarkable-O-n} and its consequences on the future wedge of R$^{3,1}$ given by \eqref{Eq:Remarkable-O-1}, valid for our choices of bases not in the general class \eqref{Eq:New-O-Expansion-k}?
    \item Will these conclusions hold for reasonable choices of initial conditions in the bulk from which we can obtain unique normalizable and regular solutions?
\end{itemize}
In fact, if the answer to the second question is positive, then obviously the answer to the first question is positive as well. This is simply due to the fact that we can explore all regular, normalizable and physically realizable solutions by explicitly solving the massless Klein-Gordon equation. We therefore explore the second question with explicit numerical simulations of the massless Klein-Gordon equation \eqref{Eq:KGMassless} in the following subsection.

To conclude, we note that any late de Sitter time behavior such as $\sigma^{-4}$ can be modified multiplicatively, e.g. by
$$f(\sigma)=\sigma^{-4}\exp(\tilde\kappa(1 - \coth(\sigma))),$$
($\tilde\kappa>0$) which gives exponentially suppressed corrections to the late de Sitter time expansion and ensures that the modified function vanishes as $\sigma\rightarrow 0$. However, our analysis of bulk regularity via the late de Sitter time expansion in the previous subsection does not allow exponential or super-exponentially suppressed corrections to be present.\footnote{Actually absence of such additional corrections can also be argued by mapping the boundary metric to R$^{3,1}$ as well and considering generic fluctuations in pure AdS$_5$ background.} 

\subsection{Validation via numerical simulations}\label{Sec:Num-1}

We can simulate the evolution of the massless bulk scalar field preserving SO(1,1) $\times$ SO(3) $\times$ $\mathbb{Z}_2$ symmetries in the metric dual to the vacuum numerically by rewriting the Klein-Gordon \eqref{Eq:KGMassless} in the characteristic form,
\begin{align}\label{Eq:Scalar-Ch}
  & -4 v \left(e^{2 \rhoL } (v+1)-v+1\right) \partial_{\rL}\dot{\Phi }+2\left(e^{2 \rhoL }(v+3) - (v-3)\right)\dot{\Phi}    \nonumber \\&+ \left(e^{2 \rhoL } (v-3) (v+1)^2-(v-1)^2 (v+3)\right) \partial_{\rL}\Phi = 0
\end{align}
   where $\dot{\Phi }$ denotes the directional derivative along the outgoing radial null geodesic, i.e.
    \begin{align}\label{Eq:dot-vac}
        \dot{f} = \partial_\rhoL f + \frac{1}{2} (1-v^2) \partial_\rL f.
    \end{align}
Given  $\Phi$ at any time-step, one can determine $\dot{\Phi}$ at that time step by viewing \eqref{Eq:Scalar-Ch} as an ordinary radial differential equation of $\dot{\Phi}$. This requires setting appropriate boundary condition at  $v = 0$ which can be obtained from the asymptotic expansion.  As we are investigating normalizable solutions, the asymptotic expansion of the scalar field is 
  \begin{align}
      \Phi(\rhoL, \rL ) = O(\rhoL)~ \rL^4 + \partial_\rhoL O(\rhoL)~ \rL^5 + \ldots 
  \end{align}
  where $O(\rhoL)$ is the v.e.v. of the dual marginal operator. Therefore, the asymptotic expansion of $\dot{\Phi}$ is
  \begin{align}
      \dot{\Phi} = 2 O(\rhoL) v^3 + \frac{7}{2}\partial_\rhoL O(\rhoL)v^4 + \ldots
  \end{align}
This can be used to set the boundary conditions which determine $\dot{\Phi}$ at each time step. From $\dot{\Phi}$, we can extract $\partial_{\rhoL}\Phi$ via \eqref{Eq:dot-vac} which can be used to find $\Phi$ at the next step.  Therefore, with an initial profile for $\Phi(v,\rhoL_{\rm in})$ at a suitable initialization time $\rhoL_{\rm in}$, we can determine the evolution of $\Phi$ uniquely. From the numerical solution of $\Phi$, the v.e.v. of the dual marginal operator $O(\varrho)$ can be extracted.
  
  \begin{figure}[ht]
     \centering
          \includegraphics[width=0.49 \linewidth]{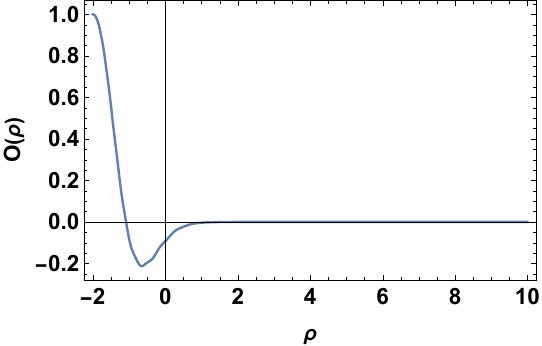}
            \includegraphics[width=0.45 \linewidth]{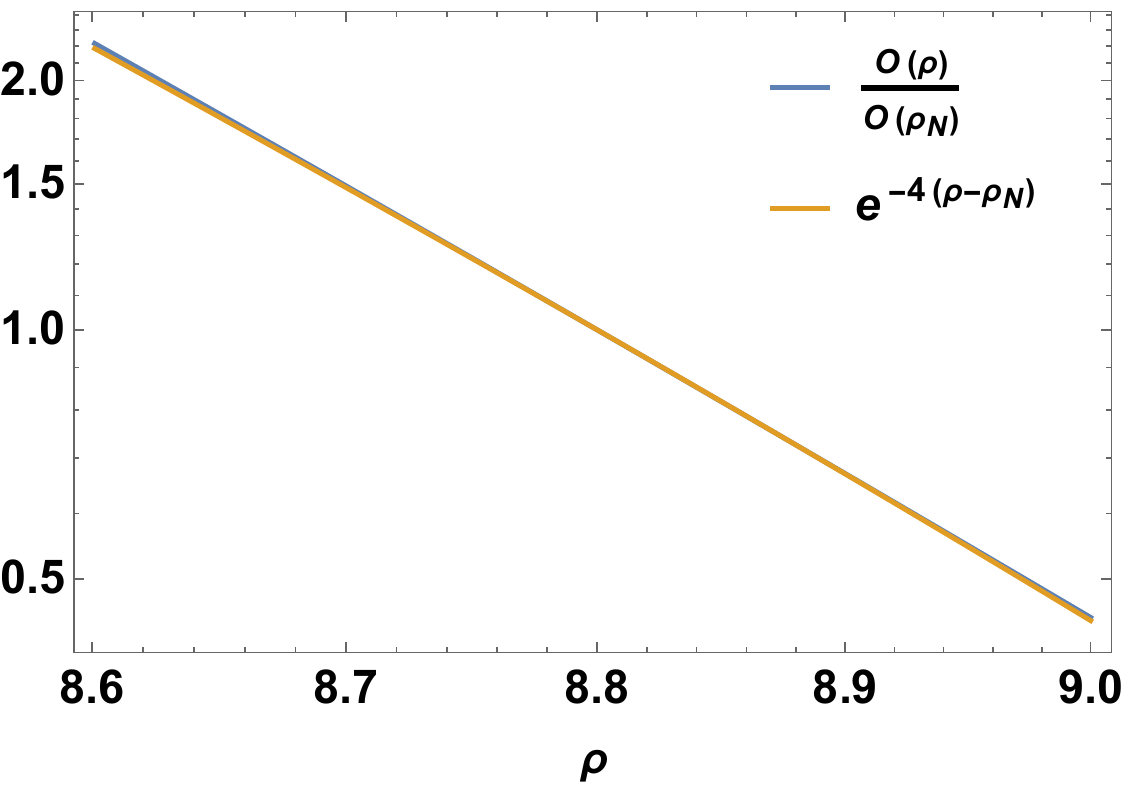}
     \caption{The left figure shows the behaviour of the v.e.v $O(\rho)$ at late de Sitter time. Here we have set $L=1$ and have suitably normalized the v.e.v. The right figure shows that the leading exponent of the v.e.v of a scalar field at late de Sitter time is $-4$, where we take $\rho_N=8.8$.  The initial condition is given by \eqref{Eq:Phi-in} with $k=1$ and the initialization time is $\rho_{\rm in} = -2$.} \label{Fig:Scalar} 
 \end{figure}

For the numerical simulation, we have done radial integration in the pseudo-spectral method via choosing Chebyshev grid with 30 points. Time steppings have been done with the fourth order Adams-Bashforth method. Each time step is 0.001 in units $L=1$.

The v.e.v. $O(\varrho)$ obtained by setting the initial profile to be 
\begin{equation}\label{Eq:Phi-in}
    \Phi(v,\rhoL_{\rm in})=k \rL^4 e^{-\rL^2}
\end{equation}
at $\rhoL_{\rm{in}} = -2$ has been shown in Fig.~\ref{Fig:Scalar} (with $k=1$). The behaviour of the v.e.v. at late de Sitter time is indeed $e^{-4 \rhoL}$ as we have shown in our analytic study. 

The behavior at early de Sitter time $\rho \rightarrow - \infty$ can in principle be explored by running the numerical simulation back in de Sitter time (via negative time stepping).  However, we find that it leads to numerical instability (in the case of pure gravity, we find that the constraints are violated quickly as we run back in de Sitter time -- more details in Sec.~\ref{Sec:Num}).  Furthermore, we find that we cannot initiate the evolution at a very large negative value of  $\varrho_{\rm in}$ as the numerical noise picked along the time evolution does not allow us to access the phase at late de Sitter time reliably. It turns out that with the parameters used in our numerical simulation, $\rho_{\rm in}$ cannot be pushed earlier than $-2$. 

Nevertheless, it is possible to map the $O(\rho)$ to $O_M(\tau, x_\perp)$, the v.e.v. on the future wedge of R$^{3,1}$ via \eqref{Eq:O-OM}å and explicitly verify if the decay of the latter is faster than any power of $x_\perp$ at large $x_\perp$ for any value of the proper time $\tau$ as stated in \eqref{Eq:Remarkable-O-1}. We recall that this is equivalent to the early $\tau$ behavior that the v.e.v. vanishes faster than any power of $\tau$ in the limit $\tau\rightarrow 0^+$ for any fixed $x_\perp$ as both the large $x_\perp$ and small $\tau$ behaviors are governed by the $\varrho\rightarrow -\infty$ limit where \eqref{Eq:Remarkable-O-n} holds. It turns out that it is easier to verify the large $x_\perp$ behavior stated in \eqref{Eq:Remarkable-O-1} without any need for extrapolating the numerical solution of the bulk scalar field beyond the domain of $\rho$ (and the corresponding domains of $\tau$ and $x_\perp$) where $O$ ($O_M$) can be extracted from the bulk scalar field. 

\begin{figure}[ht]
     \centering
     \includegraphics[width=0.4 \linewidth]{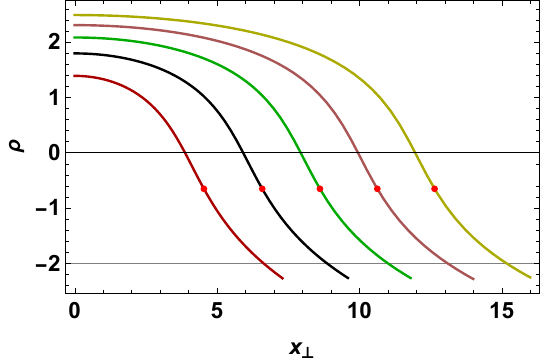}
     \includegraphics[width=0.49\linewidth]{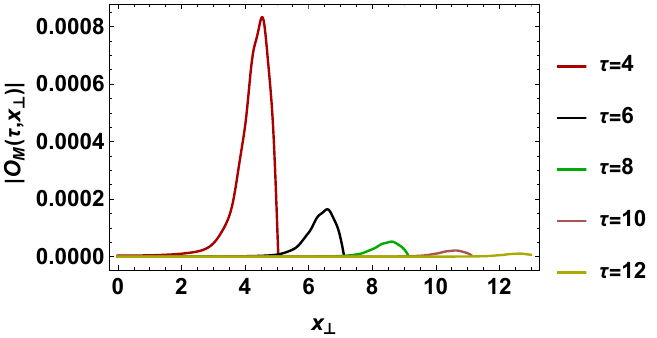}
               \includegraphics[width=0.6\linewidth]{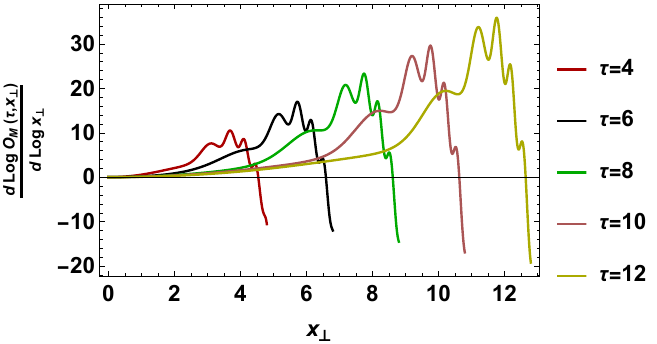}
     \caption{The plot shows the behavior of the v.e.v. $O_M$ on the future Minkowski wedge for the simulation in Fig.~\ref{Fig:Scalar}. Top left: The range of $x_\perp$ for different $\tau$ covered by the restriction $\rho > -2$ where the simulation has been performed. Red dots on the curves indicate the peaks of $O_M$ at the corresponding $\tau$. Top right: $O_M$ decays with increasing $\tau$. Bottom:  The log-derivative of $O_M(\rho)$ with respect to $x_{\perp}$ tends to $-\infty$ for large $x_\perp$ at any fixed $\tau$ as implied by \eqref{Eq:Remarkable-O-1}. } \label{Fig:Scalarlog} 
\end{figure}

As shown in the top left panel in Fig.~\ref{Fig:Scalarlog}, it follows from \eqref{Eq:rho} that limiting the initial value of $\varrho_{\rm in}$ to $-2$ implies that we cannot access the region with $x_\perp$ greater than a certain value $x_\perp^c(\tau)$ at a fixed value of $\tau$. We note that $x_\perp^c(\tau)$ increases with $\tau$ and is larger than $\tau$.  The v.e.v. on the future wedge ($O_M(\tau, x_\perp)$) obtained with the initial condition \eqref{Eq:Phi-in} set at $\varrho_{\rm in} = -2$ is plotted in the top right panel of Fig.~\ref{Fig:Scalarlog}. We note that (i) $O_M(\tau,x_\perp)$ decays at large $x_\perp$ at any $\tau$ unlike the exact analytic solutions \eqref{Eq:Sol1} and \eqref{Eq:Sol2}, (ii) the maximum value of $O_M(\tau,x_\perp)$ at any given value of $\tau$ decreases with increasing $\tau$. Therefore, unlike the exact analytic solutions \eqref{Eq:Sol1} and \eqref{Eq:Sol2}, the v.e.v. uniformly disappears on the future wedge uniformly at late proper time. Furthermore, the bottom panel of Fig.~\ref{Fig:Scalarlog} suggests that 
\begin{equation}
    \frac{\partial\ln O_M(\tau, x_\perp)}{\partial\ln x_\perp} \rightarrow -\infty \quad {\rm as}\quad x_\perp \rightarrow \infty
\end{equation}
for any fixed $\tau$ implying that the v.e.v. decays faster than any power of $x_\perp$ as stated in \eqref{Eq:Remarkable-O-1}. (We can recall that in the exact analytic solution \eqref{Eq:Sol1} $$\frac{\partial\ln O_M(\tau, x_\perp)}{\partial\ln x_\perp}\rightarrow 8\quad {\rm as}\quad x_\perp \rightarrow \infty,$$in contrast to what demonstrated in the bottom panel in Fig.~\ref{Fig:Scalarlog}.) Interestingly, we note from the bottom panel of Fig.~\ref{Fig:Scalarlog} that the maximum value of the logarithmic derivative ${\partial\ln O_M(\tau, x_\perp)}/{\partial\ln x_\perp}$ increases with increasing $\tau$ implying that, while $O_M(\tau, x_\perp)$ decays with increasing $\tau$, it also gets more narrowly peaked around $x_\perp \sim \tau$ for the solution corresponding to the initial condition \eqref{Eq:Phi-in}.

Although we can numerically validate the large $x_\perp$ behavior of $O_M$ and thus the small $\tau$ behavior stated in \eqref{Eq:Remarkable-O-1} for the initial condition \eqref{Eq:Phi-in}, we can wonder if the same may hold for any smooth arbitrary initial condition. We have verified that all these results, including the late de Sitter time behavior of $O(\rho)\sim e^{-4\varrho}$ and the large $x_\perp$ behavior of $O_M$ at any any fixed $\tau$ according to \eqref{Eq:Remarkable-O-1} hold for the initial profile \eqref{Eq:Phi-in}, as well as smooth initial profiles of the type 
 \begin{align} \label{Eq:profiles-s}
     \Phi(v,\rho_{\rm in}) = k \rL^4, \,\,\,  \Phi(v,\rho_{\rm in}) = k \frac{\rL^4}{1+\rL^2}, \,\,\,  \Phi(v,\rho_{\rm in}) = k\frac{\left(-0.625 \rL^4+0.662 
     \rL^3+0.967 \rL^2\right)}{0.18 \rL^2-0.3045 \rL -0.494}, \,\, \ldots
\end{align} 
set at any arbitrary value of the initialization time $\varrho_{\rm in}$. Such initial profiles have the property that they are generic $C^\infty$ functions in the domain $0\leq v<\infty$. Of course, as evident from the exact analytic solutions \eqref{Eq:Sol1} and \eqref{Eq:Sol2}, the large $x_\perp$ behavior should be different from that stated in \eqref{Eq:Remarkable-O-1} for special initial profiles which are rational functions of $v$ (and also $C^\infty$ functions in the domain $0\leq v<\infty$ for any finite value of the initialization $\varrho_{\rm in}$) . However, such initial profiles are not generic. For instance, we note that from \eqref{Eq:Sol1} and \eqref{Eq:Sol2} that such initial profiles where \eqref{Eq:Remarkable-O-1} is not valid have higher order poles at $v =-1$ for any value of the initialization time. It would be interesting to understand the necessary and sufficient conditions needed to violate \eqref{Eq:Remarkable-O-1}. However, we can conclude from our numerical investigations that both the large $x_\perp$ and small $\tau$ behavior of $O_M$ as stated in \eqref{Eq:Remarkable-O-1} hold for generic initial conditions set by $C^\infty$ functions in the domain $0\leq v<\infty$ at an arbitrary initialization time.

\section{Analytical study of generic holographic Gubser flow}\label{Sec:Gen}

\subsection{Generic behavior at large de Sitter time}\label{Sec:LT}

Following the simple case of the massless scalar field, we can adopt the same methodology as in \cite{Banerjee:2023djb} to find the generic late time behavior of the regular and normalizable solutions of Einstein's equations. Similar to \eqref{Eq:AnsSc}, we can postulate the ansatz for the late de Sitter time behavior of the $A$, $B$ and $C$ functions appearing in the bulk spacetime metric \eqref{Eq:ABC-metric} dual to a generic Gubser flow as follows:
\begin{align}\label{Eq:AnsGrav}
    &A(r,\rho) = \sum_{n=1}^\infty \sum_{m=0}^\infty\sigma^{- n\alpha - 2m} A_{nm}(v),\quad B(r,\rho) = \sum_{n=1}^\infty \sum_{m=0}^\infty\sigma^{- n\alpha - 2m} B_{nm}(v),\nonumber\\&C(r,\rho) = \sum_{n=1}^\infty \sum_{m=0}^\infty\sigma^{- n\alpha - 2m} C_{nm}(v).
\end{align}
Remember that $\sigma=e^\varrho.$ Unlike \eqref{Eq:AnsSc}, we have a double summation above due to the non-linear nature of Einstein's equations.

Since we aim to study the above systematic late de Sitter time expansion, we need to preserve regularity at the future horizon $r = L$ (i.e. $v=1$) at each order in this expansion. In the latter aspect, the late de Sitter time expansion is similar to fluid/gravity correspondence where the regularity of the future horizon determines the transport coefficients order by order in the gradient expansion \cite{Rangamani:2009xk}. However, we recall that the late de Sitter time expansion is not the hydrodynamic gradient expansion. For all curvature invariants to be finite at the horizon in the perturbative late de Sitter time expansion, we require that $A_{nm}(v)$,  $B_{nm}(v)$ and $C_{nm}(v)$ are $C^\infty$ at $ v =1$, i.e.
\begin{align}
&A_{nm}(v) = \sum_{p=0}^{\infty} a_{nm,p}(1-v)^p, \quad B_{nm}(v) = \sum_{p=0}^{\infty} b_{nm,p}(1-v)^p, \nonumber\\ &C_{nm}(v) = \sum_{p=0}^{\infty} c_{nm,p}(1-v)^p.
\end{align}
We can readily solve the gravitational equations in the late de Sitter time expansion as in \eqref{Eq:AnsGrav} and also in power series of $(1-v)$ at the late-time horizon $v=1$. At the leading order, $A_{10}(v)$, $B_{10}(v)$ and $C_{10}(v)$ are determined by the three constants $a_{10,0}$, $b_{10,0}$ and $c_{10,0}$ which can be chosen arbitrarily. We explicitly find that 
\begin{align}\label{Eq:Anbsol}
   A_{10}(v) &=   a_{10,0} \left(1 - (\alpha+1) (1-v) + \mathcal{O}\left((1-v)^2\right) \right), \nonumber \\  & + b_{10,0} \left(-2 \alpha (1-v) + \mathcal{O}\left((1-v)^2\right)  \right), \nonumber \\ &+ c_{10,0} \left(\frac{1}{2} \alpha(1+\alpha) (1-v) +\mathcal{O}\left((1-v)^2\right)  \right).
\end{align}
The expressions for $B_{10}(v)$ and $C_{10}(v)$ are similar. As in the case of the massless bulk scalar field, we can determine the value of $\alpha$ by simply evaluating the boundary values $A_{10}(v=0)$, $B_{10}(v=0)$ and $C_{10}(v=0)$ from their respective Taylor series about $v=1$ obtained to high orders by solving the gravitational equations. It turns out that it is sufficient to evaluate $A_{10}(v=0)$ which is explicitly
\begin{align}\label{Eq:A10nb}
     A_{10}(v=0) &= P(\alpha) (a_{10,0} + (1+\alpha) (2 b_{10,0}-c_{10,0})),
\end{align}
with $P(\alpha)$ being a polynomial of $\alpha$. For obtaining normalizable (sourceless) solutions, it is necessary (although not sufficient) to impose that $P(\alpha)=0$. We find numerically that the roots of $P(\alpha)$ are the non-negative even integers, i.e.
\begin{equation}\label{Eq:alpha-grav-1}
    \alpha = 2k, \,\, k = 0,1,2,\ldots
\end{equation}
Finally, we need to check for each of these cases whether normalizable physical solutions exist which are not just pure gauge deformations of the metric \eqref{Eq:Vac-metric} (with $A = B = C =0$) that corresponds to the vacuum of the dual theory. It turns out that only for $\alpha = 2, 4, \ldots$, we can find normalizable and physical solutions which are not pure gauge.

We note that the generic asymptotic expansion of $A$, $B$ and $C$ are given by \eqref{Eq:near_boundaryExpA}-\eqref{Eq:near_boundaryExpS}. This expansion is determined by two functions, namely $a_{(1)}(\varrho)$ and $a_{(4)}(\varrho)$ (the coefficients of $v$ and $v^4$ terms, respectively, in the Taylor expansion of $A$ about $v=0$), where $a_{(1)}(\varrho)$ is related to a proper residual gauge transformation and $a_{(4)}(\varrho)$ is essentially the expectation value of the energy density in the dual state. To match with the expansion \eqref{Eq:AnsGrav}, we need to impose
\begin{equation}\label{Eq:a4lte}
 a_{(1)}(\rhoL) =   \sum_{n=1}^\infty \sum_{m=0}^\infty\sigma^{- n\alpha - 2m} a_{(1)nm},\quad  a_{(4)}(\rhoL) =   \sum_{n=1}^\infty \sum_{m=0}^\infty\sigma^{- n\alpha - 2m} a_{(4)nm},
\end{equation}
so that $a_{(1)nm}$ and $a_{(4)nm}$ are the coefficients of $v$ and $v^4$ terms, respectively, in the Taylor expansion of $A_{nm}(v)$ about $v=0$. In order to obtain solutions which are not pure gauge, we can impose $a_{(1)nm} =0$.

Henceforth, we need to check if the solutions of $A_{10}(v)$, $B_{10}(v)$ and $C_{10}(v)$ obtained by the power series expansion about $v=1$ as described above can be matched to the asymptotic expansions about $v=0$ given by \eqref{Eq:near_boundaryExpA}-\eqref{Eq:near_boundaryExpS} with $a_{(1)}(\varrho)$ and $a_{(4)}(\varrho)$ as in \eqref{Eq:a4lte}.  We first discuss the case $\alpha = 4, 6, \ldots$ Already, the choices of $\alpha$ given by \eqref{Eq:alpha-grav-1} ensure that $A_{10}(v)$ vanishes at $v=0$. To match with \eqref{Eq:near_boundaryExpA} where $a_{(1)10} =0$, we set the coefficients of $v$ and $v^2$ terms of the Taylor expansion of $A_{10}(v)$ at $v=0$ to zero. These give two relations among the three integration constants $a_{10,0}$, $b_{10,0}$ and $c_{10,0}$ determining $A_{10}(v)$, $B_{10}(v)$ and $C_{10}(v)$, so that we can express $b_{10,0}$ and $c_{10,0}$ in terms of $a_{10,0}$. We can then exactly match with the asymptotic expansions of $A_{10}(v)$, $B_{10}(v)$ and $C_{10}(v)$ given by \eqref{Eq:near_boundaryExpA}-\eqref{Eq:near_boundaryExpS} with $a_{(1)10}=0$. From the asymptotic expansion of $A_{10}(v)$ we can obtain $a_{(4)10}$ in terms of $a_{10,0}$. The normalizable and regular solutions for $A_{10}(v)$, $B_{10}(v)$ and $C_{10}(v)$ with non-vanishing energy density at the leading order can be found explicitly in the form of rational polynomials for $\alpha = 4, 6, \ldots$ as will be reported below. The same procedure in the case of $\alpha =0$ implies that the three integration constants $a_{10,0}$, $b_{10,0}$ and $c_{10,0}$ should vanish implying that we cannot obtain non-vanishing normalizable and regular solutions in this case.

The case $\alpha =2$ is special as the above procedure of setting the coefficients of $v$ and $v^2$ terms of the Taylor expansion of $A_{10}(v)$ at $v=0$ zero gives only one relation between the three integration constants $a_{10,0}$, $b_{10,0}$ and $c_{10,0}$. In \cite{Banerjee:2023djb}, only a pure gauge normalizable solution was found at the leading order for this case. However, it turns out that a normalizable and regular solution (which is $C^\infty$ at $v=1$) which is not pure gauge exists in this case also. We will report this solution explicitly below.

Thus we conclude that the regularity at the horizon implies that the allowed values of $\alpha$ are
\begin{equation}\label{Eq:alpha-grav}
    \alpha = 2 + 2k, \,\, k = 0,1,2,\ldots.
\end{equation}
Following the same arguments as in the case of the massless scalar field, we can simplify the late de Sitter time expansion of \eqref{Eq:AnsGrav} to
\begin{equation}\label{Eq:AnsGrav1}
    A(r,\rho) = \sum_{k=0}^\infty\sigma^{- 2 - 2k} a_k(v), \quad B(r,\rho) = \sum_{k=0}^\infty\sigma^{- 2 - 2k} b_k(v), \quad C(r,\rho) = \sum_{k=0}^\infty\sigma^{- 2 - 2k} c_k(v).
\end{equation}
As in the case of the massless scalar field, at each order we obtain a contribution to $a_{(4)}(\varrho)$, and hence to the energy density. It follows that the generic late de Sitter time behavior of the energy density in holographic Gubser flow should be
\begin{equation}\label{Eq:edSGen}
    {\varepsilon}(\rho) = -\frac{3l^3}{16\pi G_N}\sigma^{-2}L^{-4} \sum_{k =0}^\infty e_k\sigma^{-2k},
\end{equation}
where ${e}_k$ are arbitrary real numbers like ${o}_k$ determining the late de Sitter expansion of the v.e.v. in \eqref{Eq:Ogen}. 

We recall that the Ward identities (or equivalently the constraints of Einstein's equations) determine the transverse pressure $P_T$ and longitudinal pressure $P_L$ in term of ${\varepsilon}(\rho)$ via  \eqref{Eq:PT} and \eqref{Eq:PL} respectively. Therefore, \eqref{Eq:edSGen} implies that at late de Sitter time
\begin{align}\label{Eq:Ratio-lt}
    &\frac{P_L(\rho)}{\varepsilon(\rho)} = 1 -\left( 4 + 2\frac{e_1}{e_0}\right)\sigma^{-2} + \mathcal{O}(\sigma^{-4}), \nonumber\\
     &\frac{P_T(\rho)}{\varepsilon(\rho)} = \left( 2 + \frac{e_1}{e_0}\right)\sigma^{-2} + \mathcal{O}(\sigma^{-4}),
\end{align}
i.e. the generic late de Sitter time behavior is free-streaming in the transverse directions exactly like in kinetic theory in the relaxation time approximation. We also note that if $e_0$ vanishes, then $P_L/\varepsilon$ and $P_T/\varepsilon$ are ill-defined at large de Sitter time. Therefore,  we expect that $e_0$ is non-zero except for the vacuum solutions where all $e_i$ vanish.

The leading order normalizable solutions in \eqref{Eq:AnsGrav1} explicitly are
\begin{align}\label{Eq:a0b0c0}
 & a_0(v) = e_0 \left(12 v^2 \left(\frac{v}{2}+\frac{\ln (v+1)}{v}-1\right)+ 6 v^3 \left(\frac{\ln (v+1)}{v}-1\right)\right), \nonumber \\
  & b_0(v) = e_0\left(\frac{3\left(\left(v^2+4 v+3\right) \ln (v+1)-3 v-5\right)}{(v+1)^2} - \frac{15 \left(v^2-2\right)}{ 2 (v+1)^2}\right), \nonumber \\
 &  c_0(v) = - e_0\left(\frac{ \left(2 v^3+9 v^2-12 \left(v^2+3 v+2\right) \ln (v+1)+24 v+25\right)}{(v+1)^2} + \frac{15 \left(3 v^2-5\right)}{3 (v+1)^2}\right),
\end{align}
where $e_0$ is the leading order contribution to the dual energy density as in \eqref{Eq:edSGen}. Note that $a_0(v)$, $b_0(v)$ and $c_0(v)$ are $C^\infty$ at $v=1$ and are $\mathcal{O}(v^4)$ at $v= 0$, consistent with the asymptotic expansions \eqref{Eq:near_boundaryExpA}-\eqref{Eq:near_boundaryExpS} (with $a_{(1)} =0$). These leading order solutions (corresponding to $\alpha = 2$) were not found in \cite{Banerjee:2023djb} although the methodology of finding the generic late de Sitter time behavior analytically was established there. 

The sub-leading terms in \eqref{Eq:AnsGrav1} explicitly are
\begin{align}\label{Eq:a1b1c1}
   & a_1(v) = e_1\frac{v^4}{(1+v)^2} +   e_0 v \left(24 (v+4) \ln (v+1)-\frac{4 v \left(17 v^2+42 v+24\right)}{(v+1)^2}\right) + \ldots, 
   \nonumber\\
   & b_1(v) = -e_1 \frac{v^4}{2(1+v)^4} + e_0 \sum_{m=1} \frac{(-1)^m (m+2) \left(-48 (m+3) H_{m+2}+m (m (m+12)+131)+216\right)}{4 (m+3)} v^{3+m} + \ldots, 
  \nonumber\\ 
& c_1(v) =e_1 \frac{2 v^5}{5 (1+v)^4} + e_0 \sum_{m=1} (-1)^m \left(\frac{(m+1) (m (m (m+17)+342)+720)}{5(m+3)}- 24 (m+2) H_{m+1}\right) v^{m+3} + \ldots,
\end{align}
Above, the dots indicate terms proportional to $e_0^2$ whose explicit forms are cumbersome. The terms proportional to $e_1$ above were explicitly found in \cite{Banerjee:2023djb}. Note that only the term proportional to $e_1$ in $a_1(v)$ contributes to the energy density.

At higher orders, terms where $e_0$ do not appear are rational polynomials of $v$. Explicitly, as found in \cite{Banerjee:2023djb},
\begin{align} 
     a_2(v) &= \frac{12 e_1 v^6}{5 (v+1)^4}+\frac{ e_2 \left(9 v^2+10 v+10\right) v^4}{10 (v+1)^4} + \ldots, \,\, \nonumber \\
     b_2(v) &= \frac{ e_1 \left(-8 v^2+4 v-15\right) v^4}{5 (v+1)^6}+\frac{ e_2 \left(-7 v^2+4 v-25\right) v^4}{20 (v+1)^6}+ \ldots, \nonumber\\ 
      c_2(v) &= \frac{4 e_1 \left(6 v^2-7 v+14\right) v^5}{35 (v+1)^6}+\frac{ e_2 \left(8 v^2+7 v+35\right) v^5}{35 (v+1)^6},\nonumber\\
a_3(v) &= -\frac{e_1^2 v^8}{14 (v+1)^6}+\frac{2 e_1 \left(7 v^2-4 v-6\right) v^6}{5 (v+1)^6} 
     +\frac{6 e_2 \left(13 v^2+14 v+21\right) v^6}{35 (v+1)^6}
     \nonumber \\
     &+\frac{ e_3 \left(29 v^4+70 v^3+126 v^2+70 v+35\right) v^4}{35 (v+1)^6} + \ldots, \nonumber \\ 
b_3(v) &= \frac{5 e_1^2 v^8}{28 (v+1)^8}+\frac{ e_1 \left(-91 v^4+68 v^3-364 v^2-28 v-105\right) v^4}{35 (v+1)^8} 
     \nonumber \\
     &+\frac{e_2 \left(-82 v^4+84 v^3-588 v^2+56 v-315\right) v^4}{70 (v+1)^8}
     +\frac{ e_3 \left(-19 v^4+20 v^3-168 v^2+28 v-140\right) v^4}{70 (v+1)^8} + \ldots, \nonumber\\ 
     c_3(v) &= \frac{e_1^2 \left(-42 v^2-250 v-225\right) v^8}{1050 (v+1)^8}
     +\frac{4 e_1 \left(17 v^4-63 v^3+78 v^2-84 v-42\right) v^5}{105 (v+1)^8} 
     \nonumber \\
     &+ \frac{2 e_2 \left(8 v^4-2 v^3+66 v^2-21 v+28\right) v^5}{35 (v+1)^8}+\frac{2 e_3 \left(8 v^4+15 v^3+90 v^2+42 v+84\right) v^5}{105 (v+1)^8} + \ldots.\label{Eq:a3b3c3}
\end{align} 
Above, $\ldots$ denote terms where $e_0$ appears (i.e. terms proportional to $e_0$, $e_0^2$, $e_0 e_1$, etc.) which are not rational functions of $v$.  

In Sec.~\ref{Sec:Num}, we will explicitly see that numerical simulations of holographic Gubser flow vindicate the above analytic results. We will verify numerically that indeed at late de Sitter time behavior the energy density behaves as $\sigma^{-2}$ at the leading order and as $\sigma^{-4}$ at the sub-leading order. Furthermore, we will verify that indeed $P_T/\varepsilon$ vanishes while $P_L/\varepsilon$ goes to unity at late de Sitter time following \eqref{Eq:Ratio-lt}.

Although the behavior of the Gubser flow in a strongly coupled large N holographic gauge theory is free-streaming in transverse directions as in the case of kinetic theories, we note that our derivation points out that it can be different if the holographic gauge theory is non-conformal or confining. To be concrete, let the holographic gauge theory have a strongly coupled ultraviolet fixed point so that we can map the Gubser flow to evolution on dS$_3$ $\times$ $\mathbb{R}$ (at the boundary). We note that the bulk metric coincides with that dual to the vacuum solution at late de Sitter time (in contrast to the Bjorken flow where the late time behavior is a perfect fluid dual to an accelerating black brane). Therefore, in a holographic confining theory for example, we can expect that the infrared dynamics of the gauge theory on dS$_3$ $\times$ $\mathbb{R}$ at the (co-moving) confining scale would be geometrically captured by the bulk and directly play a role in determining the behavior of the Gubser flow at late de Sitter time via the requirement of the existence of a smooth future horizon.

\subsection{The behavior on the full future wedge}\label{Sec:Gub-IC}
As in the case of the bulk massless scalar field, for a solution of Einstein's equations to be physical, i.e. dual to states which can be prepared via local physical operations on the vacuum, it is not only necessary that the energy density in the dual state should have the late de Sitter time expansion \eqref{Eq:edSGen} so that the corresponding bulk solutions have a regular future horizon, but also the energy density in the dual state should decay at large distances from the central axis (i.e. at large $x_\perp $) at any fixed proper time $\tau$ on the future wedge in Minkowski space. Each term in the late de Sitter time expansion \eqref{Eq:edSGen}, i.e. $\exp(-(2 + 2k)\rho)$ with $k= 0,1,2,\ldots$ blows up at large $x_\perp$ at any fixed $\tau$ as $x_\perp^{4(k+1)}/\tau^{2(k+1)}$. Therefore, as in the case of the scalar field we can use a different set of bases for the functions $A$, $B$ and $C$ appearing in the metric \eqref{Eq:ABC-metric} which are compatible with the late time expansion \eqref{Eq:edSGen} (and \eqref{Eq:AnsGrav1}) and are as follows
\begin{align}\label{Eq:New-ABC-Expansion}
    A(r,\rho) = \sum _{n=0} \left(\frac{a^a_n\left(v\right)}{\cosh^{2 n + 2}\left(\varrho\right)}+\frac{a^b_n\left(v\right)}{\left(\sinh \left(\varrho\right)+\kappa \cosh \left(\varrho\right)\right)^{2 n +2}}\right), \nonumber \\
     B(r,\rho) = \sum _{n=0} \left(\frac{b^a_n\left(v\right)}{\cosh^{2 n + 2}\left(\varrho\right)}+\frac{b^b_n\left(v\right)}{\left(\sinh \left(\varrho\right)+\kappa \cosh \left(\varrho\right)\right)^{2 n +2}}\right), \nonumber \\
     C(r,\rho) = \sum _{n=0} \left(\frac{c^a_n\left(v\right)}{\cosh^{2 n + 2}\left(\varrho\right)}+\frac{c^b_n\left(v\right)}{\left(\sinh \left(\varrho\right)+\kappa \cosh \left(\varrho\right)\right)^{2 n +2}}\right),
\end{align}
with $\kappa > 1$ and constant. Each term in the above expansion decays at large $x_\perp$, e.g. $\cosh(\varrho)^{-2(n+1)}$ decays at large $x_\perp$ as $x_\perp^{-8(n+1)}$. We note that for each term in the late time expansion \eqref{Eq:AnsGrav1}, we have two terms above, which can be determined together by simultaneously solving in the late and early de Sitter time expansion. Furthermore, we note that for $\kappa > 1$, the bases, 
$$\cosh(\varrho)^{-2(n+1)} \quad \quad \text{and} \quad \quad \left(\sinh \left(\varrho\right)+\kappa \cosh \left(\varrho\right)\right)^{-2(n+1)},$$
are bounded functions in the domain $-\infty< \varrho < \infty$ (i.e. $0< \sigma < \infty$) when $\kappa > 1$. 

It follows that the energy density can be expanded as 
\begin{align}\label{Eq:New-e-Expansion}
   \varepsilon(\rho) =-\frac{3l^3}{16\pi G_N} \sum_{k=0}^\infty  \left(\frac{e^a_k}{\cosh{(\varrho)}^{2+ 2 k}} + \frac{e^b_k}{\left(\kappa \cosh{(\varrho)}+\sinh{(\varrho)} \right)^{2 + 2 k}} \right),
\end{align}
so that we obtain the expansion \eqref{Eq:edSGen} at late de Sitter time with
\begin{equation}\label{Eq:Late-e-coefficients}
    e_0 = 4 e^a_0 + \frac{4}{(\kappa + 1)^2} e^b_0, \,\, e_1 = -8 e^a_0 + \frac{8(\kappa -1)}{(\kappa + 1)^3} e^b_0 + 16 e^a_1 + \frac{16}{(\kappa + 1)^4} e^b_1, \,\, {\rm etc.},
\end{equation}
and the expansion 
\begin{equation}\label{Eq:edSgenearly}
 {\varepsilon}(\rho) = -\frac{3l^3}{16\pi G_N}\sigma^{2}L^{-4} \sum_{k =0}^\infty e^e_k\sigma^{2k},
\end{equation}
in the limit $\rho \rightarrow -\infty$ (i.e. $\sigma \rightarrow 0$) with
\begin{equation}\label{Eq:Early-e-coefficients}
    e^e_0 = 4 e^a_0 + \frac{4}{(\kappa - 1)^2} e^b_0, \,\, e^e_1 = -8 e^a_0 + \frac{8(\kappa +1)}{(\kappa - 1)^3} e^b_0 + 16 e^a_1 + \frac{16}{(\kappa - 1)^4} e^b_1, \,\, {\rm etc.}.
\end{equation}
Similarly, we obtain late and early time expansions of $A$, $B$ and $C$ from \eqref{Eq:New-ABC-Expansion}. The late time expansions are as in \eqref{Eq:AnsGrav1} with 
\begin{equation}
    a_0(v) = 4 a^a_0(v) + \frac{4}{(\kappa + 1)^2} a^b_0(v), \,\, a_1(v) = -8 a^a_0(v) + \frac{8(\kappa -1)}{(\kappa + 1)^3} a^b_0(v) + 16 a^a_1(v) + \frac{16}{(\kappa + 1)^4} a^b_1(v), \,\, {\rm etc.}.
\end{equation}
The early time expansions are 
\begin{align}\label{Eq:Ealy-metric-Expansion}
 A(r,\rho) =\sigma^2\sum_{k=0}^\infty a^e_k(v) \sigma^{2k}, \,\, {\rm etc.}
\end{align}
with 
\begin{equation}
    a^e_0(v) = 4 a^a_0(v) + \frac{4}{(\kappa - 1)^2} a^b_0(v), \,\, a^e_1(v) = -8 a^a_0(v) + \frac{8(\kappa +1)}{(\kappa -1)^3} a^b_0(v) + 16 a^a_1(v) + \frac{16}{(\kappa - 1)^4} a^b_1(v), \,\, {\rm etc.}.
\end{equation}
We can explicitly check (as discussed below) that the expansions \eqref{Eq:New-ABC-Expansion} converge in the entire domain $-\infty < \varrho < \infty$ especially if the late de Sitter time expansion of the energy density \eqref{Eq:edSGen} has a finite truncation. Also as shown below, we can construct the full gravitational solution in the full domain $-\infty< \varrho < \infty$ in the expansions \eqref{Eq:New-ABC-Expansion} in the entire domain $-\infty < \varrho < \infty$ for a generic late de Sitter time expansion \eqref{Eq:edSGen}. Crucially, our numerical simulations of holographic Gubser flow in Sec.~\ref{Sec:Num} will validate the conclusions that we will obtain here from the analytic perturbative expansion \eqref{Eq:New-ABC-Expansion} with our choices of bases.

As in the case of the massless scalar field we can obtain $a^a(v)$, $a^b(v)$, etc. in \eqref{Eq:New-ABC-Expansion} explicitly by solving the gravitational equations perturbatively at late and early de Sitter times in power series of $\sigma^{-2}$ and $\sigma^2$, respectively. At each order, we would require (i) the solutions are normalizable, and (ii) the solutions are regular at late de Sitter time at the future horizon $v=1$. These requirements imply that we have one integration constant at each order (so that $e^a_k$ and $e^b_k$ are not independent) which determines both the corresponding coefficients $e_k$ and $e^e_k$ in the late and early de Sitter time expansions of the energy density.\footnote{We set the coefficient of $v$ in both $a^a_k(v)$ and $a^b_k(v)$ to zero to turn off the pure gauge mode at each order.} Furthermore, we find that requiring regularity at the future horizon automatically implies that the solutions are regular at $v=1$ for all time at each order. 

For $\kappa =2$, we explicitly obtain the following to leading order:
\begin{align}
  &  a^a_0(v)= -a^b_0(v) = \frac{3}{4} \Gamma_0 v ((v+2) \ln (v+1)-2 v),\nonumber \\ 
   & b^a_0(v) = - b^b_0(v) = \frac{3 \Gamma_0 \left(2 \left(v^2+4 v+3\right) \ln (v+1)-v (5 v+6)\right)}{16 (v+1)^2},\nonumber\\  
   & c^a_0(v) = -c^b_0(v)= \frac{\Gamma_0 \left(v \left(v^2+12 v+12\right)-6 \left(v^2+3 v+2\right) \ln (v+1)\right)}{4 (v+1)^2}.
\end{align}
Using \eqref{Eq:Late-e-coefficients}, \eqref{Eq:Early-e-coefficients} and $\kappa =2$, we find
\begin{equation}
e^a_0 = - e^b_0 = \frac{\Gamma_0}{8}, \,\, e_0 = \frac{4\Gamma_0}{9},\,\, e^e_0 =0.
\end{equation}
We recover \eqref{Eq:a0b0c0} at late de Sitter time and 
\begin{equation}
a^e_0(v) = b^e_0(v) = c^e_0(v)=0.
\end{equation}

At the sub-leading order, we obtain
\begin{align}
   & a^a_1(v) = \frac{\Gamma_1 v^4 }{80 (v+1)^2} \nonumber\\&+ \frac{3 \Gamma_0 v \left(32 (v+1)^2 (7 v+23) \ln (v+1)-v \left(581 v^2+1328 v+736\right)\right)}{640 (v+1)^2} + \ldots \nonumber \\
   & a^b_1(v) = -\frac{\Gamma_1 v^4}{80 (v+1)^2}\nonumber\\& + \frac{3 \Gamma_0 v \left(96 (v+1)^2 (4 v+11) \ln (v+1)-v \left(901 v^2+1968 v+1056\right)\right)}{640 (v+1)^2}+ \ldots, 
\end{align}
with $\ldots$ denoting the term proportional to $\Gamma_0^2$ when $\kappa =2$. The term proportional to $\Gamma_0^2$ cannot be written in a closed form but we can determine it in series expansion about $v=0$ or $v=1$. We recover the sub-leading terms \eqref{Eq:a1b1c1} in the late de Sitter time expansion. Crucially, we find that
\begin{equation}
    a^e_1(v)=b^e_1(v) =c^e_1(v)=0, \quad e^e_1(v) =0.
\end{equation}

From these perturbative expansions, we find that $a^e_k(v)$, $b^e_k(v)$, $c^e_k(v)$ and $e^e$ vanish order by order for any $\kappa>1$. We can conclude that 
\begin{equation}\label{Eq:e-early-sigma}
    \lim_{\rho\rightarrow-\infty} \sigma^n A(r, \rho)= \lim_{\rho\rightarrow-\infty} \sigma^n B(r, \rho)= \lim_{\rho\rightarrow-\infty} \sigma^n C(r, \rho)= \lim_{\rho\rightarrow-\infty} \sigma^n \varepsilon (\rho) =0, \,\, \forall n\geq 0.
\end{equation}
On the future wedge of R$^{3,1}$, we find after the required Weyl scaling and coordinate transformation that
\begin{equation}\label{Eq:e-early-tau}
     \lim_{\tau\rightarrow 0^+}\tau^n \varepsilon_M (\tau, x_\perp) = 0\,\,\,\, \forall n\geq 0.
\end{equation}
Therefore, as in the case of the massless scalar we find that any physical Gubser flow solution can be smoothly glued to the vacuum (pure AdS$_5$) outside the future wedge, exactly like in the case of the massless scalar field. The energy density and all its derivatives w.r.t. $\tau$ vanishes as $\tau\rightarrow 0^+$. Also, 
\begin{equation}\label{Eq:e-large-xperp}
     \lim_{x_\perp\rightarrow 0}x_\perp^n \varepsilon_M (\tau, x_\perp) = 0\,\,\,\,\forall n\geq 0,
\end{equation}
as in the case of the scalar field. We recall that the above and \eqref{Eq:e-early-tau} are connected as the limit $\rho\rightarrow-\infty$ corresponds to both large $x_\perp$ at any fixed $\tau$ and small $\tau$ at any fixed $x_\perp$. Therefore, \eqref{Eq:e-early-sigma} implies both of these, and  \eqref{Eq:e-early-tau} and \eqref{Eq:e-large-xperp} are equivalent.

The result that all the early time coefficients $e^e_k$ in the early de Sitter time expansion \eqref{Eq:edSgenearly} vanish implies that $e^a_k$ and $e^b_k$ in \eqref{Eq:New-e-Expansion} are related at each order, and both are determined uniquely by the coefficients $e_k$ of the late de Sitter time expansion \eqref{Eq:edSGen}. Using this, we find that the resummation of the series \eqref{Eq:New-e-Expansion} gives (recall the discussion is Sec. \ref{Sec:Basis})
\begin{align}\label{Eq:e-specify-k}
    \varepsilon(\rho) = E_\kappa (\sigma)
   &= -\frac{3l^3}{16\pi G_N}\Bigg(\sum_{k=0}^\infty e_k \sigma^{-(2 +2k)}\theta\left(\sigma - 1\right)\nonumber\\
   &+  \sum_{k=0}^\infty e_k\frac{(\kappa+1)^{2+2k}\sigma^{4+4k}-(\kappa-1)^{2+2k}}{\left((\kappa+1)^{2+2k}-(\kappa-1)^{2+2k}\right)\sigma^{2+2k}} \theta\left(1- \sigma\right)\theta\left(\sigma - \sqrt{\frac{\kappa-1}{\kappa +1}}\right)\Bigg).
\end{align}
We find that all $E_\kappa(\sigma)$ agree with each other for $0<\rho<\infty$ (i.e. $1<\sigma<\infty$) giving just the same late de Sitter time expansion. They differ for $\rho <0$. Note $E_\kappa(\sigma)$ is continuous in the entire domain $-\infty < \rho < \infty$ (i.e. $0< \sigma <\infty$). However, $E_\kappa(\sigma)$ can have kinks at $\sigma =1$ where $(\cosh\varrho)^{-1}$ is peaked and at $\sigma = \sqrt{(\kappa -1)/(\kappa +1)}$ where $(\kappa \cosh\varrho + \sinh\varrho)^{-1}$ is peaked. Furthermore, $E_\kappa(\sigma)$ vanishes for $0< \sigma \leq \sqrt{(\kappa -1)/(\kappa +1)}$. 

\begin{figure}[ht]
   \centering
        \includegraphics[width=0.6 \linewidth]{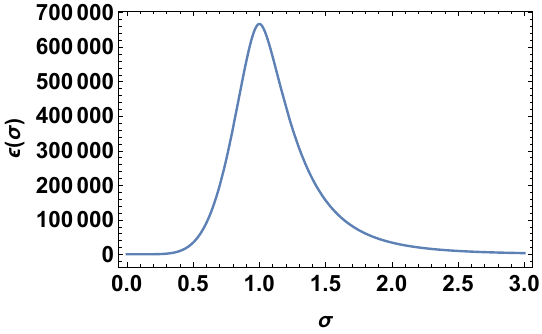}
   \caption{The function $\varepsilon(\sigma) = E_\kappa(\sigma)$ with $E_\kappa(\sigma)$ defined in \eqref{Eq:e-specify-k}, and with the choices of coefficients \eqref{Eq:e-coeffs-example}, $e_0 = 0.001$ and $\kappa =1.001$. Here $E_\kappa(\sigma)$ is suitably normalized by stripping off the factor proportional to N$^2$.}
   \label{Fig:E-example}
 \end{figure}

We note that the series \eqref{Eq:e-specify-k} should have the extra property that $\varepsilon'(\rho=0) =0$ in order that the transverse pressure $P_T(\rho)$ and longitudinal pressure $P_L(\rho)$ given by \eqref{Eq:PT} and \eqref{Eq:PL}, respectively, are finite at $\rho = 0$ (i.e. at $\sigma =1$). This is indeed borne out by explicit numerical simulations as reported in the next section. One can readily construct $E_\kappa(\rho)$ which satisfies this property and is also $C^\infty$ function in the entire domain $-\infty < \rho < \infty$ as follows. Firstly, let us choose
\begin{align}\label{Eq:e-coeffs-example}
    &e_1 = -e_0\frac{(1+\kappa^2)(1+3\kappa^2)}{(\kappa^2-1)^2}, \,\, e_2 = e_0\frac{(3+\kappa^2)(1+3\kappa^2)^2}{3(\kappa^2-1)^3},\nonumber\\
    &e_3 = -e_0\frac{1+7\kappa^2+7\kappa^4+\kappa^6}{(\kappa^2-1)^3},\,\, e_n = 0\,\, \forall n\geq 4.
\end{align}
We can verify that with these choices we can satisfy
\begin{itemize}
    \item $E'_\kappa(\sigma)$ is continuous at $\sigma = 1$ and $\sigma = \sqrt{(\kappa -1)/(\kappa +1)}$, and
    \item $E'_\kappa(\sigma =1 ) = 0$.
\end{itemize}
In Fig.~\ref{Fig:E-example} we plot $E_\kappa(\sigma)$ with the choices of coefficients \eqref{Eq:e-coeffs-example}, $e_0 = 0.001$ and $\kappa =1.001$. Similarly by adding $2n+1$ non-vanishing terms $e_{i}$ in the late de Sitter time expansion with $i>0$ and relating them to $e_0$ appropriately we can make the first n-derivatives of $E_\kappa(\sigma)$ continuous at $\sigma = 1$ and $\sigma = \sqrt{(\kappa -1)/(\kappa +1)}$, and ensure that $E_\kappa'(\rho=0) =0$. We obtain a $C^\infty$ function in the entire domain $-\infty <\rho < \infty$ in the limit $n\rightarrow\infty$.

We also note that for each $E_\kappa(\sigma)$ we can construct the bulk solutions with $A(r,\rho)$, $B(r,\rho)$ and $C(r,\rho)$ given by similar expansions that are determined by the late de Sitter time expansion coefficients $a_k(v)$, $b_k(v)$ and $c_k(v)$, respectively. We note that each choice of $\kappa$ gives a different bulk solution corresponding to the same late de Sitter time expansion. As in the case of the scalar field, we expect that one needs infinitely more data than the coefficients of the late de Sitter time expansion to recover the initial conditions set at an early value of the de Sitter time.

It is of course not clear whether the choice of bases in \eqref{Eq:New-ABC-Expansion} for the global expansion can give the most generic regular solutions. Regardless, we will show in the next section that our general conclusions, particularly \eqref{Eq:e-large-xperp} and therefore \eqref{Eq:e-early-tau} hold for numerical holographic Gubser flow set up with smooth initial data at an early de Sitter time.

\subsection{Can Gubser flow describe a jet?}\label{Sec:Gub-jet}

We have shown that the gravitational solutions dual to Gubser flow in strongly coupled large N holographic conformal field theories can be constructed in an expansion valid for all de Sitter time utilizing a class of bases. All these solutions correspond to an energy density that decays at large distances from the central axis at any value of the proper time on the future wedge of R$^{3,1}$. Our solutions have the properties \eqref{Eq:e-early-tau} and \eqref{Eq:e-large-xperp}  which are equivalent and also imply that any Gubser flow  can be smoothly glued to the vacuum outside the future wedge. Therefore a Gubser flow cannot be generated from non-trivial in-states such as colliding gravitational shock waves (in stark contrast to holographic Bjorken flow studied in \cite{Chesler:2008hg}). Since the generality of our conclusions could be limited by the choice of bases utilized in our constructions, we will study numerical gravitational solutions dual to a Gubser flow in the following section with generic smooth initial conditions at an initial value of de Sitter time, and demonstrate that our conclusions are indeed valid generally. 

However, these conclusions rely on the assumption that the Gubser flow is valid for the full medium. It is possible that the energy density has a $\theta$ dependence (recall that $\theta$ is the azimuthal angle on S$^2$ in  dS$_3$) such that the derivatives w.r.t. $\theta$ are negligible only for certain values of $\theta$ so that Gubser flow approximates only a part of the full medium where the energy density depends only on the de Sitter time $\varrho$. In this case, our conclusions may not be valid as our arguments rely on the behavior of the Gubser flow in the late de Sitter time regime which cannot be realized as we show below.

To discuss this more concretely, let us denote $qx_\perp$ and $q\tau$ as $x$ and $t$, respectively. Let us recall \eqref{Eq:theta} which gives how $\theta$ depends on $x$ and $t$, and set $\theta(x, t) = \theta_0$, a constant. Using \eqref{Eq:theta}, we can determine how $x$ depends on $t$ when the value of $\theta$ is fixed to $\theta_0$. Explicitly, $\theta(x, t) = \theta_0$ implies that 
\begin{align}
    x_{\theta_0} (t)= \begin{cases}
        \cot\theta_0 \left(-1 + \sqrt{1+ (1+ t^2)\tan^2\theta_0 }\right),\quad &{\rm for}\,\, 0 < \theta_0 < \frac{\pi}{2}, \\
        \sqrt{1+ t^2},\quad &{\rm for}\,\, \theta_0 = \frac{\pi}{2},\\
        -\cot\theta_0 \left(1 + \sqrt{1+ (1+ t^2)\tan^2\theta_0 }\right),\quad &{\rm for}\,\, \frac{\pi}{2} < \theta_0 < \pi.
    \end{cases}
\end{align}
At small $t$,
\begin{equation}
    x_{\theta_0} (t) = \tan\left(\dfrac{\theta_0}{2}\right) + \frac{1}{2}\sin\theta_0 t^2 + \mathcal{O}(t^2),
\end{equation}
and at large $t$,
\begin{equation}
    x_{\theta_0} (t) = t - \cot\theta_0 + \mathcal{O}(t^{-1}),
\end{equation}
in all these three cases. Thus $x_{\theta_0} (t)$ increases monotonically as $t$ goes from $0$ to $\infty$ with minimum value $\tan(\theta_0/2)$, and at large $t$, $x_{\theta_0}(t)\sim t$. It follows that when $\theta$ takes the value $\theta_0$, then $\varrho$ depends on $t$ as $\varrho_{\theta_0}(t)$ which can be obtained from \eqref{Eq:rho} and is explicitly as follows
\begin{align}
    &\varrho_{\theta_0}(t) = \varrho(t, x = x_{\theta_0} (t))\nonumber\\
    & =\begin{cases}
    -{\rm arcsinh}\left(\frac{1-\cot^2\theta_0\left(-1 + \sqrt{1+ (1+ t^2)\tan^2\theta_0 } \right)}{t}\right),\quad {\rm for}\,\, 0 < \theta_0 < \frac{\pi}{2}, \\
    -{\rm arccsch}(t),\quad {\rm for}\,\, \theta_0 = \frac{\pi}{2},\\
     -{\rm arcsinh}\left(\frac{1+\cot^2\theta_0\left(1 + \sqrt{1+ (1+ t^2)\tan^2\theta_0 } \right)}{t}\right) ,\quad {\rm for}\,\, \frac{\pi}{2} < \theta_0 < \pi.
  \end{cases}  
\end{align}
We note that $\varrho_{\theta_0}(t)$ is a monotonically increasing function. At small $t$, $\varrho_{\theta_0}(t)$ behaves as
\begin{equation}
\varrho_{\theta_0}(t) = \ln t+\ln\left(\dfrac{1+ \cos\theta_0}{2}\right)+\mathcal{O}(t^{-1}),
\end{equation}
and at large $t$, $\varrho_{\theta_0}(t)$ behaves as
\begin{equation}
\varrho_{\theta_0}(t) = {\rm arcsinh}(\cot\theta_0) - \frac{\csc\,\theta_0}{t} +\mathcal{O}(t^{-2}),
\end{equation}
in all three cases. As $t\rightarrow 0$, $\varrho_{\theta_0}(t)$ goes to $-\infty$ as $\ln t$. Crucially, $\varrho_{\theta_0}(t)$ reaches the maximum value ${\rm arcsinh}(\cot\theta_0)$ in the limit $t\rightarrow \infty$. When $\theta_0 = \pi/2$, i.e. on the equator of S$^2$, $\varrho_{\theta_0}(t)$ goes to $0$ when $t\rightarrow\infty$.

Let the energy density be such that the derivatives with respect to $\theta$ are small on the equator (i.e. when $\theta$ takes the value $\pi/2$). In this case, we can approximate the evolution in terms of Gubser flow in the equatorial region of S$^2$ if the energy density depends on $\varrho$ only in this region. However, at large $t$ we do not reach the large $\rho$ regime but rather the regime $\rho \sim 0$ where hydrodynamic description can be valid. Our arguments about the initial conditions for Gubser flow involved matching with the large $\rho$ expansion given by \eqref{Eq:AnsGrav1} and \eqref{Eq:edSGen}. But if the Gubser flow is restricted only to the equatorial region of S$^2$, then this large $\rho$ regime is never attained, and our conclusions about the initial conditions generating such a flow obtained should be modified too as the expansion \eqref{Eq:New-e-Expansion} need not be valid. 

It is possible that the Gubser flow conditions can be applicable to jets which approximate the Gubser flow type behavior only near the equatorial region of S$^2$ when the energy density is mapped from the future wedge of R$^{3,1}$ to dS$_3$ $\times$ $\mathbb{R}$ using \eqref{Eq:rho} and \eqref{Eq:theta}.\footnote{Here by jets we simply mean a part of the medium which behaves distinctly from the rest. Especially, this part may not hydrodynamize or do so in a distinctive way. Of course, our present discussion is in the context of a conformal gauge theory, and not QCD.} In this case, the behavior of such jets would approximate hydrodynamic behavior at large proper time (where $\rho\sim 0$ as discussed above) if they carry sufficient energy flux as discussed in Sec.~\ref{Sec:Non-Hydro}. In the future, it would be interesting to construct such dual gravitational solutions explicitly with the Gubser flow approximation valid only near the equatorial region of S$^2$ and with the remaining medium realizing the generic hydrodynamic behavior at late proper time and also the whole flow realized from gravitational shock wave collisions. It will also be interesting to see a solution where the Gubser flow can be embedded within a Bjorken flow, i.e. with the energy density not depending on the spatial transverse coordinates in most of the medium. We will discuss these issues more in Sec.~\ref{Sec:Disc}.

\section{Numerical validation}\label{Sec:Num}

In the previous section, we have analytically determined the general behavior of the holographic Gubser flow at late de Sitter time. Furthermore, we have argued that the holographic Gubser flow in the future wedge of R$^{3,1}$ can be glued smoothly to the vacuum outside the future wedge. Although the generic late de Sitter time behavior of the Gubser flow, namely that $P_T/\varepsilon \rightarrow0$ while $P_L/\varepsilon \rightarrow1$ follows from the regularity of the future horizon, the argument about the initial vacuum-like initial conditions of the Gubser flow do rely on the choice of a class of bases for the perturbative expansion which is valid for all de Sitter time. Therefore, it is important that we verify whether our conclusions hold for generic states exhibiting Gubser flow. This can be studied via numerical relativity by setting up generic smooth initial conditions at a finite value of de Sitter time. In this section, we show that our analytic results are indeed valid for generic smooth initial conditions. First, we numerically verify both the leading transverse free-streaming and sub-leading color glass condensate-like behavior of the holographic Gubser flow at late de Sitter time. Second, as in the case of the massless scalar field in Sec.~\ref{Sec:Num-1}, we verify that the asymptotic behavior of the energy density at large distances from the central axis is indeed as in \eqref{Eq:e-large-xperp} which implies the early proper time behavior \eqref{Eq:e-early-tau}.

\subsection{Numerical setup}

The numerical simulation of the holographic Gubser flow involves implementing the method of characteristics for solving Einstein's equations and adopting suitable variables which lead to reformulating these equations as a set of nested ordinary differential equations to be solved at each time step. This can be achieved by adopting the general methodology described in \cite{Chesler:2013lia}. 
It turns out that the nested equations take a simple form by redefining the variables in \eqref{Eq:ABC-metric} with which the bulk metric becomes
\begin{align} \label{Eq:newmetric}
   &{\rm d}s^2 = -\frac{2 l^2}{\rL^2} {\rm d} \rL {\rm d} \rhoL -  \at(\rL, \rhoL) {\rm d}\rho^2 + \ct(\rL, \rhoL)^2 \Big(e^{\bt(\rL, \rhoL)} \left({\rm d}\theta^2 + \sin^2{\theta} {\rm d}\phi^2 \right) + e^{-2 \bt(\rL, \rhoL)} {\rm d}\eta^2  \Big).
\end{align}
The new metric variables $\at(\rL, \rhoL)$, $\bt(\rL, \rhoL)$ and $\ct(\rL, \rhoL)$ are related to those in \eqref{Eq:ABC-metric} via
\begin{align}
 &\at(\rL, \rhoL) =   \frac{l^2}{\rL^2} (1 - \rL^2  + A\left(\rL, \rhoL \right)), \nonumber \\
 &\bt(\rL, \rhoL) = B(\rL, \rho) + \frac{2}{3} \ln \left( \cosh{\rhoL} + \rL \sinh{\rhoL} \right) - \frac{1}{3} C(\rL, \rhoL), \nonumber \\
 &\ct(\rL, \rhoL) = \frac{l}{\rL} L e^{\frac{C(\rL, \rhoL)}{6}} \left( \cosh{\rhoL} + \rL \sinh{\rhoL}\right)^{2/3}.
\end{align}
They have the following near-boundary expansions with the residual gauge fixing parameter $\tilde{a}_1$ (the coefficient of $\rL$ in the expansions of $\tilde{A}$) set to zero (and $l=1$):
\begin{align}
 &\at(\rL, \rhoL) =   \frac{1}{\rL^2} \left( 1-\rL^2 + \rL^4 {a}_4 \right) + \ldots  \label{At-nearboundary} \\
 &\bt(\rL, \rhoL) = \frac{2}{3} \ln(\cosh{\rhoL}) +\frac{1}{3} \rL \tanh{\rhoL} \left( 2 - \rL \tanh{\rhoL} +  \frac{2}{3} \rL^2 \tanh^2{\rhoL} \right) + \rL^4 \tilde{b}_4 + \ldots \label{Bt-nearboundary} \\ 
& \ct(\rL, \rhoL) = L\left(\frac{\cosh^{2/3}{\rhoL}}{\rL} + \frac{4 \sinh{\rhoL}}{6 \cosh^{1/3}{\rhoL}} - \rL \frac{\sinh^2{\rhoL}}{9 \cosh^{4/3}{\rhoL}} + \ldots\right). \label{Ct-nearboundary}
\end{align}
Consequently, Einstein's equations take the following compact form
\begin{align}
  \partial_{\rL}^2\ct &= - \frac{\ct {(\partial_{\rL}\bt)}^{2}}{2} - \frac{2 \partial_{\rL}\ct}{\rL} \label{Eq:charac1}, \\
  \partial_{\rL}d_+{\ct} &= - \frac{1}{\rL^2 \ct} \left( e^{2 \bt} + 6 \ct^2 + 6 \rL^2 d_+{\ct}\, \partial_{\rL}\ct \right) \label{Eq:charac2}, \\
 \partial_{\rL}d_+{\bt} &= \frac{e^{-\bt}}{3 \rL^2 \ct^2} - \frac{9 (d_+{\ct} \,\partial_{\rL}\bt + d_+{\bt}\, \partial_{\rL}\ct)}{6 \ct} \label{Eq:charac3}, \\
\nonumber  { \partial_{\rL}^2\at} &= \frac{6}{\rL^2 \ct}(d_+{\ct}\, \partial_{\rL}\bt + d_+{\bt} \, \partial_{\rL}\ct) - \frac{1}{3 \rL^4 \ct^2} \left( 2 e^{-\bt} + 36 \rL^2 d_+{\ct}\,  \partial_{\rL}\ct\right)    \\
   & - \frac{1}{3 \rL^2} \left( \frac{12}{\rL^2} + 6 \rL\, \partial_{\rL}\at - 9 d_+{\bt}\,  \partial_{\rL}\bt - 12 \partial_{\rL}\, d_+{\bt} \right) \label{Eq:charac4}, \\
     d_+^2{\ct} &= -\frac{1}{2} \left( \rL^2 d_+{\ct}\, \partial_{\rL}\at - (d_+{\bt})^2 \ct \right),\label{Eq:charac5}
    \end{align}
where $d_+$ denotes the directional derivative along the outgoing null geodesic, i.e.
\begin{align}\label{Eq:directional_derivative}
d_+{f}  =  \partial_\rhoL f - \frac{1}{2} \rL^2 \at ~ \partial_\rL f.\end{align}

With a given initial radial profile $\bt(v)$ and an initial value for ${a}_4$ at an initialization time $\varrho = \rhoL_{\rm in}$, one can determine a unique evolution of the full metric numerically as follows. The equations \eqref{Eq:charac1}-\eqref{Eq:charac4} can be solved sequentially via radial integration in this order yielding $\ct$, $d_+{\ct}$, $d_+{\bt}$ and $\at$, given $\bt$ and $a_{(4)}$ at each time step. We note that $d_+{\ct}$ and $d_+{\bt}$ can be used to obtain $\partial_\rhoL \bt$ and $\partial_\rhoL \ct$ with which we can evolve $\bt$ and $\ct$ to the next time step. The integration constants for the radial integration at the each time step can be determined by requiring that the boundary metric \eqref{Eq:dSds2} is unperturbed and fixing the residual gauge freedom by setting $a_{(1)} =0$ in the asymptotic expansions \eqref{Eq:near_boundaryExpA}-\eqref{Eq:near_boundaryExpS}.

The two constraints \eqref{Eq:constraint} obtained from the asymptotic expansions can be used to eliminate ${s}_4$ and  express ${a}_{(4)}'$ in terms of $b_{(4)}$. The latter expression can be used to update $\tilde{a}_{(4)}$ to the next time step. Note that \eqref{Eq:charac1} and \eqref{Eq:charac5} act as constraints. We use \eqref{Eq:charac5} also to monitor the accuracy of the numerical evolution by computing it at all radial locations. 

We have done the radial integrations using the pseudo-spectral method by discretizing the radial direction with a Chebyshev grid of 30 grid points. Time steppings have been done with the 4-th order Adams-Bashforth method with a time-step of 0.001 in units $L=1$. Numerical convergence has been checked and reported in Appendix \ref{app:convergence}.

One of the simplest initial profiles of $\bt$ is the {Gaussian-like} profile
\begin{align}\label{Eq:profile}
   \bt(\rhoL_{\rm in},v) = \tilde{b}_{4,{\rm in}} \rL^4 e^{-\rL^2}
\end{align}
which depends only on $ \tilde{b}_{4,{\rm in}}$, the initial value of $\tilde{b}_4$. We also choose other initial profiles which are rational polynomial functions of $\rL$, such as
 \begin{align} \label{Eq:profile-1}
     &\bt (\rhoL_{\rm in},v)= \tilde{b}_{4,{\rm in}}\rL^4, \,\,\,  \bt(\rhoL_{\rm in},v) = \tilde{b}_{4,{\rm in}} \frac{\rL^4}{1+\rL^2}, \nonumber\\& \bt(\rhoL_{\rm in},v) = \tilde{b}_{4,{\rm in}}\frac{ \left(-0.625 \rL^4+0.662 
     \rL^3+0.967 \rL^2\right)}{0.18 \rL^2-0.3045 \rL -0.494}, \,\, \ldots
\end{align} 
which are $C^\infty$ functions in the domain $0\leq v < \infty$.

As mentioned above, the choice of $b_{(4)}$ at initial time, and therefore $\tilde{b}_{4,{\rm in}}$, together with the choice of $a_{(4)}$ at initial time determine $a'_{(4)}$. Therefore, by choosing $\tilde{b}_{4,{\rm in}}$ appropriately, we can choose the initial value of the pressures $P_T$ (or $P_L$) via \eqref{Eq:PT} (or \eqref{Eq:PL}). To get insights into the evolution, it is useful to choose the initial values of $\varepsilon$ and $P_T$ by choosing the initial value of $a_{(4)}$ and $\tilde{b}_{4,{\rm in}}$.

\subsection{Verification of the leading and sub-leading exponents of the late de Sitter time behavior}

\begin{figure}
     \centering
          \includegraphics[width=0.49 \linewidth]{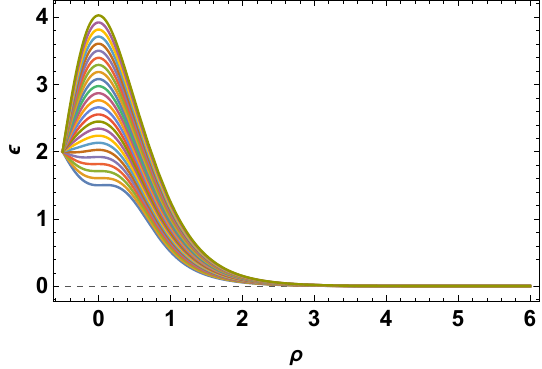}
            \includegraphics[width=0.49 \linewidth]{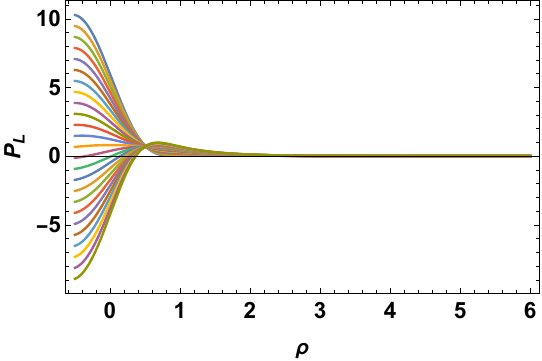}
               \includegraphics[width=0.49 \linewidth]{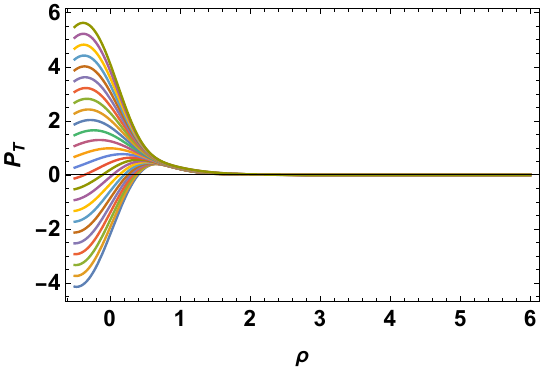}
     \caption{The evolution of the energy density $\varepsilon$, longitudinal pressure, $P_L$, and transverse pressure, $P_T$, is shown as a function of $\rho$ for various initial values $\tilde{b}_{4,{\rm in}}$ in units $L=1$. (Also $\varepsilon$, $P_T$ and $P_L$ have been normalized by stripping off the $N^2$ factor.) The initial value of the energy density $\varepsilon$ at $\rhoL_{\rm{in}} = -0.5$ is set to 2. The initial conditions are set by \eqref{Eq:profile}. The value of $\tilde{b}_{4,{\rm in}}$ varies from $-0.18$, which corresponds to the outermost blue plots, to $0.18$ corresponding to the outermost green plots. We note that larger $\tilde{b}_{4,{\rm in}}$ gives higher initial values of $P_T$ and lower initial values of $P_L$. The energies remain ordered throughout the evolution. 
     }
     \label{fig:late_time_1}
 \end{figure}

 \begin{figure}
     \centering
          \includegraphics[width=0.49 \linewidth]{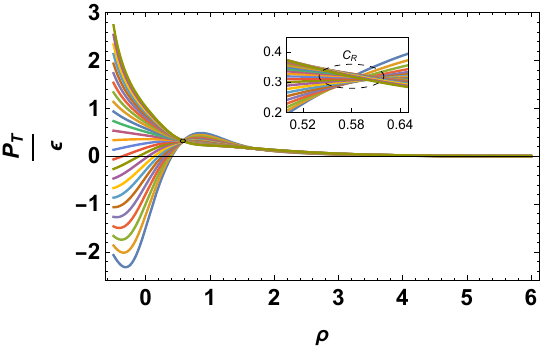}
            \includegraphics[width=0.49 \linewidth]{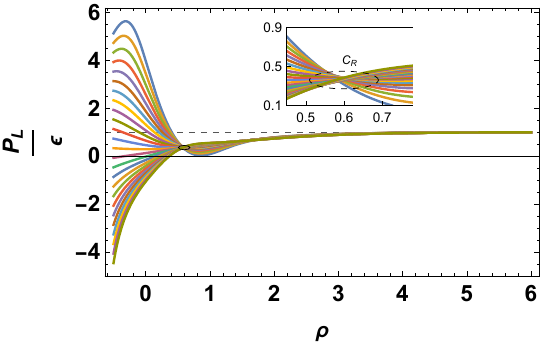}
     \caption{The free streaming behaviour of the flow in the transverse direction at the late de Sitter time where $ P_T/\varepsilon \rightarrow 0$ and $P_L/\varepsilon \rightarrow 1$. Here the initial conditions and the color codes are the same as in Fig.~\ref{fig:late_time_1}. The inset shows the bottleneck region $C_R$ where all the curves converge approximately to a (smeared) fixed point before spanning out again and reaching the free streaming behaviour. More precisely, $C_R$ is the region bounding the intersections of the curves. The dashed line in the right figure corresponds to $P_L/\varepsilon\rightarrow 1.$  }
     \label{fig:late_time_2}
 \end{figure}

\begin{figure}
     \centering
          \includegraphics[width=0.49 \linewidth]{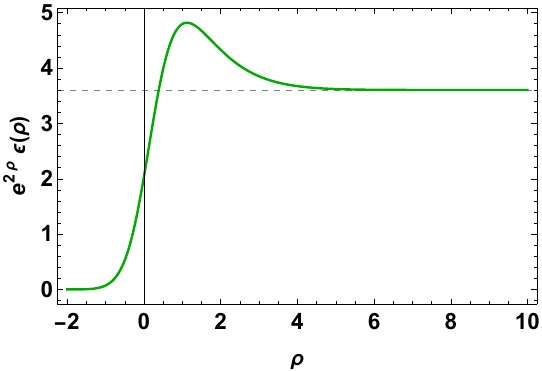}
     \caption{Here we show that the energy density multiplied with $e^{2 \rho}$ at large $\rho$ goes to the constant $e_0$. We have chosen the initial energy density to be $10^{-5}$ at initialization time $\rhoL_{\rm{in}} = -2$ and the anisotropy coefficient $\tilde{b}_{4,{\rm in}} = -0.144$ for the initial profile \eqref{Eq:profile}. The dashed line marks the value of $e_0$, which in this case is $\sim 3.602$.}
     \label{fig:c0coeff_taylor}
 \end{figure}

We can readily analyze the generic behaviour of the flow at large $\rhoL$, and verify the leading and sub-leading exponent of the late de Sitter time expansion \eqref{Eq:edSGen}. With initial radial profile of $\bt$ given by \eqref{Eq:profile} and a fixed energy density $\varepsilon = - 2$ (in units $L=1$) at initialization time $\varrho_{\rm in} = -0.5$, the (de Sitter) time evolutions of $\varepsilon$, $P_T$ and $P_L$ for varying initial values of the transverse pressure $P_T$ have been plotted in Fig.~\ref{fig:late_time_1}. The corresponding (de Sitter) time evolutions of $P_T/\varepsilon$ and $P_L/\varepsilon$ in Fig.~\ref{fig:late_time_2} confirm the general predictions of the late de Sitter time expansion \eqref{Eq:edSGen} that $P_T/\varepsilon \rightarrow 0$ and $P_L/\varepsilon \rightarrow 1$ in the limit $\varrho\rightarrow\infty$. Furthermore, as shown in Fig.~\ref{fig:c0coeff_taylor}, at late de Sitter time $e^{2\varrho}\varepsilon$ approaches a constant as expected from the leading exponent in \eqref{Eq:edSGen}.

\begin{figure}
     \centering
          \includegraphics[width=0.49 \linewidth]{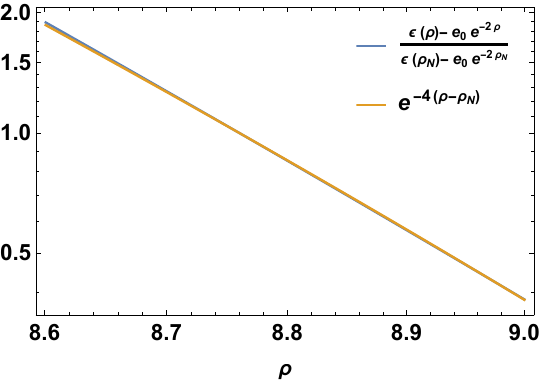}
     \caption{This figure verifies the sub-leading exponent of the large $\rho$ expansion. The initial conditions are the same as in Fig.~\ref{fig:c0coeff_taylor}. The blue plot is the normalized interpolated energy density minus the leading $e^{-2\varrho}$ falloff. The yellow plot is a fit at $\rho_N=8.8$ with $e^{-4\varrho}$ falloff. 
    }
     \label{fig:subleading_expo}
 \end{figure}

In order to verify the sub-leading behavior of $\varepsilon$ at late de Sitter time, it is convenient to choose a smaller value of the initial energy density which ensures a higher numerical accuracy. Subtracting the leading $e^{-2\varrho}$ term from $\varepsilon$ by extracting the leading coefficient $e_0$ of the late de Sitter time expansion \eqref{Eq:edSGen} as in Fig.~\ref{fig:c0coeff_taylor}, we can verify that the sub-leading term fits very well with $e^{-4\varrho}$ as demonstrated in Fig.~\ref{fig:subleading_expo}. Thus we can numerically validate the sub-leading term of the late de Sitter expansion \eqref{Eq:edSGen} of the energy density. These also hold for other types of initial conditions such as those given by \eqref{Eq:profile-1}.

\subsection{The validation of the behavior of Gubser flow on the future wedge}

In order to validate that the energy density indeed decays faster than any power on the future wedge of R$^{3,1}$ in the limits $x_\perp \rightarrow \infty$ at fixed $\tau$, and $\tau \rightarrow 0$ at fixed $x_\perp$ as stated in \eqref{Eq:e-large-xperp} and \eqref{Eq:e-early-tau} respectively, we can proceed as in the case of the massless scalar field in Sec.~\ref{Sec:Num-1}. We also recall that \eqref{Eq:e-large-xperp} and \eqref{Eq:e-early-tau} imply each other which are equivalent to the de Sitter time behavior \eqref{Eq:e-early-sigma} in the limit $\sigma\rightarrow 0$. As mentioned in Sec.~\ref{Sec:Num-1}, it is not easy to verify \eqref{Eq:e-early-sigma} directly by evolving back in de Sitter time with negative time steppings as the constraints are quickly violated. Similarly, it is also difficult to evolve to late de Sitter time starting from very early de Sitter time. However, it is possible to map $\varepsilon(\rho)$ to $\varepsilon_M(\tau, x_\perp)$ and check \eqref{Eq:e-large-xperp} directly without the need for extrapolating $\epsilon_M$ beyond the domain of $\tau$ and $x_\perp$ where it is directly evaluated. 

\begin{figure}[ht]
     \centering
           \includegraphics[width=0.4 \linewidth]{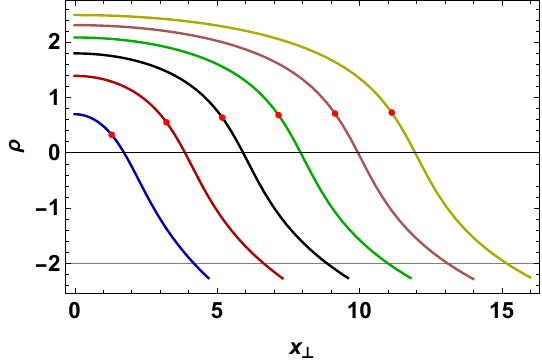}
           \includegraphics[width=0.49 \linewidth]{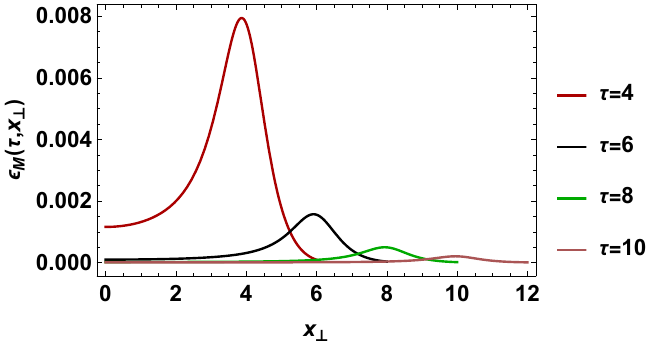}
          \includegraphics[width=0.49 \linewidth]{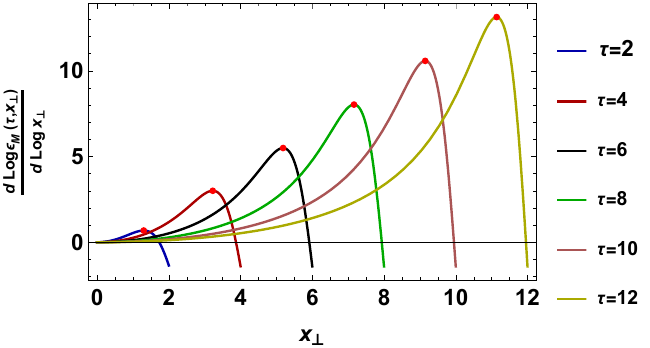}
     \caption{The behavior of $\varepsilon_M$ on the future Minkowski wedge for the simulation in Fig.~\ref{Fig:Scalar}. Top left: The range of $x_\perp$ for different $\tau$ covered by the restriction $\rho > -2$ where the simulation has been performed. Red dots on the curves indicate the peaks of $\varepsilon_M$ at the corresponding $\tau$. Top right: $\varepsilon_M$ decays with increasing $\tau$. Bottom:  The log-derivative of $\varepsilon_M(\rho)$ with respect to $x_{\perp}$ tends to $-\infty$ for large $x_\perp$ at any $\tau$ as implied by \eqref{Eq:e-large-xperp}. 
     }
     \label{fig:GaussianT}
 \end{figure}

As shown in the top left panel of Fig.~\ref{fig:GaussianT}, the restriction of the initialization time $\varrho_{\rm in}$ to $-2$ limits the maximum value of $x_\perp$ which can be accessed for a fixed value of $\tau$. However, this range extends at large $\tau$ and extends somewhat beyond $x_\perp \sim \tau$. As shown in the top right panel of Fig.~\ref{fig:GaussianT}, $\varepsilon_M(\tau, x_\perp)$ peaks around $x_\perp \sim \tau$ at large $\tau$ and then quickly decays. Furthermore, the maximum value of $\varepsilon_M$ at a fixed value of $\tau$ also decays with increasing $\tau$. Crucially, the bottom panel of Fig.~\ref{fig:GaussianT} indicates that
\begin{equation}
    \frac{\partial\ln \varepsilon_M(\tau, x_\perp)}{\partial\ln x_\perp} \rightarrow -\infty \quad {\rm as}\quad x_\perp \rightarrow \infty
\end{equation}
at any fixed value of $\tau$. We also note that the maximum value of $\partial\ln \varepsilon_M(\tau, x_\perp)/\partial\ln x_\perp$ increases with increasing $\tau$. This means that while $ \varepsilon_M$ decays with increasing $\tau$ it also gets more narrowly peaked at $x_\perp\sim\tau$. Although Fig.~\ref{fig:GaussianT} demonstrates the validity of \eqref{Eq:e-large-xperp} and thus also \eqref{Eq:e-early-tau} for the Gaussian-like initial condition \eqref{Eq:profile}, these also hold for generic smooth initial conditions such as those in \eqref{Eq:profile-1}.
 
As discussed in Sec.~\ref{Sec:Num-1}, the large $x_\perp$ and small $\tau$ behaviors stated in \eqref{Eq:e-large-xperp} and \eqref{Eq:e-early-tau} respectively can be violated even for smooth initial conditions but these are non-generic (fine tuned).

\section{Does Gubser flow hydrodynamize at intermediate time?}\label{Sec:Non-Hydro}

Having discussed the late and early de Sitter regimes of the Gubser flow explicitly, we now turn our attention to the behavior of the Gubser flow at intermediate de Sitter time, $\rho \sim 0$. We recall that in Gubser flow the medium expands along $u_d^\mu$, where \eqref{ud}. Explicitly, $\nabla\cdot u_d = \tanh \rho$ in dS$_3$ $\times$ $\mathbb{R}$. Therefore, at $\rho\sim 0$, the gradients are small and a hydrodynamic description may apply. We have discussed the generic hydrodynamic expansion of the energy density at $\rho \sim 0$ for conformal Gubser hydrodynamics in Sec.~\ref{Sec:CHydro}.

It is useful to first see what can be expected generally at $\rho\sim0$ \textit{without} assuming the validity of hydrodynamic description. As discussed before, it is necessary that $\varepsilon'(\rho =0) = 0$ for the transverse pressure \eqref{Eq:PT} and the longitudinal pressure \eqref{Eq:PL} to be finite at $\rho =0$. Clearly, there would be naked singularities in the bulk otherwise, implying that it is unphysical. Indeed, we explicitly see that $\varepsilon'(\rho =0) = 0$ in numerical simulations described in the previous section. As shown in Fig.~\ref{fig:late_time_1}, $\varepsilon'(\rho =0) = 0$ can imply that either $\varepsilon(\rho)$ has a local maxima or a local minima at $\rho = 0$, or $\rho = 0$ is an inflection point according to whether $\varepsilon''(\rho)> 0$,  $\varepsilon''(\rho)< 0$ or  $\varepsilon''(\rho) = 0$ respectively  (clearly $P_L$ and $P_T$ are finite at $\rho = 0$).
 
Furthermore, the expansion \eqref{Eq:New-e-Expansion} which is valid for all de Sitter time implies that the energy density has a Taylor expansion at $\rho =0$.  Therefore, we can expect that generally the energy density in dS$_3$ $\times$ $\mathbb{R}$ should behave as
\begin{align}\label{Eq:rho0-epsilon}
    \varepsilon(\rho) = - \frac{3l^3}{16\pi G_N} \left(\tilde{e}_0 + \tilde{e}_2 \varrho^2 + \mathcal{O}(\varrho^3)\right)
\end{align}
at $\rho \sim 0$. Above the term linear in $\rho$ vanishes since $\varepsilon'(\rho =0) = 0$. From \eqref{Eq:PT} and \eqref{Eq:PL}, it follows that
\begin{equation}
    \frac{P_T(\rho =0)}{\varepsilon(\rho =0)} = \frac{\tilde{e}_2}{\tilde{e}_0} - 1, \quad  \frac{P_L(\rho =0)}{\varepsilon(\rho =0)}= 3 - 2\frac{\tilde{e}_2}{\tilde{e}_0}.
\end{equation}
If hydrodynamics applies at $\rho = 0$, then we must have $\tilde{e}_2/\tilde{e}_0 = 4/3$, so that $P_T(\rho =0) = P_L(\rho =0)$. Note that viscous corrections are sub-dominant and do not affect $\tilde{e}_2/\tilde{e}_0$ (recall the general conformal hydrodynamic expansion of the energy density given by \eqref{Eq:hydroexp}).

To develop a phenomenological description for the Gubser flow near $\rho \sim 0$, let us denote 
\begin{equation}
    a =  \frac{P_T(\rho =0)}{\varepsilon(\rho =0)}, 
\end{equation}
so that
\begin{equation}
    \frac{\tilde{e}_2}{\tilde{e}_0} = 1 +a.
\end{equation}
Inspired by perfect fluid behavior $\varepsilon(\rho) \sim \cosh^{-8/3} \varrho$ corresponding to $a = 1/3$, we can postulate a more empirical formula
\begin{align}\label{Eq:rho0-fit}
    \varepsilon_{\rm fit}(\rho)= \varepsilon_0 \cosh^{-2-2a} \varrho,
\end{align}
which could be valid near $\rho \sim 0$ with $a$ determined by the initial conditions at an early de Sitter time. Also note that the above is consistent with \eqref{Eq:rho0-epsilon}. We do not expect that $a = 1/3$ should hold for generic initial conditions.

\begin{figure}
     \centering
          \includegraphics[width=0.49 \linewidth]{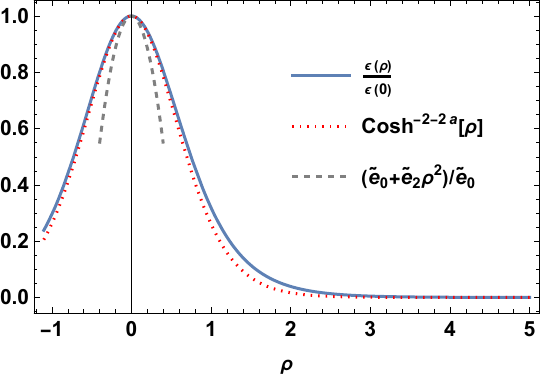}
     \includegraphics[width=0.49 \linewidth]{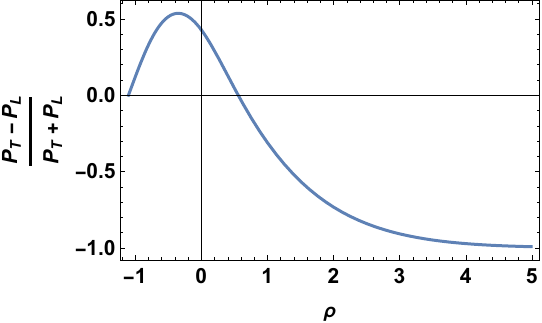}
     \caption{Left: In blue, the energy density normalized by $\varepsilon_0=\varepsilon(0)$. The grey dashed line is the Taylor expansion of the energy density to quadratic order. The red dotted line is the fit \eqref{Eq:rho0-fit} with $a\sim 0.547$. Right: Pressure anisotropy for the same system. Note that in the region $\rho \sim 0$, the anisotropy is large-ish.
     See text for initial conditions.}
     \label{fig:rho0}
 \end{figure}

Our numerical simulations demonstrate that the phenomenological ansatz \eqref{Eq:rho0-fit} is indeed valid in a range of $\rho$ around $\rho\sim 0$ to a very good approximation although it fails at large $\vert \rho\vert$. We show a typical example in Fig.~\ref{fig:rho0}. As initial data, we chose to initialize with \eqref{Eq:profile} at $\varrho_{\rm in}=-1.1$, $a_4(\varrho_{\rm in}) = -0.5$ and the profile \eqref{Eq:profile} with $\tilde{b}_{4,{\rm in}} = -\tanh{\varrho_{\rm in}}/6$. We find that although the hydrodynamic description need not be valid at $\rho \sim 0$, the emergent behavior given by \eqref{Eq:rho0-fit} is generally valid to a very good approximation in this regime \textit{even compared to a truncated power series expansion} \eqref{Eq:rho0-epsilon} (c.f. Fig.~\ref{fig:rho0}).

\begin{figure}
     \centering
          \includegraphics[width=0.46 \linewidth]{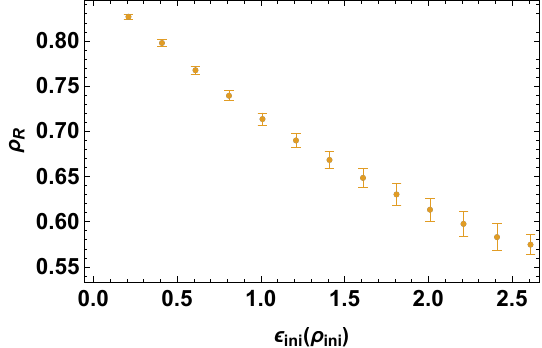}
          \includegraphics[width=0.44 \linewidth]{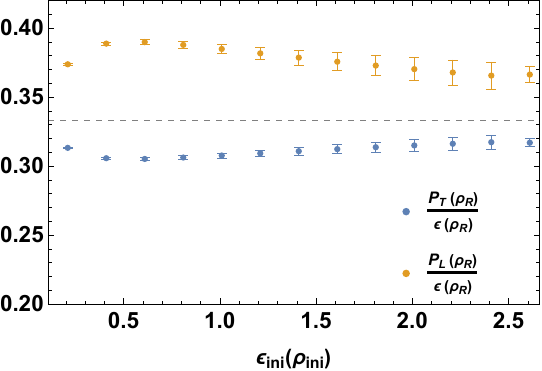}
     \caption{Here, we show the bottleneck region $C_R$ as a function of initial energy density $\varepsilon_{\rm ini}$ for initial conditions set by \eqref{Eq:profile} at $\rhoL_{\rm{in}} = -0.5$ with varying $\tilde{b}_{4,{\rm in}}$ and set $L=1$. (Note Fig. \ref{fig:late_time_2} shows $C_R$ when the initial energy density is $2$.) $C_R$ is a smeared region located approximately at $\varrho_R$ with $P_T/\varepsilon$ and $P_L/\varepsilon$ taking a narrow range of values. Left: the behaviour of the time $\rho_R$ as a function of $\varepsilon_{\rm ini}$. Right: the ratio of $P_T/\varepsilon$ and $P_L/\varepsilon$ at $\rho_R$ as a function of $\varepsilon_{\rm ini}$. The dashed horizontal line marks the $1/3$ line.}
     \label{fig:CR-region}
 \end{figure}
We find interesting emergent features at intermediate time with Gaussian-like initial conditions \eqref{Eq:profile}. As shown in Fig.~\ref{fig:late_time_1}, when the initialization time $\varrho_{\rm in}$ is negative and we initialize with a fixed positive value of the initial energy density, 
\begin{itemize}
    \item the energy densities $\varepsilon$ are ordered for different evolutions for all $\varrho > \varrho_{\rm in}$ according to the ordering of the initial value of the transverse pressure $P_T$, and 
    \item as shown in Fig.~\ref{fig:late_time_2}, both the ratios $P_T/\varepsilon$ and $P_L/\varepsilon$ evolve approximately towards a fixed point at $\varrho \sim \varrho_{R}$ (or rather the evolutions go through a very narrow bottleneck region around $\varrho \sim \varrho_{R}$ where $P_T/\varepsilon$ and $P_L/\varepsilon$ takes a very narrow range of values) with $\varrho_{R}>0$.
\end{itemize}
For Gaussian-like initial conditions \eqref{Eq:profile}, the value of $\varrho_{R}$ is set just by the initial value of the energy density when we evolve with a fixed value of the initial energy density as stated above. We readily see from Fig.~\ref{fig:rho0} that as the value of the initial energy density is increased
\begin{itemize}
    \item $\varrho_R$ approaches 0, and
    \item the narrow range of values of $P_T/\varepsilon$ and $P_L/\varepsilon$ for different evolutions at $\varrho\sim\varrho_R$ in the narrow bottleneck region approach $1/3$.
\end{itemize}
These imply that for Gaussian-like initial conditions, the behavior at $\varrho\sim 0$ becomes hydrodynamic for very large initial energy densities. 

It would be interesting to understand under which conditions the hydrodynamic behavior emerges generally at $\varrho\sim 0$, and if a large value of energy density at an early de Sitter time is always a necessary although not a sufficient condition. We leave this investigation to the future. As discussed in Sec.~\ref{Sec:Gub-jet}, this question can be relevant for understanding collective phenomenon in jets.

\section{Horizons and entropies}\label{Sec:Entropy}

Deeper insights into Gubser flow in quantum field theories can be derived from understanding irreversible entropy production. In holographic theories, these can be understood via the study of dynamical horizons, especially the event and apparent horizons. The entropy associated with a horizon is
\begin{align}\label{Eq:entropy}
    S = \frac{\Sigma_H}{4 G_N},
\end{align}
where $G_N$ is the Newton constant, and $\Sigma_H$ is the area of the spatial cross-section of the horizon.

Explicitly, if $v = v_h(\varrho)$ is the location of the bulk horizon in Gubser flow, then 
\begin{align}\label{Eq:gamma}
    \Sigma_H = \int_0^\pi {\rm d} \theta\int_0^{2\pi} {\rm d} \phi\int {\rm d}\eta \sqrt{\gamma}, \quad  \gamma = l^6 \frac{L^6e^{C(v_h(\rhoL),\rhoL )}\sin^2\theta\left(\cosh{\rhoL}+ v_h(\rhoL) \sinh{\rhoL}\right)^4}{v_h^6(\rhoL)}.
\end{align}
 Due to the infinite extent of $\eta$, it is better to define the entropy per unit rapidity which is
\begin{align}\label{Eq:ent-rap}
  \frac{dS}{ d \eta} = \pi \frac{l^3 }{G_N} \frac{L^3 e^{\frac{C(v_h(\rhoL),\rhoL )}{2}} \left(\cosh{\rhoL}+ v_h(\rhoL) \sinh{\rhoL} \right)^2}{v_h^3(\rhoL)}.
\end{align}
The spatial volume of the sphere (on which the dual field theory lives) measured in the boundary metric \eqref{Eq:dSds2} is $4\pi L^2 \cosh^2\rhoL$, and therefore the time-dependent entropy density per unit rapidity reads
\begin{align}\label{Eq:entdensity-rap}
    \frac{ds}{ d \eta} =  \frac{1}{4\pi L^2 \cosh^2\rhoL}\times \frac{1}{L}\times\frac{dS}{d \eta}.
\end{align}

As discussed before, one of the most interesting aspects of Gubser flow is that the vacuum itself has a monotonically growing entropy per unit rapidity. We will show below that despite the expansion of the boundary metric \eqref{Eq:dSds2}, the entropy density per unit rapidity reaches a constant value at late de Sitter time in Gubser flow for arbitrary initial conditions when the future wedge is mapped to dS$_3$ $\times$ $\mathbb{R}$. 
\begin{figure} [h!]
     \centering
          \includegraphics[width=0.48 \linewidth]{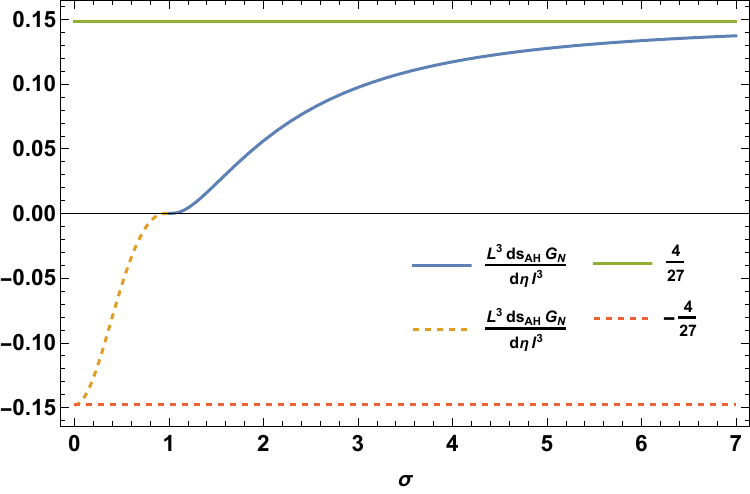}
     \includegraphics[width=0.462 \linewidth]{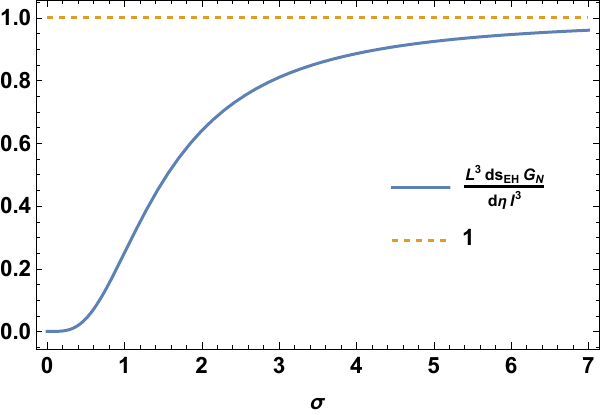}
     \caption{The apparent (left panel) and event (right panel) horizon entropy densities per unit rapidity in the vacuum as function of $ \sigma = e^\rhoL$ computed analytically.}
     \label{fig:vacentropy}
 \end{figure}

\subsection{Event horizon and its entropy}
The location of the event horizon $\rL_{\rm EH}(\rhoL)$ can be determined by the radial null geodesic equation
\begin{align}\label{Eq:geodesic}
    \frac{{\rm d} v_{{\rm EH}}(\rhoL)}{{\rm d} \rhoL } + \frac{1 - v_{{\rm EH}}^2 (\rhoL)+A(v_{{\rm EH}}(\rhoL),\rhoL)}{2} = 0
\end{align}
with the condition that $v_{\rm EH}(\rhoL = \infty) =1$, since the horizon coincides with that of the solution dual to the vacuum \eqref{Eq:Vac-metric} in the limit $\rhoL\rightarrow\infty$. In the state dual to the vacuum, the event horizon is at $v_{\rm EH}(\rhoL) =1$ and has a constant Hawking temperature $(2\pi L)^{-1}$ although it is not a Killing horizon.

In the state dual to the vacuum, the entropy density per unit rapidity associated with the event horizon is (see also Fig.~\ref{fig:vacentropy})
\begin{align}\label{Eq:vaceng}
   \frac{G_N}{l^3} L^3 \frac{d s_{\rm EH}}{ d \eta} &= \frac{1}{4 \cosh^2\rhoL}  \left(\cosh \rhoL + \sinh\rhoL \right)^2\nonumber\\
    &=\frac{1}{4} \left(1+ \tanh \rhoL  \right)^2.
\end{align}
Clearly, this entropy requires an algebraic understanding as the metric is itself locally AdS$_5$. As mentioned before, the bulk metric dual to the vacuum \eqref{Eq:Vac-metric} breaks the $\rho\rightarrow -\rho$ symmetry of the boundary metric. This is why a monotonically growing entropy can exist. It is interesting to compare with the case of the \textit{topological black hole} of Casini, Huerta and Myers \cite{Casini:2011kv} which is also locally AdS but covers only a causal wedge at the boundary. Therefore its entropy can be used to compute the entanglement and Renyi entropies of ball shaped regions at the boundary. It can be expected that the entropy \eqref{Eq:vaceng} can be associated with the observers on the future wedge co-moving with de Sitter time. It has also been shown in \cite{Maldacena:2012xp} that the vacuum entanglement entropy in de Sitter space can scale with the spatial volume.

\begin{figure} [ht]
     \centering
     \includegraphics[width=0.67 \linewidth]{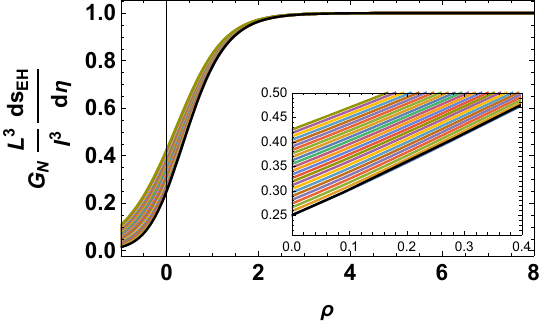}
     \caption{The numerically computed event horizon entropy density per unit rapidity for evolutions of the type shown in Fig. \ref{fig:late_time_1} with the corresponding color codes to the ordering of $\tilde{b}_{4,{\rm in}}$. The initial energy density is set to $0.2$ and initialization time is $\rhoL_{\rm{in}} = -1$. The bold black line denotes the event horizon entropy density per unit rapidity associated with the vacuum state. }
     \label{fig:entropy}
 \end{figure}
 
The bulk dual of a generic holographic Gubser flow admits the late-time expansion \eqref{Eq:AnsGrav1}, and thus the location of the event horizon given by \eqref{Eq:geodesic} admits a perturbative solution with the condition $v_{\rm EH}(\rhoL = \infty) =1$. Explicitly, the perturbative expansion of $v_{\rm EH}$ at late de Sitter time is
\begin{align}
    v_{{\rm EH}}(\rhoL) = 1 + \sum_{m=0}^\infty v_{{\rm EH},m} e^{- (2 + 2m)\rhoL}.
\end{align}
Using the explicit perturbative solution for the metric given by \eqref{Eq:a0b0c0}-\eqref{Eq:a3b3c3}, we obtain
\begin{align}
    v_{{\rm EH},0} = e_0 (\ln 8 - 2), 
\end{align}
and therefore
\begin{align}
 L^3  \frac{G_N}{l^3} \frac{ds}{d \eta} = 1 - \left(2+ e_0\left(\frac{9}{4} -\ln 8 \right) \right) e^{- 2 \rhoL} + \ldots
\end{align}

If $e_0$ is positive, then the energy density at leading order is negative, and $ds/d\eta$ is less than in the vacuum at late de Sitter time. On the other hand if $e_0$ is negative, then the energy density at leading order is positive, and $ds/d\eta$ is larger than that in the vacuum at late de Sitter time. 

From the numerical solutions described in Sec.~\ref{Sec:Num}, we can also compute the time-dependent event horizon entropy explicitly by solving \eqref{Eq:geodesic} numerically. The event horizon entropy density per unit rapidity grows monotonically and eventually approaches the vacuum value as shown in Fig.~\ref{fig:entropy}. Note this is non-trivial as the entropy density need not be monotonic although the total entropy could be so when the spatial volume (of S$^2$) is expanding.  Furthermore, for solutions where the energy density always remains positive, the event horizon entropy density per unit rapidity is always greater than that of the vacuum as demonstrated in Fig.~\ref{fig:entropy}.

\subsection{Apparent horizon and its entropy}

The apparent horizon is a null hypersurface separating trapped and untrapped regions of spacetime with a vanishing expansion along outgoing null geodesics.  In case of the Gubser flow particularly, the apparent horizon is of the form $v= v_{\rm AH}(\varrho)$ by virtue of the symmetries. It satisfies
\begin{equation}\label{Eq:Apphor}
    d_+ \ln \sqrt{\gamma} = 0,
\end{equation}
with $d_+$ being the derivative along the outgoing null geodesic defined in \eqref{Eq:directional_derivative}) and $\gamma$ is the spatial volume element \eqref{Eq:gamma}.

In the bulk metric \eqref{Eq:Vac-metric} dual to the vacuum, the apparent horizon satisfies the cubic equation
\begin{equation}\label{Eq:AppV2}
    v_{\rm AH} \left(v_{\rm AH}^2-5\right) \sinh{\rhoL}-\left(v_{\rm AH}^2+3\right) \cosh{\rhoL}= 0.
\end{equation}
This equation has a single real solution. We note that the equation is symmetric under $v_{\rm AH}\rightarrow - v_{\rm AH}$ and $\rhoL \rightarrow - \rhoL$. It follows that $v_{\rm AH}(-\varrho) = - v_{\rm AH}(\varrho)$. It is easy to see that in the limit $\varrho\rightarrow \infty$, $v_{AH}\rightarrow 3$ while in the limit  $\varrho\rightarrow 0_+$, $v_{AH}\rightarrow \infty$. Thus the apparent horizon moves from $\infty$ to 3 as $\varrho$ is taken from to 0 to $\infty$, and for negative $\varrho$ it exists only on the unphysical sheet. In the physical sheet, the apparent horizon exists only for $\varrho >0$, and as expected is enclosed by the event horizon $v_{\rm EH}(\varrho) = 1$.

The apparent horizon entropy density per unit rapidity for the vacuum state of Gubser flow can be readily computed from \eqref{Eq:ent-rap} and \eqref{Eq:entdensity-rap} (with $v_h(\varrho)$ replaced by $v_{\rm AH}(\varrho)$). The analytic expression is rather long. We have plotted it in Fig.~\ref{fig:vacentropy}. We see that the apparent horizon entropy density per unit rapidity for the vacuum state increases from 0 to a constant value $4/27$ as $\varrho$ increases from $0$ to $\infty$. If we continue the apparent horizon to the unphysical sheet, then the apparent horizon entropy per unit rapidity is negative and in fact odd under $\varrho \rightarrow -\varrho$. The apparent horizon entropy density per unit rapidity for the vacuum state for $\varrho <0$ has been shown in dotted orange line in Fig. \ref{fig:vacentropy}.

At late de Sitter time, the apparent horizon entropy density per unit rapidity for the vacuum state is
\begin{align}
 L^3  \frac{G_N}{l^3}  \frac{ds}{ d \eta} =  \frac{4}{27} - \frac{5}{9} e^{-2 \rhoL} + \ldots .
\end{align}

 \begin{figure} [ht]
     \centering
     \includegraphics[width=0.67 \linewidth]{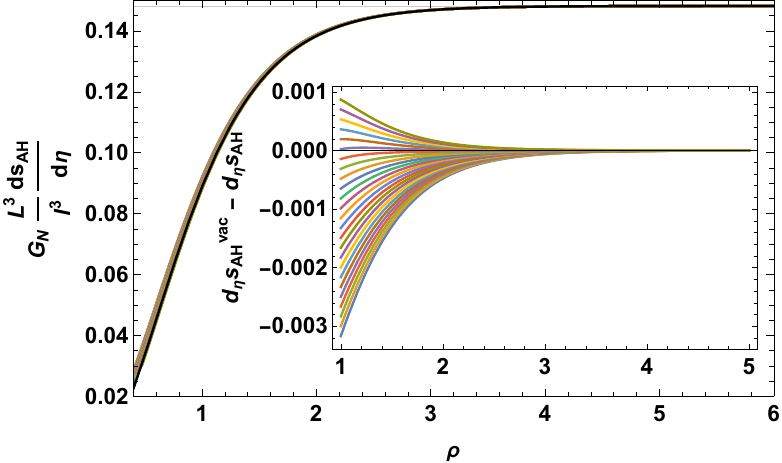}
     \caption{The numerically computed apparent horizon entropy density per unit rapidity for evolutions of the type shown in Fig.~\ref{fig:late_time_1} with the color codes corresponding to the ordering of $\tilde{b}_{4,{\rm in}}$. The initial energy density is set to $0.01$ at an initialization time of $\rhoL_{\rm{in}} = -0.001$. The bold black line denotes the apparent horizon entropy density per unit rapidity associated with the vacuum state. The inset shows the deviation of the apparent horizon entropy density per unit rapidity from the vacuum state.}
     \label{fig:entropy2}
 \end{figure}

The apparent horizon entropy clearly requires a distinct algebraic understanding from the entropy associated with the event horizon. The latter entropy density is always positive and larger than the former. Furthermore, as discussed in \cite{Chandrasekaran:2022cip}, the entropy in de Sitter space can be negative. Since there is non-trivial time-dependent entropy associated with the Gubser flow in the vacuum state itself, it provides an opportunity for concrete understanding the physical meanings of both these entropies. For related discussions, see \cite{Engelhardt:2017aux,Joshi:2017ump,Jejjala:2023zxw}.

For a generic state exhibiting Gubser flow, the apparent horizon can be computed in the perturbative late de Sitter time expansion (as in the case of the event horizon) which is given by
\begin{align}
  v_{Ah}(\rhoL) = 3 + \frac{3}{2} \left( 1 + \frac{3}{2} e_0 (8 \ln 2 - 3)\right) e^{-2 \rhoL} + \ldots.
\end{align}
The late de Sitter time expansion of the apparent horizon entropy density per unit rapidity turns out to be
\begin{align}
 L^3  \frac{G_N}{l^3} \frac{ds}{d \eta} = \frac{4}{27} - \left(\frac{5}{9} + e_0 \frac{(39 -32 \ln 2)}{36} \right)e^{-2 \rhoL}  + \ldots
\end{align}
As in the case of the event horizon, we find that the entropy density per unit rapidity is larger than the vacuum when $e_0 <0$ (i.e. for positive energy density) and smaller than the vacuum when $e_0 >0$ (i.e. for negative energy density).

From the numerical solutions described in Sec.~\ref{Sec:Num}, we can also compute the time-dependent apparent horizon entropy explicitly by solving \eqref{Eq:Apphor} numerically. The apparent horizon entropy density per unit rapidity also grows monotonically as shown in Fig.~\ref{fig:entropy2}. (Recall that this is non-trivial as the entropy density need not be monotonic, although the total entropy could be so when the spatial volume (of S$^2$) is expanding.) However, unlike the case of the event horizon, the apparent horizon entropy density per unit rapidity can be smaller than that of the vacuum at early de Sitter time even if the energy density is positive as shown in Fig.~\ref{fig:entropy2}.

\section{Discussion}\label{Sec:Disc}

Conformal holographic Gubser flow provides a concrete instance of the gauge-gravity duality for a non-trivial time-dependent evolution which does not thermalize and exhibits novel universal behavior in certain domains irrespective of the initial conditions. The latter is attained  in the central region of the flow and at late proper time in the future wedge, corresponding to free-streaming in the transverse directions. Furthermore, the sub-leading corrections in this domain also have universal characteristics. The most remarkable property of the conformal holographic Gubser flow in the large N limit, which we have established in this work, is that it is self-supported on the future wedge, meaning that it can be glued to the vacuum state outside of the future wedge and cannot be generated from non-trivial initial states like colliding shock waves. 

It is important to understand which operators can create a Gubser flow from the vacuum. For this purpose, we can readily use our perturbative solutions which are valid in the full future wedge in large N holographic theories. In fact, for a first understanding we can consider the simpler case of the bulk massless scalar field satisfying the linear Klein-Gordon equation and SO(3) $\times$ SO(1,1) $\times$ $\mathbb{Z}_2$ symmetries as described here for investigating the time-dependent modular Hamiltonian of the dual state on the lines of \cite{Hamilton:2005ju,Hamilton:2006az}, and how the latter behaves in the limit of vanishing proper time ($\tau\rightarrow 0$). Although the v.e.v. of the marginal operator and all its derivatives w.r.t. to the proper time vanishes in this case, the modular Hamiltonian should be non-trivial in this limit. Therefore, the creation of a Gubser flow would typically require a very non-trivial large operator smeared over the transverse plane ($z=0$) at $\tau\rightarrow 0$. The construction of the modular Hamiltonian in some simple examples should give a clear understanding of the operators which can create such flows in the large N limit. 

An explicit understanding of the bulk reconstruction (see \cite{Harlow:2018fse,Kibe:2021gtw} for reviews) in holographic conformal Gubser flow in terms of operators of the dual gauge theory also opens new doors to a deeper understanding of some non-perturbative aspects of quantum field theories. Clearly the Gubser flow is a non-trivial generalization of a decaying particle state which is strictly not part of the in-states of a quantum field theory. More specifically, it is a transient phenomenon embedded within a vacuum to vacuum transition process. Therefore, its understanding within the framework of algebraic quantum field theory can lead to a generalized notion of S-matrix in a non-perturbative way in quantum field theories. Furthermore, as discussed in Sec.~\ref{Sec:Entropy}, the Gubser flow provides a great opportunity to understanding the entropy associated with event and apparent horizons in terms of algebraic quantum field theory. 

It would be also interesting to understand if the Gubser flow can lead to a more fundamental non-perturbative understanding of jet-like phenomena with collective behavior. As discussed in Sec.~\ref{Sec:Gub-jet}, if the Gubser flow holds as an approximation in part of the medium instead of the whole, it can be embedded within more non-trivial evolution (such as those generated from scattering of shock waves) and mimicking jets within the medium. Furthermore, as pointed out in Sec.~\ref{Sec:Gub-jet} and Sec.~\ref{Sec:Non-Hydro}, the late proper time behavior in such cases could be closer to usual hydrodynamics especially if there is sufficient energy flux in the jet. It would be interesting to construct such bulk solutions explicitly by embedding the Gubser flow within non-trivial flows like Bjorken flow which can be generated from shock wave scattering. The decoding of such solutions in the dual field theory should lead us to understand how localized operators can create Gubser flow within a part of the medium and give rise to jet-like behavior. In such cases, these types of jets could be understood as novel transient phenomena embedded within non-trivial transitions between asymptotic states.

The study of how Gubser flow can be embedded within a non-trivial evolving medium undergoing hydrodynamization would be also important for phenomenological understanding of jets in medium and collective flow within the jets. Our present results, in any case, strongly suggest that the Gubser flow should not be applied to understand the collective behavior of the full medium formed in heavy-ion collisions but be perhaps applied to some types of jets.

It would also be interesting to understand how Gubser flow behaves in holographic non-conformal and confining gauge theories \cite{Craps:2013iaa,Gursoy:2015nza,Attems:2016ugt,Attems:2016tby}, and also in semi-holographic theories which can take into account both weakly interacting ultraviolet and strongly interacting infrared degrees of freedom \cite{Iancu:2014ava,Banerjee:2017ozx,Kurkela:2018dku,Ecker:2018ucc, Mondkar:2021qsf}\footnote{In this regard, we can note that semi-holographic hydrodynamic attractors which have been studied in \cite{Mitra:2020mei,Mitra:2022xtb} show that the initial conditions on the attractor surface and also the generic initial behavior in phenomenologically relevant evolutions are determined by the perturbative weakly coupled sector. Therefore, it would be interesting to understand the initial conditions which can generate a semi-holographic Gubser flow.}. In particular, we would like to understand (i) if the universal behavior at late de Sitter time (in the central region and at late proper time on the future wedge) can reveal directly the mechanisms of confinement and fundamental aspects of the strong interactions in the infrared, and (ii) if the Gubser flow can be generally smoothly glued to the vacuum. Furthermore, the investigation of jet-like behavior which can be described by Gubser flow in these setups can {provide insight into} the phenomenology of jets in heavy-ion collisions.

We also plan to extend the method for determining out-of-equilibrium holographic Schwinger-Keldysh correlation functions of Bjorken flow \cite{Banerjee:2022aub} via the horizon cap approach to the Gubser flow. This would be crucial for a deeper understanding of the properties of the universal transverse free-streaming phase of the holographic Gubser flow, particularly in terms of how the usual thermal fluctuation-dissipation relations generalize in this phase. Computations of the holographic entanglement and Renyi entropies, (non-)saturation of the quantum null energy conditions, etc.~should also give new insights into understanding how quantum thermodynamics\footnote{It has been shown the quantum null energy condition can be used to understand quantum thermodynamics in relativistic field theories \cite{Kibe:2021qjy,Banerjee:2022dgv}.} can imply that such novel non-hydrodynamic phases are attained irrespective of the initial conditions in presence of special symmetries such as SO(1,1) $\times$ SO(3) $\times$ $\mathbb{Z}_2$ symmetries of the Gubser flow.

Finally, we note that our result that the Gubser flow in the future wedge can be glued smoothly to vacuum in the large N limit can be used also a toy model for understanding reconstruction of black hole interiors. Particularly, it would be interesting to go beyond the large N limit and understand in concrete terms whether the dynamics within the future wedge can be reconstructed by asymptotic observers due to formation of \textit{islands} (see \cite{Raju:2020smc,Kibe:2021gtw} for reviews).

\begin{acknowledgments}
We would like to thank Jorge Casalderrey-Solana, Suresh Govindarajan, Sa\v so Grozdanov, David Mateos, Ren\'e Meyer, Giuseppe Policastro and  Javier Subils for discussions. T.M. has been supported by an appointment to the JRG Program at the APCTP through the Science and Technology Promotion Fund and Lottery Fund of the Korean Government, by the Korean Local Governments -- Gyeong\-sang\-buk-do Province and Pohang City -- and by the National Research Foundation of Korea (NRF) funded by the Korean government (MSIT) (grant number 2021R1A2C1010834). A.M.  acknowledges support from Fondecyt grant 1240955. A.S. has been supported by funding from Horizon Europe research and innovation programme under the Marie Skłodowska-Curie grant agreement No.~101103006 and the project N1-0245 of Slovenian Research Agency (ARIS). 
\end{acknowledgments}

\appendix

\section{Demonstration of numerical convergence}\label{app:convergence}

\begin{figure}[ht]
     \centering
          \includegraphics[width=0.49 \linewidth]{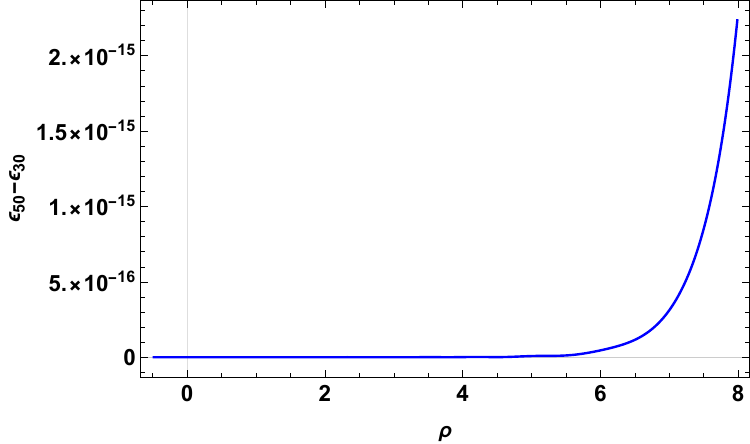}
            \includegraphics[width=0.49 \linewidth]{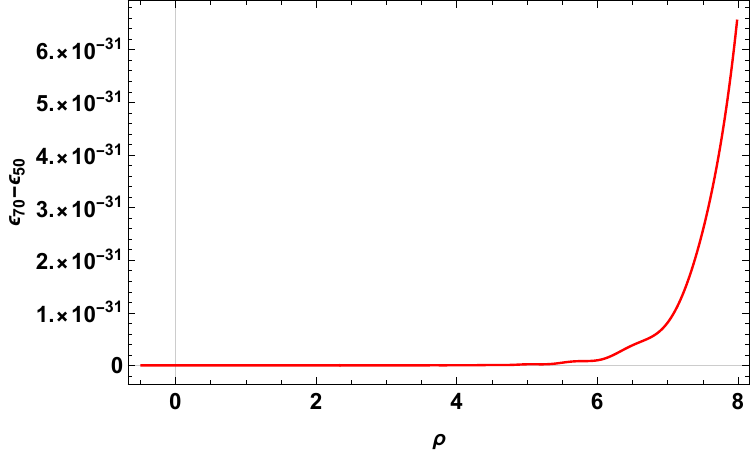}
     \caption{This figure illustrates the convergence of numerically evaluated physical quantities when the number of Chebyshev grid points is increased, maintaining the same timestep. Left: the difference in energy density computed using 50 and 30 gridpoints for the Chebyshev grid. Right: the same as in the left with 70 and 50 grid points instead.}
     \label{fig:conv}
 \end{figure}

We have verified that various numerically evaluated quantities such as energy density, pressures, and areas of apparent and event horizons converge with the increase in the number of Chebyshev grid points for a fixed value of timestep. This is demonstrated for the energy density in Fig.~\ref{fig:conv}. We have also verified that the constraint equation \eqref{Eq:charac5} is satisfied to better accuracy as the number of grid points is increased, keeping the timestep the same.

\bibliographystyle{jhep}
\bibliography{gubser}

\end{document}